\documentclass[preprint2]{aastex}

\newcommand{\myemail}{adam.ruzicka@univie.ac.at}

\usepackage{amsbsy}

\begin{document}


\title{Rotation of the Milky Way and the formation of the Magellanic Stream}

\author{Adam R\accent23u\v{z}i\v{c}ka\altaffilmark{1} and Christian Theis\altaffilmark{2}}
\affil{Institut f\"{u}r Astronomie der Universit\"{a}t Wien, T\"{u}rkenschanzstrasse 17, A-1180 Wien, Austria}
\email{\myemail}
\and
\author{Jan Palou\v{s}}
\affil{Astronomical Institute of the Academy of Sciences of the Czech Republic, Bo\v{c}n\'{i} II 1401, 14131 Praha}
\altaffiltext{1}{also Astronomical Institute of the Academy of Sciences of the Czech Republic, Bo\v{c}n\'{i} II 1401, 14131 Praha}
\altaffiltext{2}{present address: Planetarium Mannheim, Wilhelm--Varnholt--Allee 1 (Europaplatz), 68165 Mannheim, Germany}

\begin{abstract}
We studied the impact of the revisited values for the LSR circular velocity
of the Milky Way (Reid et al. 2004) on the formation of the Magellanic Stream.
The LSR circular velocity was varied within its observational uncertainties as a free parameter of the interaction
between the Large (LMC) and the Small (SMC) Magellanic Clouds and the Galaxy.
We have shown that the large--scale morphology and kinematics
of the Magellanic Stream may be reproduced as tidal features, assuming the recent values of the proper motions
of the Magellanic Clouds (Kallivayalil et al. 2006).
Automated exploration of the entire parameter space for the interaction was performed to
identify all parameter combinations that allow for modeling the Magellanic Stream.
Satisfactory models exist for the dynamical mass of the
Milky Way within a wide range of $0.6\cdot10^{12}$\,M$_\odot$ to $3.0\cdot10^{12}$\,M$_\odot$
and over the entire 1--$\sigma$ errors of the proper motions of the Clouds.
However, the successful models share a common interaction scenario.
The Magellanic Clouds are satellites of the Milky Way, and in all cases two close
LMC--SMC encounters occurred within the last 4\,Gyr at $t<-2.5$\,Gyr
and $t \approx-150$\,Myr, triggering the formation of the Stream and of the Magellanic Bridge, respectively.
The latter encounter is
encoded in the observed proper motions and inevitable in any model of the interaction.
We conclude that the tidal origin of the Magellanic Stream implies the introduced
LMC/SMC orbital history, unless the parameters of the interaction are revised substantially.
\end{abstract}


\section{Introduction}\label{intro}
Modeling the evolution of the Magellanic Clouds and their interaction with the Milky Way is a challenging task. We have to deal
with a unique system of nearby galaxies which have been subject to various detailed observational surveys.
The resulting amount of data has established an extended set of constraints for every theoretical study of the
formation and evolution of the Magellanic Clouds. On the other hand, namely the information about the kinematics and the morphology of the large--scale
structures associated with the Clouds have proved useful to reduce significantly the parameter space of the interaction between
the Large Magellanic Cloud (LMC), the Small Magellanic Cloud (SMC) and the Milky Way.

Undoubtedly, the spatial velocities of the Magellanic Clouds are the most critical input parameters for models of
their interaction~\citep{Ruzicka09}. Regarding the difficulties accompanying the observational
measurements of the proper motions of the LMC and the SMC, it is not surprising that the results have always suffered from
large uncertainties~\citep[see, e.g.][]{Kroupa97, Pedreros02}. Thus, the most important parameters were also the most uncertain ones.

A common feature of models of the Magellanic Clouds (see Section~\ref{old_models})
are bound orbits around the Galaxy, with several revolutions over the Hubble time.
However, recent proper motion measurements by~\cite{KalliLMC, KalliSMC} have introduced entirely surprising results.
Besides a substantial reduction of the errors, they claim significantly increased galactocentric velocities.
The subsequent detailed analysis by~\cite{Besla07}
has revealed a new value of the LMC galactocentric velocity $\upsilon_\mathrm{LMC}\approx 350$\,km\,s$^{-1}$, which is close to the local
escape velocity. Thus, the scenario with several perigalactic approaches has become rather unlikely.

Through a detailed automated exploration of the parameter space,~\cite{Ruzicka09}
have shown that the new data by~\cite{KalliLMC, KalliSMC} rule out tidal stripping as the process
possibly dominating the formation of the Magellanic Stream -- a gaseous trailing tail emanating from the Magellanic Clouds,
crossing the South Galactic Pole and stretched over $\approx 100^\circ$ of the plane of sky.
\cite{Mastropietro09} later claimed that the efficiency of ram pressure
increased due to the higher relative velocity of the Clouds and the hot halo gas. This might compensate
for the shorter timescale for the interaction, and allow for the creation of the Magellanic Stream of a spatial extent comparable
to the observations.

The analysis of the orbital history of the LMC by~\cite{Shattow09}
allowing for the modified values of the LSR
circular velocity $\Theta_0$ and the LSR angular rotation rate $\Omega_0$~\citep{Reid04} has introduced promising results.
The combination of changes to both, the kinematics and dynamics of the interaction,
may result in the decrease of the total energy of
the LMC--Milky Way pair, and keep the LMC within the tidal radius of the Galaxy for several Gyrs.

Following the findings by~\cite{Shattow09}, we have investigated the dynamical evolution of the Magellanic Clouds and their interaction with the Galaxy in terms
of a fully automated exploration of the parameter space for the interaction.
The goal of this study is to show whether
the up--to--date observational estimates of the LMC/SMC proper motions together with the revisited values for
the solar galactocentric motion and distance allow for the reproduction of the large--scale distribution
of neutral hydrogen (H\,I) in the LMC--SMC--Milky Way system (the Magellanic System)
due to the gravitational tidal stripping.
The approach applied was introduced by~\cite{Ruzicka07, Ruzicka09}, and it is based on the evolutionary optimization
(genetic algorithm)
of the model input according to its ability to reproduce the H\,I observations of the Magellanic System~\citep{Bruens05}.

The procedure devised by~\cite{Ruzicka09} was followed with a couple of modifications.
The ranges of the proper motions of the Clouds were redefined in order
to accommodate not only the data by~\cite{KalliLMC, KalliSMC} but also the results by~\cite{Piatek08}
who reprocessed independently the original raw proper motion data (Figure~\ref{LMC_SMC_pm}).
The parameter space of the LMC--SMC--Milky Way interaction
was also extended by an additional dimension representing the LSR circular velocity.
A more detailed description of the studied parameter space will be provided in Section~\ref{parameters}.

The reader familiar with the papers by~\cite{Ruzicka07, Ruzicka09} may skip Section~\ref{method}
which introduces the technical details of our numerical model and the principles of the automated exploration
of parameter spaces. In Section~\ref{solarvel} the influence of the LSR circular velocity on the interaction
between the Magellanic Clouds and the Galaxy
is studied. Section~\ref{results} offers our results which are later summarized in Section~\ref{summary},
followed by a discussion in Section~\ref{discussion}. Finally, the most challenging
open questions related to the formation of the large--scale Magellanic structures are mentioned in
Section~\ref{openquestions}.

\section{Models of the Magellanic System}\label{old_models}
It has been a common practice in the theoretical studies of
the Magellanic System to adopt several reasonable assumptions on the orbital paths of the Clouds
to reduce the volume of the current LMC/SMC velocity space.
In most cases the Magellanic Stream has been used as a tracer of the past orbits of the Magellanic
Clouds~\citep[e.g.][]{Fujimoto76, Gardiner94, Connors06}.

This approach has allowed for remarkably good models of the large--scale distribution of H\,I in the Magellanic System.
\cite{Connors06} devised a complex tidal model of the interaction between the Clouds and the Milky Way offering an impressive
reproduction of the kinematics of H\,I in the Magellanic Stream.
An over--density was present at the tip of the Stream, however,
demonstrating the common problem of the tidal stripping.

Alternatively, ansatz explaining the origin of the Stream by the ram pressure
stripping of the LMC/SMC neutral hydrogen has been introduced as an explanation. The advanced
numerical models by~\cite{Bekki05} and~\cite{Mastropietro05} simulated the distribution of H\,I at the main body of the LMC and the global
morphology of the Magellanic Stream. \cite{Bekki05} described the LMC as a system of self--gravitating
particles. To account for the dissipative processes in the gaseous medium, the method of sticky particles was employed.
The SPH code by~\cite{Mastropietro05} allowed the redistribution of H\,I in the Clouds to occur
at timescales of the order $10^8$\,yr only.

\section{Parameter space of the Magellanic System}\label{parameters}

Various aspects of modeling observed interacting galaxies were discussed to a great
detail in the studies by~\cite{Theis99} or~\cite{Theis01}. Obviously, even
an interaction of two galaxies described in a strongly simplified manner
involves well over 10 parameters necessary to describe the structure, the total mass and the mass distribution
of both galaxies, and also their initial positions and velocities. In principle,
the values of the parameters may be derived from observational data.
However, regarding distances of galaxies, acquiring observational data of
the resolution and variety sufficient to uniquely determine the parameters of their interactions
is very difficult. Unfortunately, this is true even for the satellites of the Milky Way.
Thus, modeling interacting galaxies involves inevitably an exploration of the
parameter space for the interaction to determine the parameter combinations, i.e. the evolutionary scenarios,
compatible with the observed kinematics and morphology of the galactic system.

\subsection{Key parameters}\label{key_parameters}
The number of parameters required to describe the interaction involving the Milky Way, the LMC and the SMC
depends on the physical model adopted. We have employed a restricted N--body numerical code assuming
the gravitational interaction only. Even such a simplified view of the interaction involves
$\approx 25$ independent parameters including the initial conditions of the LMC and the SMC motion, their total masses,
parameters of mass distributions, particle disk radii, and orientation angles, and also
the parameters defining the gravitational potential of the Milky Way, i.e. the flattening parameter of the dark matter (DM) halo,
the LSR circular velocity and the LSR angular rotation rate (see Section~\ref{solarvel}).
Some of the parameters were constrained by theoretical studies (including scale radii $\epsilon$ of the LMC/SMC halos,
the Coulomb logarithm $\Lambda$ for the dynamical friction in the Milky Way halo, and the halo flattening
parameter $q$ for the model of the gravitational potential of the Galaxy). Their mean values
and searched errors were discussed in~\cite{Ruzicka07}. This section introduces the observationally estimated
parameters of the LMC--SMC--Galaxy interaction (see also Table~\ref{table_1}).

It has been shown by~\cite{Ruzicka09} that the current proper motions of the Magellanic
Clouds are the most critical parameters of their interaction with the Milky Way. Figure~\ref{LMC_SMC_pm}
shows the proper motions of the LMC and the SMC as estimated by the latest analysis
by~\cite{KalliLMC, KalliSMC} and~\cite{Piatek08} based on the measurements by the HST. 
For convenience, the LMC/SMC proper motion vectors are decomposed into two components related to the local directions
of west ($\mu_\mathrm{W}$) and north ($\mu_\mathrm{N}$). The reason for involving two results
derived from a single data--set is the discrepancy between the studies by~\cite{KalliSMC} and~\cite{Piatek08}
regarding the SMC western proper motion $\mu_\mathrm{W}$. The resulting proper motion space for
the Magellanic Clouds explored in this study is a union of the sets defined by the 1--$\sigma$ errors
of the data by~\cite{KalliLMC, KalliSMC} and~\cite{Piatek08} for the LMC and the SMC, respectively
(Figure~\ref{LMC_SMC_pm}):
\begin{eqnarray} 
\mu_\mathrm{W}^\mathrm{lmc} & = & \langle -2.11,-1.92 \rangle\,\mathrm{mas\,yr^{-1}} \nonumber \\  
\mu_\mathrm{N}^\mathrm{lmc} & = & \langle +0.39,+0.49 \rangle\,\mathrm{mas\,yr^{-1}} \nonumber \\
\mu_\mathrm{W}^\mathrm{smc} & = & \langle -1.34,-0.69 \rangle\,\mathrm{mas\,yr^{-1}} \nonumber \\ 
\mu_\mathrm{N}^\mathrm{smc} & = & \langle -1.35,-0.99 \rangle\,\mathrm{mas\,yr^{-1}}.
\label{hires_pm} 
\end{eqnarray}

Detailed analysis by~\cite{Ruzicka09} has quantified the sensitivity of the dynamical
evolution of the Magellanic System to the variation in different free parameters. Except the LMC/SMC current
spatial velocities and positions, the remaining parameters had a rather weak impact on the
interaction. As long as the parameters are independent adding another parameters to the original set
does not change the conclusion by~\cite{Ruzicka09} regarding their importance.
However, once parameter independence cannot be guaranteed,
it is a good practice to keep the dimensionality of the parameter space as high as possible. Therefore,
we will introduce the LSR circular velocity as a new dimension of the Magellanic parameter space
established in the way similar to~\cite{Ruzicka07, Ruzicka09}.

Unlike the proper motion of the Clouds their LSR radial velocities could be measured with high accuracy.
Following~\cite{vandermarel02} we set \(\upsilon_\mathrm{rad}^\mathrm{lmc}=262.2\pm 3.4\)\,km\,s\(^{-1}\). The
SMC radial velocity error was estimated by~\cite{Harris06} as
$\upsilon_\mathrm{rad}^\mathrm{smc}=146.0\pm 0.60$\,km\,s$^{-1}$.

The heliocentric position vector of the LMC was adopted from~\cite{vandermarel02}, i.e.
the equatorial coordinates are
$(\alpha_\mathrm{lmc}, \delta_\mathrm{lmc})=(81.90^\circ\pm 0.98^\circ, -69.87^\circ\pm 0.41^\circ)$,
its distance modulus is ${(m-M)}_\mathrm{lmc}=18.5\pm 0.1$.
The equatorial coordinates of the SMC were set to the ranges
$(\alpha_\mathrm{smc}, \delta_\mathrm{smc})=(13.2^\circ\pm 0.3^\circ, -72.5^\circ\pm 0.3^\circ)$
\citep[see][and references therein]{Stanimirovic04}.
A range of distance determinations for the SMC was provided by~\cite{Vandenbergh00}
and we used his resulting distance modulus ${(m-M)}_\mathrm{smc}=18.85\pm 0.10$.

Several observational determinations of the LMC disk plane orientation have been published
so far~\citep[see, e.g.][]{Lin95, vandermarel02}. In our parameter study the LMC inclination $i$
and position angle $p$ and their errors agree with~\cite{vandermarel02},
i.e. $i=34.7^\circ\pm 6.2^\circ$ and $p=129.9^\circ\pm 6.0^\circ$.
As the SMC misses a well defined disk, the orientation and the position angle usually refer to the
SMC "bar" defined by~\cite{Gardiner96}. Based on the estimates by~\cite{Vandenbergh00} or~\cite{Stanimirovic04}
we adopted the error ranges $i=60^\circ\pm 20^\circ$ and $p=45^\circ\pm 20^\circ$ for the SMC initial disk inclination
and position angle, respectively.

\cite{Gardiner94} analyzed the H\,I surface contour
map of the Clouds to estimate the initial LMC and SMC disk radii
$r_\mathrm{disk}^\mathrm{lmc}$ and $r_\mathrm{disk}^\mathrm{smc}$, respectively.
With the use of their work and of the study by~\cite{Bruens05} we
varied the LMC/SMC disk radii within the ranges of $10.5\pm1.5$\,kpc (LMC)
and $6.5\pm1.5$\,kpc (see Table~\ref{table_1}).
Regarding the absence of a clearly defined disk of the SMC and
possible significant mass redistribution in the Clouds during their
evolution, the results require careful treatment.

Current total masses $m_\mathrm{lmc}$ and $m_\mathrm{smc}$ follow the estimates by~\cite{Vandenbergh00}.
The masses of the Clouds are functions of time and evolve due to the LMC--SMC exchange of matter
and as a consequence of
the interaction between the Clouds and the MW. Our test--particle model does not allow for a reasonable treatment
of a time--dependent
mass loss. Therefore, the masses of the Clouds are considered constant in time and their initial values at the starting epoch
of simulations are approximated by the current LMC and SMC masses.

The gravitational potential of the Milky Way is modeled by the superposition of three static components
including the axially symmetric logarithmic potential~\citep{Binney87} of the DM halo, the Miyamoto--Nagai
potential of the Galactic disk, and the Hernquist bulge.
Since the logarithmic halo is the dominant component of our model, its flattening parameter was treated as a free parameter.
The flattening $q$ was varied within the range
of $\langle0.71,1.30\rangle$. The lower limit is based on the condition $q>1/\sqrt{2}$ required to avoid negative mass densities
in the logarithmic halo. The upper limit was set up to introduce a convenient symmetry with respect to the spherical shape ($q=1$).
The scaling velocity factor $\upsilon_0$ of the logarithmic model~\citep[more details in][]{Binney87} is a function of
the LSR circular velocity and the solar galactocentric distance.

The freedom in the values of the parameters
$q$ and $\upsilon_0$ introduced a spread in the total mass of the Galaxy
$m_\mathrm{MW} = \langle 0.59,5.90\rangle\cdot 10^{12}$\,M$_\odot$ within the radius of 250\,kpc.
The estimates of the total mass of the Milky Way exceeding $\approx 3\cdot 10^{12}$\,M$_\odot$ have
neither observational not theoretical support~\citep[see e.g.][]{Binney87, Li08}, but this fact will be taken
into account when the results are interpreted.

\subsection{Rotation curve of the Milky Way}
Unlike the studies by~\cite{Ruzicka07, Ruzicka09} we have treated the LSR circular velocity $\Theta_0$ as a free parameter.
This decision was motivated by recent findings by~\cite{Reid04}. Their measurements of the proper
motion of Sag\,A$^*$ yielded the estimate for the LSR circular velocity $\Theta_0=236\pm15$\,km\,s$^{-1}$.
At the same time, they give a revisited value of the LSR angular rotation rate $\Omega_0$ of 29.45\,km\,s$^{-1}$\,kpc$^{-1}$.
These two quantities then define the variations of the solar galactocentric distance, as $R_0 = \Theta_0 / \Omega_0$.

\cite{Reid04} performed very precise position measurements of the famous radio source Sag\,A$^*$ with
respect to two background extragalactic radio sources. The data collected over a period of 8\,years
yielded values of the apparent motion of Sag\,A$^*$. This quantity is a composition of the reflected
solar galactocentric motion and of the peculiar motion of Sag\,A$^*$ which is quite small.
The fraction of the apparent motion corresponding to Solar rotation around the Galactic Center
may be treated as a combination of the Local Standard of Rest (LSR) circular velocity $\Theta_0$ and the deviation
of the solar motion from the circular orbit~\citep{Dehnen98}. Finally,~\cite{Reid04} were able to determine the latter
which yielded an estimate for the LSR circular velocity $\Theta_0=236\pm15$\,km\,s$^{-1}$ assuming the solar distance
to the Galactic center ($R_0$) of $8.0\pm0.5$\,kpc. Notably, the uncertainty of $\Theta_0$
originating from the measurement error was only 1\,km\,s$^{-1}$. The rest appears due to the
uncertainty of $R_0$~\citep{Reid04}. Calculating the LSR angular rotation rate
$\Omega_0 \equiv \Theta_0/R_0$ yields a value with a very small uncertainty: $\Omega_0 = 29.45\pm0.15$\,km\,s$^{-1}$\,kpc$^{-1}$.
The corresponding relative error is only $0.15/29.45=5\cdot10^{-3}$ which is by one order of magnitude
below the relative error of any of the parameters examined in our study. Therefore, the LSR angular rotation rate
$\Omega_0$ will be treated as a constant in this paper, i.e. $\Omega_0=29.45$\,km\,s$^{-1}$\,kpc$^{-1}$.

Although we are already familiar with the impact of most of the discussed parameters on the behavior
of the restricted N--body model of the LMC--SMC--Milky Way interaction~\citep[see][]{Ruzicka09}, this does not apply on the
LSR circular velocity.
Even though it is a natural choice, taking the 1--$\sigma$ error box of $\Theta_0$
by~\cite{Reid04} for the range explored by the parameter study is
arguably not a good practice. We want to avoid the case of important  models being localized at either end of the studied parameter
range. Such a coincidence would then require an extension of the studied range of the LSR circular velocity followed
by another time--consuming run of the automated parameter search. The described procedure is necessary to uncover and understand the behavior
of the interacting system in such a parameter region of apparently special features. Therefore, the
range of the new input parameter $\Theta_0$ was obtained by extending the 1--$\sigma$ uncertainty of the LSR
circular velocity as published in~\cite{Reid04} by 10\,km\,s$^{-1}$, i.e. $\Theta_0 = \langle 210, 260 \rangle$\,km\,s$^{-1}$,
corresponding to the range of the Solar galactocentric distance $R_0$
of $\langle7.13,8.83\rangle$\,kpc.

\section{Method}\label{method}
This paper focuses on the gravitational interaction between the Magellanic Clouds and the Milky Way.
Although hardly any doubts exist regarding the presence of the hydrodynamical processes,
as the LMC/SMC neutral hydrogen interacts with the ambient hot gas of the Galactic halo,
the tidal stripping is intrinsically involved in every model assuming the gravitational interaction.

\subsection{Restricted N--body model}
The model itself is an advanced version of the scheme by~\cite{Gardiner94}: it is a restricted N--body (i.e. test particle)
code describing the gravitational interaction between the Galaxy and its dwarf companions. The potential of
the Milky Way is dominated by the flattened dark matter halo and dynamical friction is exerted on the Magellanic
Clouds as they move through the halo~\citep{Binney77}. The LMC and the SMC are represented by Plummer spheres, initially
surrounded by test--particle disks. For further details see~\cite{Ruzicka07, Ruzicka09}.

\subsection{Searching the parameter space}
The exploration of the parameter space for the interaction involving the Galaxy and the Magellanic Clouds
was performed by a genetic algorithm. Genetic algorithms belong to the class of evolutionary optimizers
that mimic the selection strategy of natural evolution. \cite{Holland75} was the first one who proposed the
application of genetic algorithms on optimization problems in mathematics. Recently,
the performance of genetic algorithms was studied for galaxies in interaction~\citep[see, e.g.][]{Theis99}.
\cite{Theis01} analyzed the parameter space
of two observed interacting galaxies -- NGC\,4449 and DDO\,125. Genetic algorithms turned out to be very robust tools
for such a task if the routine comparing
the observational and modeled data is appropriately defined. The approach by~\cite{Theis01} was later adopted and improved
in order to explore the interaction of the Magellanic Clouds and the Galaxy~\citep{Ruzicka07, Ruzicka09}.

The comparison between the model and
observations became more efficient by involving an explicit search for the shapes in the data~\citep{Ruzicka09}. Also
the significant system--specific features (such as a special geometry and kinematics) were taken into account,
further improving the performance of the genetic algorithm for exploration of the LMC--SMC--Galaxy interaction.
More detailed information is to be found in Section~\ref{fitness} of this paper.

\subsection{Fitness function}\label{fitness}
The automated search of the parameter space is driven by a routine comparing the modeled
and observed distributions of H\,I associated with the Magellanic System~\citep{Bruens05}.
The match is measured by the \textit{fitness function} ($f$) which is a function
of all input parameters, as every parameter set determines the resulting simulated
H\,I data--cube. Our fitness function returns a floating--point number between 0.0
(complete disagreement) and 1.0 (perfect match) and consists of four different comparisons,
including search for structures and analysis of local kinematics. For details on
the fitness function used for this study see Appendix~\ref{appendixA}.

The fitness function $f$ is the only part of the genetic algorithm that reflects the nature of the studied
problem. Although the automated evolutionary optimization is a robust method applicable on a remarkable
variety of systems, its actual performance and efficiency are critically dependent on the proper
choice for the fitness function. If the function is defined in a sensible way, the number
of parameter combinations examined by the genetic algorithm is minimized.
However, the performance of the genetic algorithm also depends on its convergence rate expressed as
$d \overline f(i)/d i$, where $i=1,2,...$ stands for the number of the actual generation
and $\overline f(i)$ is the mean value of the fitness function in the $i$--th generation. The
convergence rate is generally proportional to the ratio of the generation size $N_\mathrm{gen}$
and of the dimension $n$ of the studied parameter space over which the fitness function is defined.
Unfortunately, the value of $N_\mathrm{gen}/n$ is very low if genetic algorithm--based
optimizer is applied on the parameter spaces of interacting galaxies. The number of parameters involved
is always high (see Section~\ref{parameters}). On the other hand, the maximum generation size has to be limited in order
to keep the total computational time requested by the parameter search reasonable.

In order to overcome such a difficulty we searched the parameter space repeatedly
in a fixed number of optimization steps, i.e. generations of models, and localized $\approx 100$
high--fidelity models.
Such a procedure is not likely to reveal the global maximum of the fitness function, but
it results in a map of distribution of good models over the entire parameter space.
In principle, every region of the parameter space allowing for the satisfactory
reproduction of the large--scale structures associated with the Magellanic Clouds may be identified.

The reader might ask why there is such emphasis placed on
the fitness function itself if it, in fact, does not seem to provide any
physical information about the interacting system of the Galaxy and the Magellanic Clouds.
Indeed, the function $f$ serves
primarily as a driver to the genetic algorithm. However, the search for
good models of the observed Magellanic System is efficient only if relevant astrophysical data
are supplied as the input to $f$. As already mentioned, our study deals with detailed morphological
and kinematic information from the 21\,cm survey by~\cite{Bruens05}
and with the corresponding modeled data. The fitness function then makes a link between the observable
data and the initial state of the Magellanic System.

\section{The role of the Solar galactocentric velocity and distance}\label{solarvel}
The conclusions by~\cite{Reid04} regarding the LSR angular rotation rate $\Omega_0$ and
the LSR circular velocity $\Theta_0$ have
a significant impact on both the kinematics and the dynamics of the Magellanic System.

The dynamics of the
interaction between the Magellanic Clouds and the Galaxy is influenced by the change of the mass distribution
and of the total mass of the Milky Way.
The value of $\Omega_0$ by~\cite{Reid04} implies that the IAU standards for $\Theta_0^\mathrm{IAU}$
and for the Solar galactocentric distance $R_0^\mathrm{IAU}$ cannot hold at the same time anymore, as
\begin{displaymath}
\Omega_0^\mathrm{IAU} = \frac{\Theta_0^\mathrm{IAU}}{R_0^\mathrm{IAU}} = \frac{220\,\mathrm{km\,s^{-1}}}{8.5\,\mathrm{kpc}}=25.88\,\mathrm{km\,s^{-1}\,kpc^{-1}},
\end{displaymath}
while~\cite{Reid04} expect that $\Omega_0 = 29.45\,\mathrm{km\,s^{-1}\,kpc^{-1}}$.
Therefore, a rescaling of the rotation curve of the Galaxy occurs if the results by~\cite{Reid04} are taken into account, which
necessarily means a rescaling of the mass distribution in the Galaxy.

The current positions of the Magellanic Clouds in the phase space have been measured with respect to the
phase space position of the Sun. In order to solve the equations of motion for the interaction between the
Clouds and the Milky Way, the galactocentric positions and velocities of the LMC and the SMC are needed, i.e.
the knowledge in the LSR circular velocity $\Theta_0$ and the LSR galactocentric distance $R_0$
is necessary (see Figure~\ref{magsystem}).
Thus, it is obvious that the results by~\cite{Reid04} may influence the kinematics of the Magellanic Clouds.

\subsection{Model of the Milky Way}
The rotation curve of a galaxy 
\begin{equation}
\upsilon^2(R) = R\left|\frac{\partial\Phi_\mathrm{tot}(R,z)}{\partial R}\right|_{z=0}
\label{rot_curve}
\end{equation}
introduces a fundamental constraint on the models of its overall potential $\Phi_\mathrm{tot}(R,z)$.
In our study, the following three component model of the potential
of the Milky Way has been used:
\begin{equation}
\Phi_\mathrm{tot}(R,z) = \Phi_\mathrm{L}(R,z) + \Phi_\mathrm{MN}(R,z) + \Phi_\mathrm{B}(r),
\label{total_potential}
\end{equation}
where
\begin{equation}
\Phi_\mathrm{L}(R,z) = \frac{1}{2}\upsilon_0^2\ln\left(R_\mathrm{c}^2 + R^2 + \frac{z^2}{q^2}\right)
\label{log_potential}
\end{equation}
is the logarithmic model of the Galactic DM halo,
\begin{equation}
\Phi_\mathrm{MN}(R,z) = -\frac{GM_\mathrm{d}}{\sqrt{R^2 + \left(b + \sqrt{z^2 + c^2}\right)^2}}
\label{mn_potential}
\end{equation}
is the Miyamoto--Nagai potential of the Galactic disk with the total mass $M_\mathrm{d}$, and
\begin{equation}
\Phi_\mathrm{B}(r) = -G\frac{M_\mathrm{b}}{r + a}
\label{hern_potential}
\end{equation}
is the Hernquist formula for the spherically symmetric potential of the Milky Way bulge with the mass $M_\mathrm{b}$.

At the galactocentric distance where the Magellanic Clouds have resided, i.e. $D\gtrapprox 50$\,kpc, the
gravitational field of the Galaxy is dominated by its DM halo component. The role of the
DM halo becomes more prominent with the increasing distance to the galactic center, because $\nabla\Phi_\mathrm{L} \sim 1/r$
while $\nabla\Phi_\mathrm{MN} \sim 1/r^2$ and also $\nabla\Phi_\mathrm{B} \sim 1/r^2$.
The necessary adjustment of the Milky Way potential
$\Phi_\mathrm{tot}$ required by the Equation~(\ref{rot_curve}), if the LSR circular velocity $\Theta_0$ is changed, was achieved
by varying the velocity constant $\upsilon_0$ in Equation~(\ref{log_potential}).

The choice for the remaining parameters in the Equations~(\ref{log_potential}), (\ref{mn_potential}), and~(\ref{hern_potential})
follows~\cite{Fellhauer06} and~\cite{Ruzicka07, Ruzicka09} and so we have set $a=0.7$\,kpc, $b=6.5$\,kpc, $c=0.26$\,kpc,
$M_\mathrm{b}=3.4\,10^{10}$\,M$_\odot$, $M_\mathrm{d}=10^{11}$\,M$_\odot$, and $R_\mathrm{c}=12.0$\,kpc.

The value of the characteristic radius $R_\mathrm{c}$ determining the concentration of the Galactic halo
is five times smaller
than the present galactocentric distances of the Magellanic Clouds, and by the order of 10 smaller compared
to the typical LMC/SMC galactocentric distances over the last several Gyr. Hence, its impact on the
density profile of the region of the DM halo of the Milky Way, where the Clouds have resided, is weak.
Regarding the total mass of the Galaxy enclosed within a given radius, it is determined
by the velocity parameter $\upsilon_0$, as
\begin{equation}
m_\mathrm{MW} = \int_\mathrm{halo} \rho_\mathrm{L}(R,z,q)d^3\mathbf{r} + M_\mathrm{b} + M_\mathrm{d},
\label{loghalo_dens}
\end{equation}
where the density of the logarithmic halo of the Galaxy $\rho_\mathrm{L}(R,z) \sim \upsilon_0^2$,
while $\rho_\mathrm{L}(R,z) \sim 1/R_\mathrm{c}^2$. We did not consider the scale radius
$R_\mathrm{c}$ as a free parameter because its influence on the mass distribution and the total
mass of the DM halo of the Milky Way is significantly lower compared to the velocity parameter $\upsilon_0$.

The flattening
$q$ of the DM halo potential entered our simulations of the Magellanic System as a free parameter (see Sec.~\ref{parameters}).
The velocity parameter $\upsilon_0$ of the logarithmic halo was considered a function of the LSR circular velocity
$\Theta_0$. The corresponding formula comes out of the Equations~(\ref{rot_curve}) and~(\ref{total_potential})
after substituting for $\Phi_\mathrm{L}$, $\Phi_\mathrm{MN}$, and $\Phi_\mathrm{B}$:
\begin{eqnarray}
\upsilon_0^2 &=& \Theta_0^2\left(1 + \frac{R_\mathrm{c}^2}{R_0^2}\right) - G\left(R_\mathrm{c}^2 + R_0^2\right) \nonumber \\
&& \cdot\left[\frac{M_\mathrm{d}}{\left(R_0^2 + (b + c)^2\right)^{1.5}} + \frac{M_\mathrm{b}}{R_0(R_0 + a)^2}\right],
\label{loghalo_const}
\end{eqnarray}
where $R_0 = \Theta_0 / \Omega_0$.

The analysis of the Equation~(\ref{loghalo_const}) shows that if
\begin{equation}
a<\Theta_0/\Omega_0
\label{param_a}
\end{equation}
and
\begin{equation}
(b+c-\sqrt{1.5}R_\mathrm{c})(b+c+\sqrt{1.5}R_\mathrm{c}) < 0.5\Theta_0^2/\Omega_0^2
\label{param_bcR}
\end{equation}
then
\begin{displaymath}
\frac{\partial\upsilon_0^2}{\partial\Theta_0} > 0.
\end{displaymath}
It can be easily seen that the conditions~(\ref{param_a}) and~(\ref{param_bcR}) are always satisfied for the values
of the parameters $a$, $b$, $c$, and $R_\mathrm{c}$ assumed in our study.
As we have Equation~(\ref{loghalo_dens}) for the logarithmic halo~\citep{Binney87},
the total mass of the Milky Way enclosed within an
arbitrary radius $r$ must grow if the LSR circular velocity $\Theta_0$ is increased.

\subsection{Galactocentric positions and velocities of the Magellanic Clouds}
A convenient transformation of the position and velocity vectors of the Magellanic Clouds from the heliocentric spherical coordinates to the galactocentric
Cartesian frame (Figure~\ref{magsystem}) was derived by~\cite{vandermarel02}. Let $(l,b)$ and $(\alpha,\delta)$ be the corresponding
pair of the galactic and equatorial coordinates of the center of mass of either of the Magellanic Clouds, respectively. If the Cloud's present heliocentric distance is
$D$, its galactocentric Cartesian coordinates are
\begin{equation}
r^i = R_\odot^i + Du^i_0,\quad i=0,1,2,
\label{position_transf}
\end{equation}
where $\mathbf{R_\odot}$ is the Solar galactocentric position vector and $\mathbf{u_0}$
is the heliocentric unit position vector of the Cloud.

The heliocentric spatial velocity of an object in space is usually expressed in terms of
its proper motion and the line--of--sight systemic velocity $\upsilon_\mathrm{sys}$.
The proper motions in the directions of west and north are defined as~\citep{vandermarel02}
\begin{equation}
\mu_\mathrm{W} = -\cos\delta\frac{d\alpha}{dt},\quad \mu_\mathrm{N} = \frac{d\delta}{dt}.
\label{def_propmot}
\end{equation}
The transformation from the heliocentric velocity coordinates $\mu_\mathrm{W}$, $\mu_\mathrm{N}$, and $\upsilon_\mathrm{sys}$
to the galactocentric velocity components may be then expressed as
\begin{equation}
\upsilon^i = \upsilon^i_\odot + \upsilon_\mathrm{sys}u^i_0 + D\mu_\mathrm{W}u^i_1 + D\mu_\mathrm{N}u^i_2,
\label{velocity_transf}
\end{equation}
where
\begin{eqnarray}
\mathbf{u_0} &=& (\cos{l}\cos{b}, \sin{l}\cos{b}, \sin{b}) \nonumber \\
\mathbf{u_1} &=& -\frac{1}{\cos\delta}\frac{\partial\mathbf{u_0}}{\partial\alpha} \nonumber \\
\mathbf{u_2} &=& \frac{\partial\mathbf{u_0}}{\partial\delta},
\label{unit_vectors}
\end{eqnarray}
and $\boldsymbol{\upsilon_\odot}$ is the Solar galactocentric velocity vector.

Taking Equations~(\ref{position_transf}) and~(\ref{velocity_transf}) into account together with the fact that
$\mathbf{R_\odot} = (-\Theta_0/\Omega_0, 0, 0)$ and $\boldsymbol{\upsilon_\odot} = (0, \Theta_0, 0)$ (see Figure~\ref{magsystem}),
it is obvious that the current galactocentric positions and velocities of the Magellanic Clouds depend
on the values of the LSR circular velocity $\Theta_0$ and of the LSR angular rotation rate $\Omega_0$.
If either of these two parameters is varied, a change to the galactocentric position vectors of the Clouds occurs due to the change
of their $x$--components:
\begin{equation}
r^0 = -\Theta_0/\Omega_0 + Du^0_0.
\label{pos_x}
\end{equation}
Similarly, the galactocentric spatial velocities of the LMC and the SMC would be influenced, because their $y$--components involve
the explicit dependence on $\Theta_0$:
\begin{equation}
\upsilon^1 = \Theta_0 + \upsilon_\mathrm{sys}u^1_0 + D\mu_\mathrm{W}u^1_1 + D\mu_\mathrm{N}u^1_2.
\label{vel_y}
\end{equation}

Note that the phase space configuration of the system involving the Sun, the Magellanic Clouds
and the Galactic Center is such that $\upsilon^1 \sim \Theta_0$, while $\vert\upsilon^1\vert \sim -\Theta_0$. As a consequence,
the magnitude of the galactocentric velocity
\begin{equation}
\vert\boldsymbol\upsilon\vert^2 \sim (\Theta_0 + \upsilon_\mathrm{sys}u^1_0 + D\mu_\mathrm{W}u^1_1 + D\mu_\mathrm{N}u^1_2)^2
\label{vel_magnitude}
\end{equation}
of the LMC decreases if the LSR circular velocity is increased. Figure~\ref{lmcsmc_vel-solarvel} reveals clearly that the same conclusion holds
for the SMC.


Variation of the LSR circular velocity $\Theta_0$ and of the LSR angular rotation rate $\Omega_0$ modifies
both the magnitudes and the directions of the position and velocity vectors of the Clouds.
Specifically, increase in the LSR circular velocity
causes the reduction of the magnitude of the LMC/SMC galactocentric velocities.

\section{Results}\label{results}
The complex nature of the interaction involving the Magellanic Clouds and the Milky Way results
in a high--dimensional parameter space.
We have studied the parameter space by employing genetic algorithms as robust optimization tools characterized
by reliability and low sensitivity to local extremes~\citep{Theis01}. This method was adopted
to analyze the performance of pure tidal models for the Magellanic System.

Regarding the insufficient convergence rate of our genetic algorithm (see Section~\ref{fitness}),
the automated exploration of the parameter space was performed repeatedly and the properties
of the resulting high--fidelity models were analyzed statistically. We have identified
the regions of the parameter space where promising models of the LMC--SMC--Milky Way
interaction exist.

This paper focuses on the influence of the LSR circular velocity $\Theta_0$ on the formation
of the Magellanic Stream provided the current spatial velocities of the Magellanic Clouds respect
the recent proper motion data by~\cite{KalliLMC, KalliSMC} and by~\cite{Piatek08}.
In particular, we want to address the question of how to choose this parameter in a way
that agrees with the recent measurements by~\cite{Reid04} and allows
for the reproduction of the observed kinematics and spatial extent of H\,I in the Magellanic Stream.

\subsection{Exploration of the parameter space}\label{parameters_general}
With the use of the genetic algorithm, $\sim 10^6$ parameter combinations, i.e. individual restricted N--body
simulations, were examined in total, and $\sim 100$ sets providing the highest fitness models
were collected. In this section the features of these models will be discussed with respect
to the parameters influencing the motion of the Magellanic Clouds.

We have treated the LSR circular velocity as a free parameter in order to analyze its impact
on the tidal interaction in the Magellanic System. The automated exploration of the modified
parameter space revealed a significant qualitative change regarding the preferred proper motions
of the Clouds, compared to the work by~\cite{Ruzicka09}.

They have done a detailed analysis
of how the models of the LMC--SMC--Galaxy
interaction depend on different parameters. It was clearly shown that the interacting
system is very sensitive to the parameters defining the past positions of the Clouds'
orbits in phase space. In the work by~\cite{Ruzicka09} the set of the critical parameters involved
the current heliocentric positions and velocities of the Magellanic Clouds and, to some extent,
the flattening parameter of the Milky Way halo. However, introducing the new value of the LSR angular
rotation rate by~\cite{Reid04} and treating the LSR circular velocity as a free parameter has changed the old picture.

Figure~\ref{pm_fitness} shows the distribution of the local peaks of the fitness function $f$
(i.e. the fitness of every model identified by the genetic algorithm) over the studied ranges of the
western and northern components of the proper motions for both Magellanic Clouds.
In contrast to~\cite{Ruzicka09}, there is no apparent preference for specific values
of the proper motions for either of the Clouds. Models of a very similar quality
have been located over the entire studied proper
motion ranges of the LMC and the SMC. Notably, the high--fitness models fall into either
of two groups clearly separated by the fitness values of their members. The majority of the
models has the fitness exceeding the value of 0.55 while most of the rest are described
by the relation $f<0.50$. Only $\sim 5\%$ of all the models are located within the fitness
range of $0.50<f<0.55$. Later in Section~\ref{goodmodel} the lower limit for the fitness of the satisfactory models will be
precisely established and discussed. For the moment we will just mention that the almost empty belt
between the fitness values of 0.50 and 0.55 separates the acceptable models of the Magellanic System
from those reproducing the H\,I observational data by~\cite{Bruens05} insufficiently.

\subsection{LSR circular velocity and the halo flattening as parameters}\label{velocity_flattening}
It can be easily seen from Equations~(\ref{log_potential}) and~(\ref{loghalo_const}) that
the LSR circular velocity $\Theta_0$ and the flattening $q$ of the Galactic DM halo determine
the total mass of the Milky Way.
We have focused on the distribution of all the high--fidelity models of the LMC--SMC--Galaxy interaction
over the $q-\Theta_0$ plane of the studied parameter space. Figure~\ref{loghalomassmap} visualizes
the result. The actual value of the total enclosed mass of the Milky Way does not
introduce any limitation regarding the ability of our tidal model to reproduce the observed
H\,I large--scale Magellanic structures. Such behavior exists due to the fact that
local peaks of the fitness function (i.e. good models) of a similar quality have been localized
over the entire range of the LSR circular velocity.

The distribution of the genetic algorithm fits
with respect to the flattening parameter $q$ is quite different. A strong preference for oblate ($q<1$)
configurations of the gravitational potential of the Milky Way halo exists. Such a result agrees
with the conclusions by~\cite{Ruzicka07}. In the next paragraphs the role of the LSR circular
velocity and the flattening of the Galactic halo will be addressed regarding the impact
of these parameters on the orbital history of the Magellanic Clouds.

\subsection{Galactocentric velocity of the LMC and the mass of the Milky Way}\label{anticorrelation}
We have revealed an interesting anti--correlation in the output of the automated
parameter search by the genetic algorithm. Figure~\ref{lmcvel_solarmass} depicts the
relation between the magnitude of the current galactocentric velocity of the LMC and the total
mass of the Milky Way for all the high--fitness models. The general trend of the dependence,
i.e. the galactocentric velocity of the LMC decreasing as the Galactic mass increases,
is driven by the LSR circular velocity $\Theta_0$ due to the way it links the Equations~(\ref{loghalo_dens}),
(\ref{loghalo_const}) and~(\ref{vel_magnitude}), and together with the fact that the mass of the Milky Way is
dominated by the DM halo.

Equation~(\ref{loghalo_dens}) may be expressed as
\begin{equation}
m_\mathrm{MW} = \upsilon_0^2(\Theta_0) g(q) + M_\mathrm{b} + M_\mathrm{d},
\label{parametric_mass}
\end{equation}
and for the galactocentric velocity we get
\begin{equation}
\vert\boldsymbol\upsilon_\mathrm{i}\vert = h(\Theta_0, \alpha_\mathrm{i},\delta_\mathrm{i}, (m-M)_\mathrm{i},
\mu_\mathrm{W}^\mathrm{i}, \mu_\mathrm{N}^\mathrm{i}, \upsilon_\mathrm{rad}^\mathrm{i}),
\label{parametric_vel}
\end{equation}
where $\mathrm{i}=\mathrm{lmc}$. Equations~(\ref{parametric_mass}) and~(\ref{parametric_vel}) define
the parametric representation of a curve in the $m_\mathrm{MW}-\vert\boldsymbol\upsilon_\mathrm{lmc}\vert$ plane.
If all parameters except the LSR circular velocity $\Theta_0$ are assumed constant and
$\Theta_0$ is varied within the range of $\langle 210,260\rangle$\,km\,s\(^{-1}\), the parametric equations
yield the curves depicted in Figure~\ref{lmcvel_solarmass}. The curves are associated with randomly selected
high--fitness models by taking their parameter values for the input to
Equations~(\ref{parametric_mass}) and~(\ref{parametric_vel}). Apparently, the curves are close to linear.
Their slope is controlled by the LSR circular velocity and they are
well aligned with the distribution of models in the $m_\mathrm{MW}-\vert\boldsymbol\upsilon_\mathrm{lmc}\vert$ plane
which provides the key
to the Figure~\ref{lmcvel_solarmass}. The distribution of models in the plot is primarily
controlled by the LSR circular velocity which sets the linearly decreasing trend of the
velocity--mass dependence. The vertical (velocity) and horizontal (mass) placement of the
curves depends on the heliocentric positions and velocities of the LMC and on the flattening of the
DM halo of the Milky Way respectively.

\subsection{Orbits of the Magellanic Clouds}\label{orbits}
Figure~\ref{orbits_solarvel_q} will help us to understand how the variations of the flattening
parameter q and of the LSR circular velocity are reflected in the past orbits of the Clouds.
One of the parameters was varied within its entire range considered in this study,
while the second one was fixed to the value corresponding to a high--fitness model of the interaction.

Figure~\ref{orbits_solarvel_q} reveals a qualitative difference in the way
the specified parameters affect the motion of the Magellanic Clouds. The influence of the
LSR circular velocity is significantly stronger than the role of the flattening of the Galactic potential.
This is not surprising.
While the flattening parameter only has an impact on the distribution of mass of the Galaxy,
variations in the LSR circular velocity also lead to changes of the LMC/SMC galactocentric velocities.
Moreover, these effects amplify each other regarding their impact on the orbits
of the Clouds (see Section~\ref{solarvel}). If $\Theta_0$ is increased, the total mass of the Milky Way increases
implying a higher gravitational potential energy of the Clouds with respect to the Galaxy. Regarding
Equation~(\ref{vel_magnitude}) and the related discussion, the increasing LSR circular velocity reduces the
galactocentric velocities of the Clouds, i.e. their kinetic energy. Thus, the total energy of the
LMC--Galaxy and the SMC--Galaxy pairs changes with $\Theta_0$ rapidly.

The lower row of Figure~\ref{orbits_solarvel_q} shows that the orbits of the Magellanic
Clouds projected to the plane of sky are always offset from the position of the Magellanic Stream,
regardless of the choice for the $[q,\Theta_0]$ pair. Closer look at Figure~\ref{model_0078} reveals
that such a deviation is controlled by the northern component
$\mu_\mathrm{N}^\mathrm{lmc}$ of the LMC proper motion and cannot be removed unless the results
by~\cite{KalliLMC, KalliSMC} and~\cite{Piatek08} are revisited. The particles forming the
leading and trailing streams in our restricted N--body model (and generally in every tidal model)
essentially follow the orbits of their progenitors. This purely tidal approach is thus limited.
As long as the LMC/SMC projected orbits are offset from the observed Magellanic Stream, it is
difficult to model the full extent of the large--scale structure of the Stream in the position--position--LSR radial velocity space
(see Section~\ref{stream_morphology}).

\cite{Choi07} have demonstrated that tidal tails originating in massive satellite galaxies become
offset from their progenitors' orbits to some level. However, unless additional physical processes introduce
the dissipation of energy and angular momentum of the particles in the tidal tails, such a phase--space deviation
between the stream and the orbit of its progenitor is limited. Figure~\ref{orbits_radvel_maglong_700} shows
how this is reflected by the kinematics of the Magellanic Clouds and the Stream. The
high--fitness models are split into the groups of $f>0.60$ and $f<0.50$, respectively. It is a common
feature of the best models ($f>0.60$) that the LSR radial velocity along the LMC orbital track
agrees well with the famous linear velocity profile of the Magellanic Stream~\citep{Bruens05}.

Despite the previously discussed behavior of tidal models, the apparent offset of the
recent ($-300\,\mathrm{Myr} \lesssim t \leq 0$) LMC/SMC orbits
from the Magellanic Stream was reproduced partially
by our simulations. It did not occur due to a large deviation between the tidal particle tail and the
orbits of the Clouds. The modeled projection of the Magellanic Stream to the plane of sky
is still well aligned with the previous revolutions of the LMC/SMC orbits (see Figure~\ref{model_0078}),
but the preceding orbital cycles
are projected to different positions in the plane of sky reflecting the geometry and physical properties
of the LMC--SMC--Milky Way system.


The subsequent revolutions of the orbits of the Magellanic Clouds are shifted
with respect to each other in our good models of the interaction with the Milky Way. This
is to be attributed to the projection effect of the configuration of the system involving
the observer (Sun), the Galactic Center and the Magellanic Clouds (Figure~\ref{magsystem}).
The flattening of the Galactic potential must be taken into account, as it
prevents the LMC from having an orbit confined within a 2\,D plane. Figure~\ref{loghalo_force} shows
the intensity of the radial and the axial components of the gravitational field
of the logarithmic halo. The ratio of the components depends
strongly on the position $[R,z]$ with respect to the Galactic Center. That leads to
the formation of a 3D orbit of the LMC. We did not mention the SMC, because its low mass makes it
dependent not only on the gravitational field of the Galaxy, but also on the attraction by the LMC.

The previous studies of the dynamical evolution of the Magellanic Clouds
assumed a good alignment of the LCM/SMC orbits with the Magellanic Stream in the position--position--LSR radial velocity
space~\citep[see, e.g.][]{Gardiner96, Mastropietro05, Connors06}. However, such an assumption
is at odds with the HST proper motion measurements for the Clouds~\citep{Besla07}. We have seen that
nothing changes even for the revisited values of $\Theta_0$ and $\Omega_0$ by~\cite{Reid04}.
What it represents to modeling the large--scale features of the Magellanic Clouds was briefly discussed
in this section and it will be further addressed later in Sections~\ref{stream_morphology} and~\ref{stream_formation}.

\subsection{LMC--SMC encounters}\label{encounters}
The Magellanic Clouds share a common low--density gaseous envelope~\citep[see][]{Bruens05} which is
usually considered one of the signs indicating that the Clouds have formed a gravitationally bound couple
even for several Gyr.
Gravitational binding of the Clouds lasting several Gyr
turned out to be extremely rare in our simulations. However, the proper motion data by~\cite{KalliLMC, KalliSMC}
and by~\cite{Piatek08} imply that the distance between the LMC and the SMC is very likely to have been shorter
than today for several hundred Myr.
This may be sufficient to explain the presence of the H\,I envelope,
as our simulations have shown that an envelope surrounding the
Magellanic Clouds may be formed rather quickly on the specified time--scale.

The information about the LMC--SMC distance in the recent past comes directly
from the combination of Figures~\ref{velocityscalarproduct} and~\ref{reldist_700}. We have calculated
the scalar product of the present galactocentric velocity vectors of the Magellanic Clouds for
all the high--fitness models. Figure~\ref{velocityscalarproduct} shows the fitness of these models
as a function of the actual value of the scalar product. The current LMC and SMC galactocentric
velocity vectors are nearly parallel to each other in all cases and their angular deviation does not exceed
$20^\circ$ with the mean value of only $\approx 10^\circ$. It is natural to ask whether
the spatial separation of the Clouds is increasing or decreasing at present. Figure~\ref{reldist_700}
offers the answer. If the time dependence of the distance between the Clouds is plotted over the last
700\,Myr for the models of $f>0.55$, one may clearly see that the LMC and the SMC have been receding
from one another for $\approx 100$\,Myr.

To quantify the rate at which the LMC--SMC separation $D_\mathrm{l-s}$ changes we have calculated
the function $d D_\mathrm{l-s}/d t$ at the present time $t=t_0=0$\,Gyr.
If the function $d D_\mathrm{l-s}/d t$ is evaluated for the models plotted in Figure~\ref{reldist_700},
it returns the values between 50\,kpc\,Gyr$^{-1}$ and 120\,kpc\,Gyr$^{-1}$.

The preceding paragraphs lead to the conclusion that the orbital history of the Magellanic Clouds in
the high--fitness models
always involves a close encounter between the LMC and the SMC at the time of
$-250\,\mathrm{Myr} \lesssim t \lesssim -80\,\mathrm{Myr}$. It is quite interesting to
see whether such behavior is outstanding or typical within the
LMC/SMC proper motion ranges established by the recent HST measurements.

Figure~\ref{relvelmap} illustrates how the rate at which the current spatial separation of the Magellanic Clouds
changes depends on the proper motion of the SMC. We have examined two cases. In one case, the proper motion components of the
LMC were fixed to the values of a selected very good model ($f > 0.60$). The model was chosen in such a way that its
LMC proper motion components are close to the midpoint values of the intervals established by Equations~(\ref{hires_pm}).
The SMC proper motion was varied within the entire ranges for $\mu_\mathrm{W}^\mathrm{smc}$ and $\mu_\mathrm{N}^\mathrm{smc}$
involved in our study.
In all cases, the Magellanic Clouds were found to be receding from one another (Figure~\ref{relvelmap}, left plot), i.e.
\begin{equation}
\frac{d D_\mathrm{l-s}}{d t}(\mu_\mathrm{W}^\mathrm{smc}, \mu_\mathrm{N}^\mathrm{smc}) > 0.
\label{relvelmap_eq}
\end{equation}
In the other case, we have kept the values of all the parameters equal to those ones of the mentioned
high--fitness model, but the LMC proper motion was modified to reach the minimum galactocentric
velocity. Calculating the rate $d D_\mathrm{l-s}/d t$ then yielded the result depicted
in the right hand plot of Figure~\ref{relvelmap}. Obviously, no change occurred and
Equation~(\ref{relvelmap_eq}) still holds.

Thus, we have seen that strong support exists for a remarkable feature of the kinematics of the Magellanic Clouds.
The recent estimates of the LMC/SMC proper motions by~\cite{KalliLMC, KalliSMC} and by~\cite{Piatek08}
are very likely to introduce a close ($D_\mathrm{l-s} \approx 10$\,kpc) past encounter between the Magellanic Clouds
at the time of $\approx -100$\,Myr for an arbitrary choice for the LMC/SMC proper motions.

It is reasonable to expect a long--term gravitational binding between the Clouds of the order of $10^9$\,yr
to be a natural condition for repeated encounters in the LMC--SMC system. The corresponding
alignment of the past orbital tracks of the Clouds in the phase space requires a very special setup of the
parameters for the model of the interacting system, making such a case very unlikely. However,
Figure~\ref{reldist_4000} reveals a surprising picture of the LMC/SMC orbital motion
in the deep past. If the separation between the Clouds for the best models ($f>0.60$) is plotted as a function of time,
similar behavior is found in all cases. After the previously discussed close approach a couple of Myr ago,
the mutual distance of the Magellanic Clouds reaches $150$\,kpc, exceeding the value of $200$\,kpc for some
configurations. The Clouds become unbound shortly after their most recent encounter.
Nevertheless, every model contains a second close approach at the time $t<-2.5$\,Gyr with the separation
of the Magellanic Clouds dropping below $\approx 15$\,kpc. As we will show later, this encounter is the event
triggering the formation of an extended trailing tidal tail (the Magellanic Stream) and of its leading
counterpart (the Leading Arm).


\subsection{Redistribution of mass in the Magellanic System}\label{massredistribution}
We have stated in Section~\ref{parameters_general} that only a subset of the models identified
by the genetic algorithm in the parameter space may be considered satisfactory reproductions
of the observational data~\citep{Bruens05}. In general, an acceptable model
of the interaction between the Magellanic Clouds and the Milky Way is supposed to
reproduce the extent of the observed Magellanic Stream in the 3\,D space composed
of the LSR radial velocity and of the position in the plane of sky. The second large--scale
H\,I structure associated with the Clouds -- the Leading Arm -- is not used to
separate good models from the unsuccessful ones.

The Leading Arm appears in tidal models of the LMC--SMC--Galaxy interaction as a natural counterpart
of the Magellanic Stream~\citep[see][]{Gardiner96, Connors06}, but its acceptable reproduction
by the means of numerical simulations remains a challenge. Our parameter study has not
changed that picture -- the modeled distribution and kinematics of H\,I in the Leading Arm region and around the
main LMC and SMC bodies remains similarly unsatisfactory over the parameter space,
especially failing to reproduce the observed
morphology of the Leading Arm. In terms of the fitness function, the value of $f$ never exceeds $\approx 0.35$ if calculated
only for the Leading Arm. 

Reliable modeling of the evolution of the central regions of the Magellanic Clouds currently eludes
the level of sophistication of our numerical code. The inner parts of the Clouds
contain high--density baryonic mass and in order to study their evolution a detailed model
involving self--gravity and hydrodynamical processes which account for the mass exchange cycle
between stars and gas, is required. Our restricted N--body code overestimates
the amount of gas tidally stripped from the LMC/SMC central regions because
the tides are not balanced by self--gravity and the dissipative behavior of gas. As a consequence,
the modeled column density is lower than observed values in the inner 5\,kpc of the
Clouds. This effect is stronger in the case of the SMC as it is the dominant source of matter
for the Magellanic Stream and the Leading Arm (see Figures~\ref{good_bad_model} or \ref{lostparticles}). The particle
distribution in the LMC is affected by tidal heating by the Milky Way and
the original particle disk is turned into a 3\,D spherical structure.

\subsection{Definition of a good model}\label{goodmodel}
The Magellanic Stream turned out to be the appropriate indicator of the quality of our models.
Its formation is very sensitive to the choice of the model parameters and critically influences the
resulting fitness.  Figure~\ref{good_bad_model}
illustrates these facts. While the model of $f=0.61$ is able to fit the basic
features of the Magellanic Stream both in the projected H\,I distribution and the LSR radial velocity profile,
a typical simulation representing the model group of $f<0.50$ places the Magellanic Stream to the
position--position--LSR radial velocity space incorrectly. The Stream is extended insufficiently
both in position and the LSR radial velocity (upper row of Figure~\ref{good_bad_model}),
and slope of the simulated profile of the LSR radial velocity along the Stream exceeds
the observed one already at the Magellanic longitude of $\approx -30^\circ$ (lower right plot
of Figure~\ref{good_bad_model}).
Generally speaking, the described behavior of the modeled trailing stream is responsible
for the resulting fitness of a given model, and was used to define the desired threshold level
of the fitness value.

The obvious lack of models within the fitness range of $(0.50, 0.55)$, clearly visible in
Figures~\ref{pm_fitness} and~\ref{velocityscalarproduct}, was combined with the specified
features of the high--fitness models. This yielded the desired threshold level of the
fitness function quite naturally. Figure~\ref{pm_fitness} shows that the fitness of the model
establishing the upper limit to the mentioned gap in the distribution of the fitness values
lies slightly below $0.55$ (it is $0.546$ actually). However, the transition between the good
and unsatisfactory models is always gradual, as we have seen, and thus the value $f=0.55$
can be taken for the threshold value of the fitness function with no loss of generality.

\subsection{Morphology and kinematics of the Magellanic Stream}\label{stream_morphology}
In order to discuss our results regarding the reproduction of the large--scale distribution
of H\,I associated with the Magellanic Clouds, we compared a selected high--fidelity model
with the low--resolution data--cube of the H\,I observations by~\cite{Bruens05}. The simulated
3\,D data--cube depicted in Figure~\ref{model_0078} (already shown in Figure~\ref{good_bad_model})
was produced by a model with
the following values of the LSR circular velocity and the flattening of the Galactic halo:
$\Theta = 232$\,km\,s$^{-1}$, $q = 0.81$. These values yield the total enclosed mass of the Milky Way
within the radius of $250$\,kpc of $2.20\cdot10^{12}$\,M$\odot$.

Assuming the recent measurements of the LMC and the SMC proper motions, our tidal model was able
to produce a trailing tail similar to the observed Magellanic Stream regarding
both the LSR radial velocity profile and the extent in the projected position.
While the LSR radial velocities measured along the Magellanic Stream are
reproduced very well,
the far tip of the simulated stream at the position of $b\gtrsim -60^\circ$
and $l\approx 70^\circ$ is displaced compared to the observations (Figure~\ref{model_0078}).
Such a displacement appears in every high--fitness model identified by the genetic algorithm
in the studied parameter space.

The column density of H\,I in the Magellanic Stream decreases rather smoothly towards its far tip.
This feature exists also in our high--fidelity simulations of the tidally induced formation of the
large--scale Magellanic structures. The middle row of Figure~\ref{model_0078} shows how
the mean H\,I column density on the line perpendicular to the great circle of the Magellanic 
longitude~\citep[defined in][]{Wannier72} changes with the longitude. The simulated
profile of the column density (right hand plot) has a steeper slope than the observed Stream and its gradual decrease
ends at the Magellanic longitude of $\approx -70^\circ$ compared to $\approx -90^\circ$
for the observations.

The plots in the lower row of Figure~\ref{model_0078} also offer the view of the area in the plane
of sky where the H\,I structure called the Leading Arm is located. \cite{Bruens05} introduced
the division of the structure into three systems of gaseous clumps named LA\,I, LA\,II and LA\,III.
Their approximate positions in the Galactic coordinates $[b,l]$ are $[-15^\circ,300^\circ]$ (LA\,I),
$[15^\circ,290^\circ]$ (LA\,II), and $[15^\circ,270^\circ]$ (LA\,III). The simulation shown
in Figure~\ref{model_0078} was able
to reproduce the part LA\,III and a fraction of the clump LA\,II. Some of our high--fitness
models offer quite a good match of the overall kinematics and spatial extent
for both observed structures LA\,II and LA\,III. Nevertheless, no model reproducing
both the system LA\,I and the Magellanic Stream at the same time was found.

\subsection{Formation of the Magellanic Stream}\label{stream_formation}
It is interesting to take a look at the composition of the simulated Magellanic Stream,
and at the process of its build--up over the course of time. The lower left plot of Figure~\ref{good_bad_model}
reveals that most of the particles forming the Magellanic Stream in the selected high--fitness
model are of the SMC origin. A number of the LMC particles are spread over the full
extent of the Stream as well. This is a typical result and it is based on the fact that the
LMC particles are more tightly bound due to the significantly higher mass of the LMC compared to the SMC --
the LMC:SMC mass ratio is 10:1 in our simulations.

It has already been discussed by~\cite{Connors06} that the Magellanic Stream might consist
of several filaments projected to the same region in the plane of sky. Such filaments
would be the tidal debris spread along the past orbits of the Clouds. Our study has confirmed
such a scenario. Figure~\ref{model_0078} shows the projected orbits of the Magellanic Clouds
from $t=-3$\,Gyr to the present, plotted over the contours of the integrated column density of H\,I (lower right plot).
The SMC has crossed the area currently occupied by the Magellanic Stream three times,
while two passages have occurred in case of the LMC. If we disentangle the projected orbital paths,
we will find out that the previous orbital revolutions appear more diagonal in our plot.
They are placed to lower Galactic longitudes. The resulting trailing tail is thus a mixture of particles
released from the LMC/SMC due to the tidal stripping during several epochs
covering the time range from present to $\approx-3.3$\,Gyr when the redistribution
matter was triggered by a close LMC--SMC encounter.

The complex structure of the
modeled Stream can be seen in the lower plot of Figure~\ref{lostparticles}. The LMC/SMC particles
are color coded according to the epoch when they were stripped from the Clouds. The Stream and the
Magellanic Bridge are dominated by the SMC particles, while most of the LMC particles remain
in the vicinity of the main LMC body and have been gradually tidally heated due to the gravitational
field of the Milky Way.

The past orbital history of the Magellanic Clouds for the discussed high--fitness model
can be seen in the lower plot of Figure~\ref{lostparticles}. For the last $4$\,Gyr
the Clouds have moved within the Galactocentric radius of 200\,kpc, undergoing two close
($D_\mathrm{l-s}\approx10$\,kpc) encounters at the times of $-3.3$\,Gyr and $-0.15$\,Gyr, respectively.
The impact of the encounters on the distribution of the LMC/SMC particles was different.

To quantify the influence of the encounters in the LMC--SMC--Galaxy system,
we have counted the number of particles shifted from the original orbits
of the radius $r_i$ to the distance of
$2 r_i$ from the center of their progenitor. If this function is evaluated for the time bins of $500$\,Myr
from $t=-4$\,Gyr, we obtain the upper plot of Figure~\ref{lostparticles}. Apparently,
the number of the disturbed particles was increased already in the time bin of $\langle-3.5,-3.0\rangle$\,Gyr
as a consequence of the first LMC--SMC close approach. This level was later only mildly
decreasing, as more of the disturbed particles were stretched in the tidal field of the Milky
Way, satisfying the above stated definition of strongly affected particles.

The second close LMC--SMC encounter was significantly more dramatic, disturbing more than $10\%$
of all particles. This event is the origin of the common envelope surrounding both the Clouds,
and of the turbulent gaseous filament connecting the main bodies of the Clouds, named
the Magellanic Bridge~\citep{Bruens05}. Substantial fraction of the particles
contributed to the Magellanic Stream as well.

The results summarized in the preceding paragraphs and depicted in Figure~\ref{lostparticles} have
a remarkable consequence regarding the formation history of the Magellanic Stream.
Over the last $\approx3$\,Gyr it was supported
at an almost constant rate by the gas tidally stripped from the Clouds.
However, this rate was increased by a factor $\approx20$ due to a recent
encounter between the Clouds at $t=-0.15$\,Gyr. We would like to mention that
on the qualitative level these results are an appropriate description
of all the high--fitness models found in the parameter space for the
LMC--SMC--Milky Way interaction.

We have found that the high--fitness simulations prefer the oblate ($q<1$) shape of the
logarithmic halo of the Galaxy over the prolate configuration (see Figure~\ref{loghalomassmap}),
but several reasonable models of $q$ as high as $1.10$ exist. As long as good models of the interacting system exist for
the prolate halo, this shape of the Galactic gravitational potential cannot be
ruled out. However, we have shown that such a possibility is substantially
less likely than the case of $q<1$. Moreover, only two acceptable models with
significantly prolate ($q \gtrsim 1.10$) DM haloes of the Milky Way
have been identified for a reasonable total mass of the halo, i.e. below the limit of
$\approx 3\cdot12\,\mathrm{M_\odot}$ (see Section~\ref{key_parameters}).

The impact of the flattening parameter on the phase--space positioning of the LMC/SMC
orbits has been addressed in Section~\ref{orbits} of this paper. The shape of the
Galactic halo also influences the redistribution of the matter associated with the
Magellanic Clouds due to the resulting tidal field of the Milky Way. This
effect is the key to the preference for the oblate configuration of the DM halo
of our Galaxy. We have calculated the tidal force exerted by the Milky Way
on two points separated by $\Delta r = 10$\,kpc and moving around the
Galactic Center on circular polar orbits of the radii $50$\,kpc, $100$\,kpc and $150$\,kpc.
Figure~\ref{loghalo_tides} shows the result. In general, the
oblate configuration offers the most efficient tidal stripping, improving the
probability that a sufficient redistribution of the LMC/SMC particles occurs
within the available time--scale of several Gyr.

\section{Summary and Conclusions}\label{summary}
We have performed a detailed exploration of the parameter space for the interaction involving
the Magellanic Clouds and the Milky Way. The method applied was similar to the approach by~\cite{Ruzicka09},
but the results were quite different as a new free parameter was involved.

The recent work by~\cite{Shattow09} lead to the conclusion that the modified values of the
LSR circular velocity and of the LSR angular rotation rate~\citep{Reid04} introduce
a significant change to the orbital history of the Magellanic Clouds.
We have followed their study in order to show whether the large--scale distribution of H\,I in the Magellanic system
can be simulated successfully even for the increased values of the LMC/SMC proper motions
measured recently by~\cite{KalliLMC, KalliSMC} and by~\cite{Piatek08} if the revisited
view of the galactic rotation~\citep{Reid04} is adopted.

The study by~\cite{Ruzicka09} ruled out the tidal models of the LMC--SMC--Galaxy interaction
because they were unable to reproduce the basic kinematic and morphological features of
the most prominent structures originating in the Magellanic Clouds, i.e. of the Magellanic Stream
and the Leading Arm~\citep{Bruens05}. Their automated analysis of the parameter space for the
interaction has shown clearly that the recent proper motion measurements imply
the past LMC/SMC orbits that allow for efficient tidal stripping of gas from the Clouds.

Unlike~\cite{Ruzicka09}, we have taken into account the observational data by~\cite{Reid04} and treated the LSR
circular velocity as a free parameter varied within the range $\Theta_0 = \langle210,260\rangle$\,km\,s$^{-1}$.
Following~\cite{Reid04}, a modified value of the LSR angular rotation rate $\Omega_0$
of $29.45$\,km\,s$^{-1}$\,kpc$^{-1}$ 
was adopted as well, yielding the studied range of the Solar galactocentric distance $R_0$
of $\langle7.13,8.83\rangle$\,kpc.
As shown in Section~\ref{solarvel} of this paper, the new values of $\Theta_0$ and $\Omega_0$ affect
the galactocentric positions and velocities of the Magellanic Clouds, as well as the mass of the Galaxy.

The exploration of the modified parameter space for the interaction has revealed a remarkable qualitative
change regarding the features of the resulting candidates for acceptable tidal models.
The expectations of~\cite{Shattow09} have been confirmed because satisfactory reproduction
of the large--scale Magellanic structures was possible for the LMC/SMC proper motion data by the HST.
Moreover, such quality models have been localized for a large number of the proper motion
combinations (Figure~\ref{pm_fitness}). This is due to the fact that the proper motions no longer
play the exclusive role in establishing the actual 3\,D galactocentric motion of the Clouds.

The LMC/SMC velocity (and position) vectors are determined not only by their proper motions but also by the
LSR circular velocity and the LSR angular rotation rate. The current phase space
coordinates of the Clouds are linked with the mass distribution of the Galaxy due to $\Theta_0$
and $\Omega_0$. As a consequence, a great variability and freedom have been introduced in the parameter space
concerning the options to choose the orbits of the Magellanic Clouds allowing for the efficient
tidal redistribution of mass resulting in the formation of the Magellanic Stream.

All the high--fitness models of the interaction localized in the studied parameter space involved
two close LMC--SMC encounters within the last $4$\,Gyr. The first one occurred at the time
$t<-2.5$\,Gyr and triggered the evolution of the Magellanic Stream. This encounter
caused the tidal heating of the outer regions of the original LMC/SMC particle discs.
Subsequently, the disturbed particles were spread along the orbital paths of the Clouds due to the
tidal stripping by the gravitational field of the Galaxy.

The latter encounter was placed only as recently as $-150$\,Myr
($-250\,\mathrm{Myr} \lesssim t \lesssim -80\,\mathrm{Myr}$), but its impact on the LMC/SMC particles
was at least by a factor of 10 stronger compared to the first encounter (for more rigorous discussion
see Section~\ref{stream_formation}). This event was also the beginning of the formation of the
filament connecting the Clouds, composed of turbulent gas and young stars, called
the Magellanic Bridge~\citep{Bruens05}. It is a notable fact that the recent encounter between the
Clouds is a rather general feature of the interaction. Figure~\ref{relvelmap} demonstrates that
it is likely an intrinsic property of the recent proper motion data. The studies
by~\cite{KalliLMC, KalliSMC} and by~\cite{Piatek08} seem to introduce a LMC--SMC approach to the
distance of $\approx10$\,kpc at the time of $\approx-150$\,Myr.

\section{Discussion}\label{discussion}
Although it is usually assumed that a long--term
gravitational binding between the Clouds has existed, we have confirmed the previous findings
by~\cite{Ruzicka07, Ruzicka09} that such a condition is not necessary regarding the tidally induced
redistribution of mass associated with the Clouds leading to the formation the Magellanic Stream.
Figure~\ref{reldist_4000} shows clearly that after the first encounter, the spatial separation
of the Clouds usually exceeds $200$\,kpc.

The assumption of the Clouds being gravitationally bound is supported by several more observational indications.
First of all, the Magellanic Clouds are surrounded by a low--density gaseous envelope which is a natural
feature associated with two interacting bodies sharing a common history. However, we have found that the second
of the mentioned LMC--SMC encounters is able to produce such a diffuse gaseous structure. 

The gravitational binding of the Clouds is also often substantiated by their unique composition which is
quite outstanding within the neighborhood of the Milky Way. Most of the satellites of the Galaxy are dwarf spheroidal
while the LMC and the SMC are highly unevolved gas--rich irregular galaxies. The likelihood that such a couple
was formed by chance in the Local Group of Galaxies is very small, obviously. These issues are serious,
but they cannot be addressed within the framework of our parametric study.

It is one of the most surprising results of the genetic algorithm--based exploration of the parameter space that
good models have been localized independently of the total mass of the Milky Way (see Figure~\ref{loghalomassmap}).
This behavior of the interacting system occurred due to the LSR circular velocity that links the present
phase--space positions of the Clouds with the actual enclosed mass of the galactic DM halo. The consequence of such a complex
nature of the system is nicely manifested by Figure~\ref{lmcvel_solarmass}.

However, even the remarkable reduction of the uncertainties of the LMC/SMC proper motions achieved
by~\cite{KalliLMC, KalliSMC} and by~\cite{Piatek08} has left error boxes large enough to accommodate virtually every
request for the proper motions to compensate the actual choice for the LMC circular velocity.
Hence, the current phase--space positions of the Clouds can be selected in order to make the tidal model work over
a wide range of the enclosed Galactic mass. Reasonable simulations of the Magellanic Stream have been
made over the range of $0.6\cdot10^{12}$\,M$_\odot$ to $3.0\cdot10^{12}$\,M$_\odot$ for the total mass
of the Milky Way enclosed within the radius of $250$\,kpc.

It cannot be omitted that as the Galactic mass decreases due to the decrease of the LSR circular velocity,
the density profile of the DM halo steepens in the inner region of the Galaxy. This is caused by the
rescaling of the rotation curve of the Milky Way yielded by the relation $\Omega_0=\Theta_0/R_0=\mathrm{const.}$
Such behavior is helpful in respect to our goals, but it must be treated with care.

\cite{Besla07} have pointed out that various observations have put constraints on the upper limit
for the total mass enclosed within $\approx50$\,kpc from the Galactic Center. However,
taking such findings into account does not disprove our results.
The limits on the total mass in the inner part of the Milky Way
can be considered simply as additional constraints reducing the number of acceptable parameter combinations.

The redistribution of the matter associated with the Magellanic Clouds followed a scenario similar to that one
by~\cite{Gardiner94}. The formation of the Magellanic Stream (and of its natural counterpart -- the Leading Arm)
was triggered by a LMC--SMC encounter $2.5$\,Gyr ago. However, the Stream as old as $\approx2$\,Gyr is at odds
with the recent conclusions by~\cite{Stanimirovic08} or by~\cite{Bland-hawthorn09}. They argue that the interaction
of the relatively cold H\,I of the Stream with the hot ambient halo of the Milky Way results in the thermal
fragmentation~\citep{Stanimirovic08} and the ablation~\citep{Bland-hawthorn09} of the H\,I clouds.
Such a decay yields the maximum survival time of the H\,I clouds not exceeding $\approx1$\,Gyr.
However, our model introduces a continuous replenishment of the Stream gas, and a strong boost to this process
due to the second LMC--SMC encounter at $\approx-0.15$\,Gyr.

We also have to point out the fundamental difference regarding the phase--space structure of the Magellanic Stream
if the tidal and ram pressure models are compared. \cite{Mastropietro05} have shown that the ram pressure--induced
Stream possesses a rather compact phase--space structure. On the other hand, we have seen that the tidal mechanism leads
to the Magellanic Stream composed of several filaments of different ages. These filaments are the remnants of the
past LMC/SMC orbital revolutions and might lie in significantly different distances even as large as
$150$\,kpc (see Figures~\ref{model_0078} and~\ref{lostparticles}). Since the study by~\cite{Stanimirovic08} 
assumed the Stream to lie not further than $50$\,kpc from the Galactic
Center, their results cannot fully be applied to our models. The decay of the H\,I clouds due to the interaction with 
the hot gaseous halo of the Milky Way might be slower at larger radii as the density of the ambient
Galactic medium falls with the galactocentric distance.

It is a challenging problem for the tidal models of the Magellanic System to explain the absence
of stars in the Stream. Whilst evolutionary scenarios taking ram pressure stripping as the
dominant process in the formation of the Magellanic Stream explain the problem naturally, tides
affect both stars and gaseous clumps. Hence, the missing stars pose a serious issue regarding the tidal
origin of the Stream.

The study by~\cite{Bekki09} has shown by the means of self--consistent
numerical simulations that the key parameter determining the
stellar content in a tidal model of the Magellanic Stream is the relative extent of gas and stars
in the SMC. Assuming that the radii of the H\,I and stellar distributions in the SMC are of
$2 \leq r_\mathrm{H\,I}/r_\mathrm{stars} \leq 4$, \cite{Bekki09}
have successfully reproduced the observed composition of the Magellanic Stream. The corresponding
stellar distribution was disturbed by the recent ($t \approx -200$\,Myr) LMC--SMC encounter and contributed to
the extended SMC/LMC halos, but no stars appeared in the Magellanic Stream.
The assumption on $r_\mathrm{H\,I}/r_\mathrm{stars}$ has a solid observational support, as the H\,I diameters of gas--rich galaxies
are observed to be significantly larger than their optical disks~\citep[e.g.][]{Broeils94}.

Stars and gaseous clumps are treated identically as test particles in our model and they satisfy the same
equations of motion. Whether a given particle represents a star or an H\,I cloud depends on its initial
distance from the SMC (LMC) center and on the assumed radius of the stellar body. This quantity was
estimated through the use of observational and theoretical constraints summarized in the previous paragraph
and was set to $0.5 r_\mathrm{disk}$ (see Table~\ref{table_1}).

The recent LMC--SMC encounter
strongly affected the test particles within the radius of the stellar body, but similar to~\cite{Bekki09}
most of the particles contributed to the common envelope of the Clouds.
However, $<5$\,\% of such particles were moved to the Stream and may thus be considered its stellar contamination.
Unfortunately, this is inevitable in a restricted N--body simulation. Our model overestimates the amount of
matter stripped from the SMC center due to its over--simplified description, where
self--gravity and dissipative gas dynamics will act against the tidal stripping as significant restoring forces
(see Section~\ref{massredistribution}).

\section{Open questions}\label{openquestions}
The values of the LSR circular velocity by~\cite{Reid04} certainly represent a progress towards the resolution
of the difficulties arising from the recent HST proper motions of the Magellanic Clouds. However, we have
not revealed a parameter set allowing for the correct modeling of the far tip of the Magellanic Stream.
An offset of a model from the observations was present (Figure~\ref{model_0078}). This occurred because
the LMC/SMC orbits have not crossed the corresponding region of the position--position--LSR radial velocity
space. Nevertheless, we should keep in mind the simplicity of our model. Neither the dissipative hydrodynamical
processes were included, nor did we account for the alternative scenarios such as the recent idea
by~\cite{Nidever08}.

They have analyzed the kinematics of the Magellanic Stream and the Leading Arm, with the focus on the
transition regions between these structures and the main bodies of the Clouds. Their results suggest
that both the Magellanic Stream and the structure LA\,I~\citep{Bruens05} might be evolutionarily related
to the region in the south--east of the LMC where a massive star formation takes place. It is possible that
such a star forming activity has caused a strong blowout of gas which was later redistributed by the tidal
or ram pressure stripping.

The scenario proposed by~\cite{Nidever08} might be the complementary process to the tidal stripping
concerning the origin of the Leading Arm. While the tidal model succeeded reproducing the clumps
LA\,II, LA\,III, the gas blow--out would transport gas to the position of the complex LA\,I.
In general, such a process is able to eject gas with an arbitrary direction of its momentum with respect
to the LMC motion.
It might allow for filling the regions of the phase--space unreachable by the tidal/ram pressure
models with gaseous matter.

\acknowledgments
We are grateful for the support by the FWF Austrian Science Fund (the grant P20593--N16),
by the Academy of Sciences of the Czech Republic (the Junior research grant KJB300030801), and
by the project M\v{S}MT LC06014 Center for Theoretical Astrophysics. Our thanks go
also to Christian Br\"{u}ns who kindly provided excellent observational data,
and to Matthew Wall for his advanced C++ library of genetic algorithms.

\appendix
\section{Fitness function}\label{appendixA}
We have further improved the fitness function devised by~\cite{Ruzicka07} in order to handle
the weaknesses of their scheme revealed later in the work by~\cite{Ruzicka09}. The first component
of our fitness function is still based on the approach by~\cite{Theis01} who proposed a generally applicable
technique of comparing the relative intensities of the corresponding pixels in the modeled and observed
data--cubes.

Both modeled and observed H\,I column
density values are scaled relative to their maxima to introduce
dimensionless quantities. Then, we get
\begin{equation} 
f_1 = \frac{1}{N_{\upsilon} \cdot N_x \cdot N_y}\sum\limits_{i=1}^{N_{\upsilon}} \sum\limits_{j=1}^{N_y} \sum\limits_{k=1}^{N_x} 
\frac{1}{1 + \left|\sigma_{ijk}^\mathrm{obs} - \sigma_{ijk}^\mathrm{mod}\right|}, 
\label{fitness1} 
\end{equation}
where $\sigma_{ijk}^\mathrm{obs}$, $\sigma_{ijk}^\mathrm{mod}$ are
normalized column densities measured at the position $[j, k]$ of the
$i$--th velocity channel of the observed and modeled data, respectively.
$N_{\upsilon}$ is the number of separate LSR radial velocity channels in our data.
$(N_x \cdot N_y)$ is
the total number of positions on the plane of sky for which observed and modeled H\,I column density values are available.

The second component of the fitness function performs a search for structures
in the position--position--velocity space of the data--cube. The modeled
distribution of H\,I is compared to its observed counterpart regardless of the
exact levels of the H\,I radio emission. It combines the enhancement of structures in the data by their Fourier
filtering~\citep[see][]{Ruzicka07} with the subsequent check for empty/non--empty pixels in both data--cubes.
The corresponding component of the fitness function is defined as follows:
\begin{equation} 
f_2 = \frac{\sum\limits_{i=1}^{N_{\upsilon}} \sum\limits_{j=1}^{N_y} \sum\limits_{k=1}^{N_x}
\mathrm{pix}_{ijk}^\mathrm{obs} \cdot \mathrm{pix}_{ijk}^\mathrm{mod}} 
{\max\left(\sum\limits_{i=1}^{N_{\upsilon}} \sum\limits_{j=1}^{N_y} \sum\limits_{k=1}^{N_x}
\mathrm{pix}_{ijk}^\mathrm{obs}, 
\sum\limits_{i=1}^{N_{\upsilon}} \sum\limits_{j=1}^{N_y} \sum\limits_{k=1}^{N_y}
\mathrm{pix}_{ijk}^\mathrm{mod}\right)},
\label{fitness2} 
\end{equation}
where $\mathrm{pix}_{ijk}^\mathrm{obs}\in\{0,1\}$ and $\mathrm{pix}_{ijk}^\mathrm{mod}\in\{0,1\}$ indicate whether there is matter
detected at the position $[i, j, k]$ of the 3\,D data--cube of the observed and modeled Magellanic System, respectively.

\cite{Ruzicka07} showed that the search for structures
significantly improves the performance of the genetic algorithm if the structures of interest occupy only a small fraction of the system's
entire data--cube ($<10$\% in the case of the Magellanic Stream and the Leading Arm).

The fitness function $f_2$ is a measure for the agreement of the shape in the data. No attention is paid to
the actual H\,I column density values. However, the efficiency of such a search for structures
becomes lower with the increasing
resolution of the compared data--cubes, as the relative number of non--empty pixels usually decreases in such a case, and
the function $f_2$ returns very small values. To resolve the described difficulty, we have proposed a new component
of the fitness function which is based on Equation~(\ref{fitness2}). In fact, we have introduced another simplification level
for the view of the properties of the studied system. Similarly to Equation~(\ref{fitness2}) the number the overlapping
non--empty pixels in the observed and modeled data--cubes is counted. This procedure is performed repeatedly
for a series of $N$ data--cubes of an increasing resolution. It is significantly easier to achieve a good match
in case of low--resolution data, but the sensitivity to structures is rather poor, and only global positioning
of the system as a whole is evaluated. High--resolution data provide more detailed information on the system,
but the corresponding complexity prevents the comparison from driving the genetic algorithm efficiently. The newly
designed component of the fitness function may be described as
\begin{equation}
f_3 = \frac{\sum\limits_{m=0}^{N} 2^m \sum\limits_{i=1}^{N_{\upsilon}^m} \sum\limits_{j=1}^{N_y^m} \sum\limits_{k=1}^{N_x^m}
\mathrm{pix}_{ijk}^\mathrm{obs} \cdot \mathrm{pix}_{ijk}^\mathrm{mod}}
{\sum\limits_{m=0}^{N} 2^m \sum\limits_{i=1}^{N_{\upsilon}^m} \sum\limits_{j=1}^{N_y^m} \sum\limits_{k=1}^{N_x^m}
\mathrm{pix}_{ijk}^\mathrm{obs}},
\label{fitness3}
\end{equation}
where $N_x^m$, $N_y^m$, and $N_{\upsilon}^m$ are the dimensions of the $m$--th data--cube and
\begin{displaymath}
N_x^m < N_x^{m+1} \wedge N_y^m < N_y^{m+1} \wedge N_{\upsilon}^m < N_{\upsilon}^{m+1}.
\end{displaymath}
Obviously, we have introduced a weighting factor of $2^m$ in Equation~(\ref{fitness3}). It reflects the above discussion
and strongly emphasizes the models that are able to satisfy the high--resolution observational data.

\cite{Ruzicka07} also recommended and
successfully applied a system--specific comparison.
In the case of the Magellanic Clouds, the very typical linear radial velocity
profile of the Stream including its high negative velocity tip was considered important.
The slope of the LSR radial velocity
function is a very specific feature, strongly dependent on
the features of the orbital motion of the Clouds. We have
slightly modified the original definition for the LSR radial velocity
check by~\cite{Ruzicka07}, and the fourth fitness function component is defined as
\begin{equation}
f_4 = \frac{1}{1 + \sum\limits_{i=\mathrm{x},\mathrm{y},\mathrm{\upsilon}}
\left|\frac{\mathrm{pix}^\mathrm{obs}_i(\upsilon_\mathrm{min}) - \mathrm{pix}^\mathrm{mod}_i(\upsilon_\mathrm{min})}{N_i}\right|}, 
\label{fitness4}
\end{equation}
where $\mathrm{pix}^\mathrm{obs}_i(\upsilon_\mathrm{min})$ and $\mathrm{pix}^\mathrm{mod}_i(\upsilon_\mathrm{min})$ are the
pixels with the minima of the observed LSR
radial velocity profile of the Magellanic Stream and its model, respectively.
The resulting fitness function $f$ combines the above defined components in the following way:
\begin{equation} 
f = \frac{1}{4}\sum\limits_{i=1}^4 f_i.
\label{fitness_total}
\end{equation}

Our definition of the fitness function is different from the approach by~\cite{Ruzicka09}. Moreover, the
additional fitness component $f_3$ has been introduced by Equation~(\ref{fitness3}). Therefore,
the value of the fitness function returned for the given model needs further analysis
to clarify whether the identified local peaks
of the fitness function correspond to satisfactory models. This means that every system has its
own threshold value of the fitness function which needs to be defined.

\bibliographystyle{apj}
\bibliography{ms}

\clearpage
\begin{deluxetable}{lll}
\tabletypesize{\small}
\tablecaption{Free parameters of the LMC--SMC--Milky Way interaction \label{table_1}}
\tablewidth{0pt}
\tablehead{
\colhead{Parameter} & \colhead{Value} & \colhead{Comment}
}
\startdata
$(\alpha_\mathrm{lmc}, \delta_\mathrm{lmc})$ & $(81.90^\circ\pm 0.98^\circ, -69.87^\circ\pm 0.41^\circ)$ & Equatorial coordinates\\
$(\alpha_\mathrm{smc}, \delta_\mathrm{smc})$ & $(13.2^\circ\pm 0.3^\circ, -72.5^\circ\pm 0.3^\circ)$ & \\
${(m-M)}_\mathrm{lmc}$ & $18.5\pm 0.1$ & Distance moduli\\
${(m-M)}_\mathrm{smc}$ & $18.85\pm 0.10$ & \\
$\mu_\mathrm{W}^\mathrm{lmc}\mathrm{[mas\,yr^{-1}]}$ & $\langle -2.11,-1.92 \rangle$ & Proper motion components\\
$\mu_\mathrm{N}^\mathrm{lmc}\mathrm{[mas\,yr^{-1}]}$ & $\langle +0.39,+0.49 \rangle$ & \\
$\mu_\mathrm{W}^\mathrm{smc}\mathrm{[mas\,yr^{-1}]}$ & $\langle -1.34,-0.69 \rangle$ & \\
$\mu_\mathrm{N}^\mathrm{smc}\mathrm{[mas\,yr^{-1}]}$ & $\langle -1.35,-0.99 \rangle$ & \\
$\upsilon_\mathrm{rad}^\mathrm{lmc}\mathrm{[km\,s^{-1}]}$ & $262.2\pm 3.4$ & Line--of--sight velocities \\
$\upsilon_\mathrm{rad}^\mathrm{smc}\mathrm{km\,s^{-1}}$ & $146.0\pm 0.60$ & \\
$r_\mathrm{disk}^\mathrm{lmc}\mathrm{[kpc]}$ & $\langle 9.0, 12.0 \rangle$ & Initial radii of the particle disks \\
$r_\mathrm{disk}^\mathrm{smc}\mathrm{[kpc]}$ & $\langle 5.0, 8.0 \rangle$ & \\
$(i_\mathrm{lmc}, p_\mathrm{lmc})$ & $(34.7^\circ\pm 6.2^\circ, 129.9^\circ\pm 6.0^\circ)$ & Inclination and position angles of the particle disks \\
$(i_\mathrm{smc}, p_\mathrm{smc})$ & $(60^\circ\pm 20^\circ, p=45^\circ\pm 20^\circ)$ & \\
$\Theta_0\mathrm{[km\,s^{-1}]}$ & $\langle 210, 260 \rangle$ & LSR circular velocity\\
$q$ & $\langle 0.71, 1.30\rangle$ & Flattening of the DM halo of the Milky Way \\
\enddata
\end{deluxetable}
\clearpage

{
\onecolumn
\begin{figure}
\includegraphics[angle=90, scale=.55, clip]{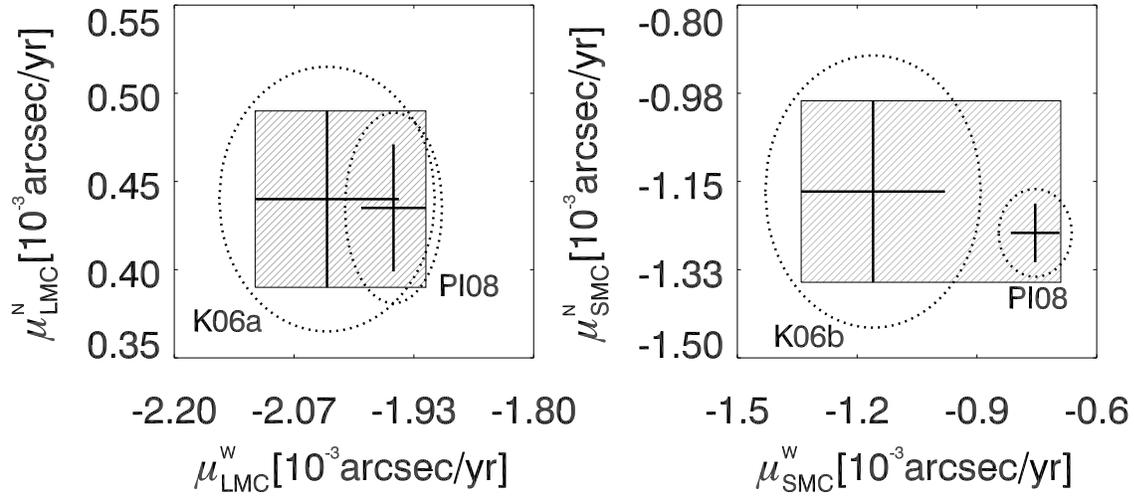}
\caption{The 2\,D projections of the Magellanic parameter space
to the ($\mu^\mathrm{N}$, $\mu^\mathrm{W}$)--plane for both the LMC (left plot) and the SMC.
The gray fillings mark the proper motion ranges explored by genetic algorithm. The labels
indicate the proper motions as expected by the studies by~\cite{KalliLMC}(K06a), \cite{KalliSMC}(K06b),
and~\cite{Piatek08}(PI08). The ellipses show the 68.3\,\% confidence regions.
\label{LMC_SMC_pm}}
\end{figure}
}
\clearpage

{\twocolumn
\begin{figure}
\includegraphics[bb=30 190 577 713,angle=0,scale=.39, clip]{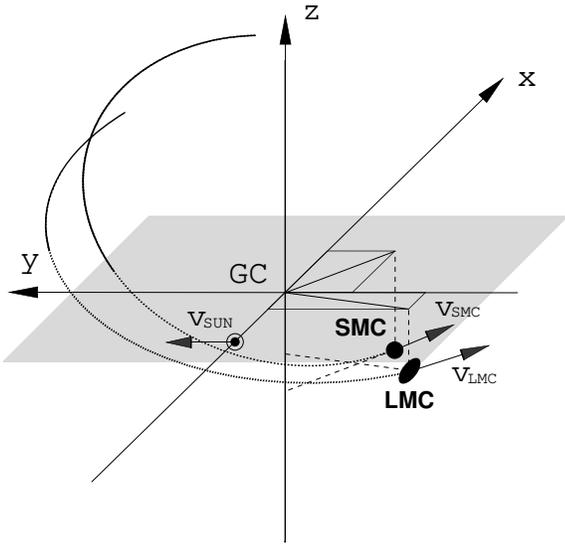}
\caption{View of the Magellanic System. The positions of the Sun and of the Magellanic Clouds are shown in
a galactocentric Cartesian frame. The frame was selected so that its z--axis coincides with the axis of the
Milky Way disk and points towards the Northern Galactic Pole. The current Solar position
vector is $\mathbf{R_\odot}=(-\Theta_0/\Omega_0, 0, 0)$\,kpc, and the Sun is moving in the direction of the y--axis, i.e.
$\boldsymbol{\upsilon_\odot}=(0, \Theta_0, 0)$\,km\,s$^{-1}$. The Magellanic Clouds are
$\approx$20\,kpc far from each other, and their present day velocity vectors are almost parallel.
\label{magsystem}}
\end{figure}
}
\clearpage

{\twocolumn
\begin{figure}
\includegraphics[bb=0 0 530 715,angle=90,scale=.3, clip]{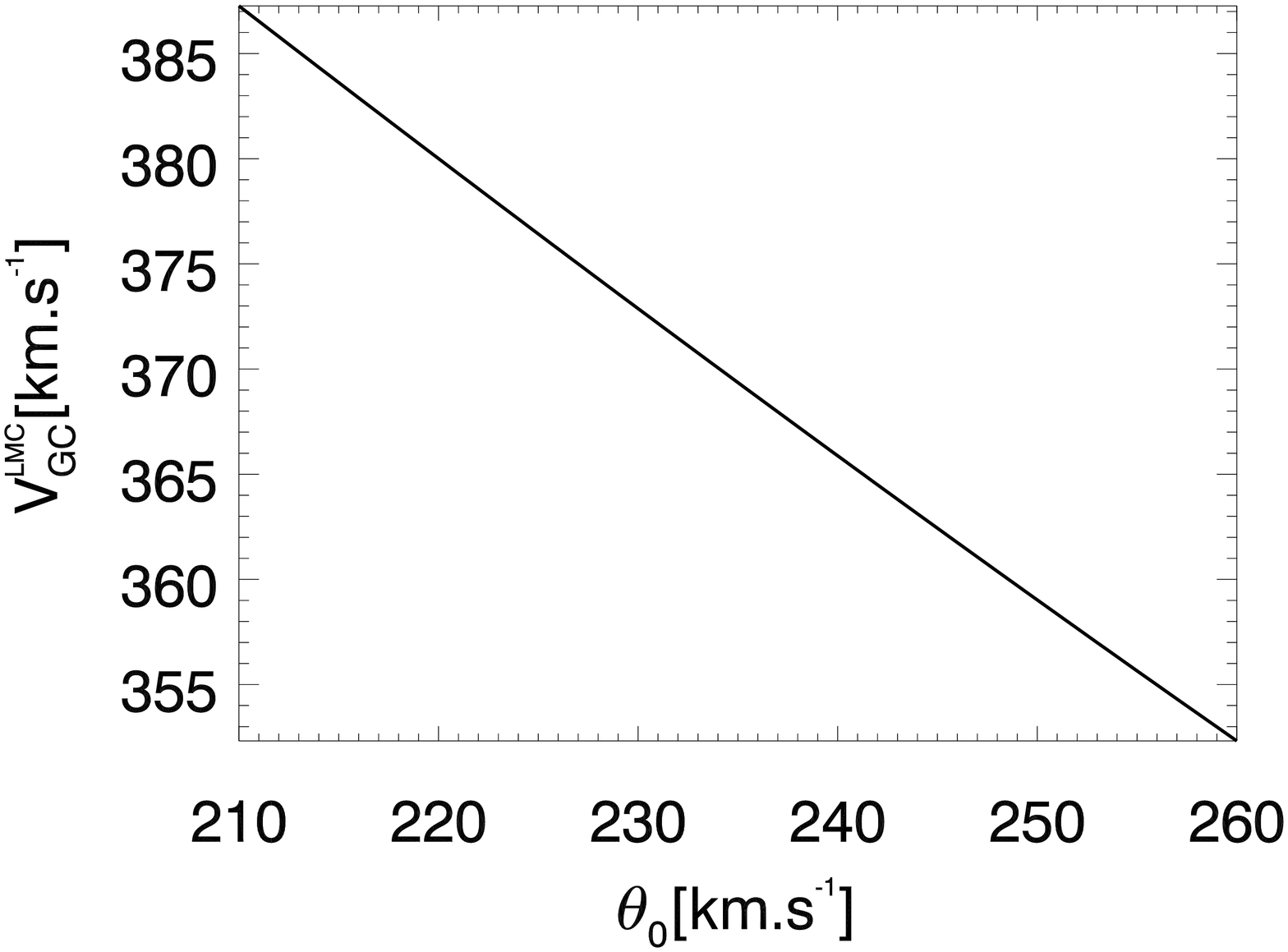}\\
\includegraphics[bb=0 0 530 715,angle=90,scale=.3, clip]{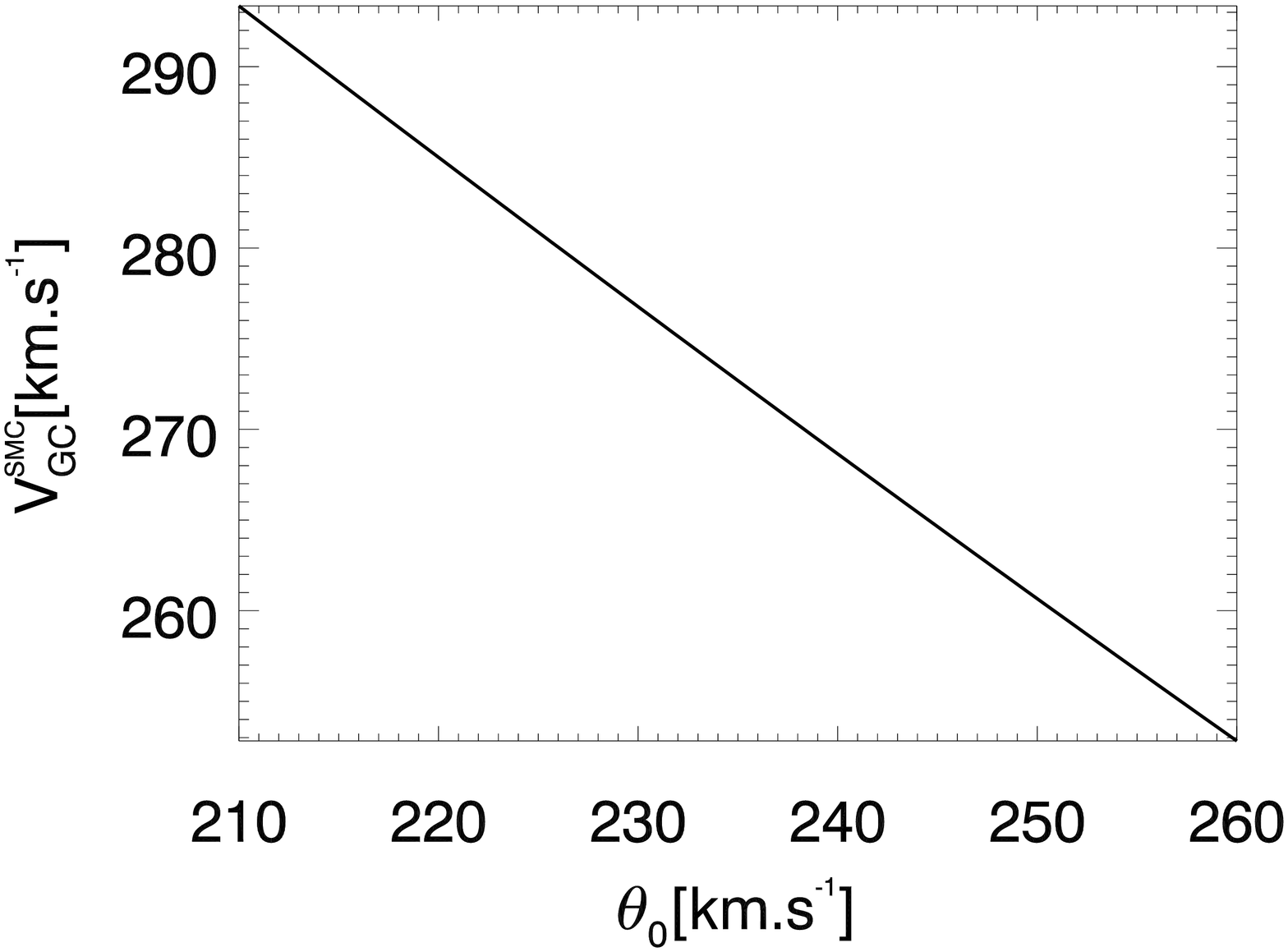}
\caption{Galactocentric velocities of the Magellanic Clouds. The magnitude of the velocity vector
of the LMC as a function of the LSR circular velocity is displayed in the upper plot. The lower plot
shows the same relation for the SMC. Both plots assume
the galactocentric coordinate frame depicted in Figure~(\ref{magsystem}) and the LSR angular rotation rate
$\Omega_0 = 29.45$\,km\,s$^{-1}$\,kpc$^{-1}$~\citep{Reid04}.
Both Clouds slow down with respect to the Galactic Center if the LSR circular velocity is increased.
\label{lmcsmc_vel-solarvel}}
\end{figure}
}
\clearpage

{\onecolumn
\begin{figure}
\includegraphics[bb=38 0 490 750,angle=90,scale=.31, clip]{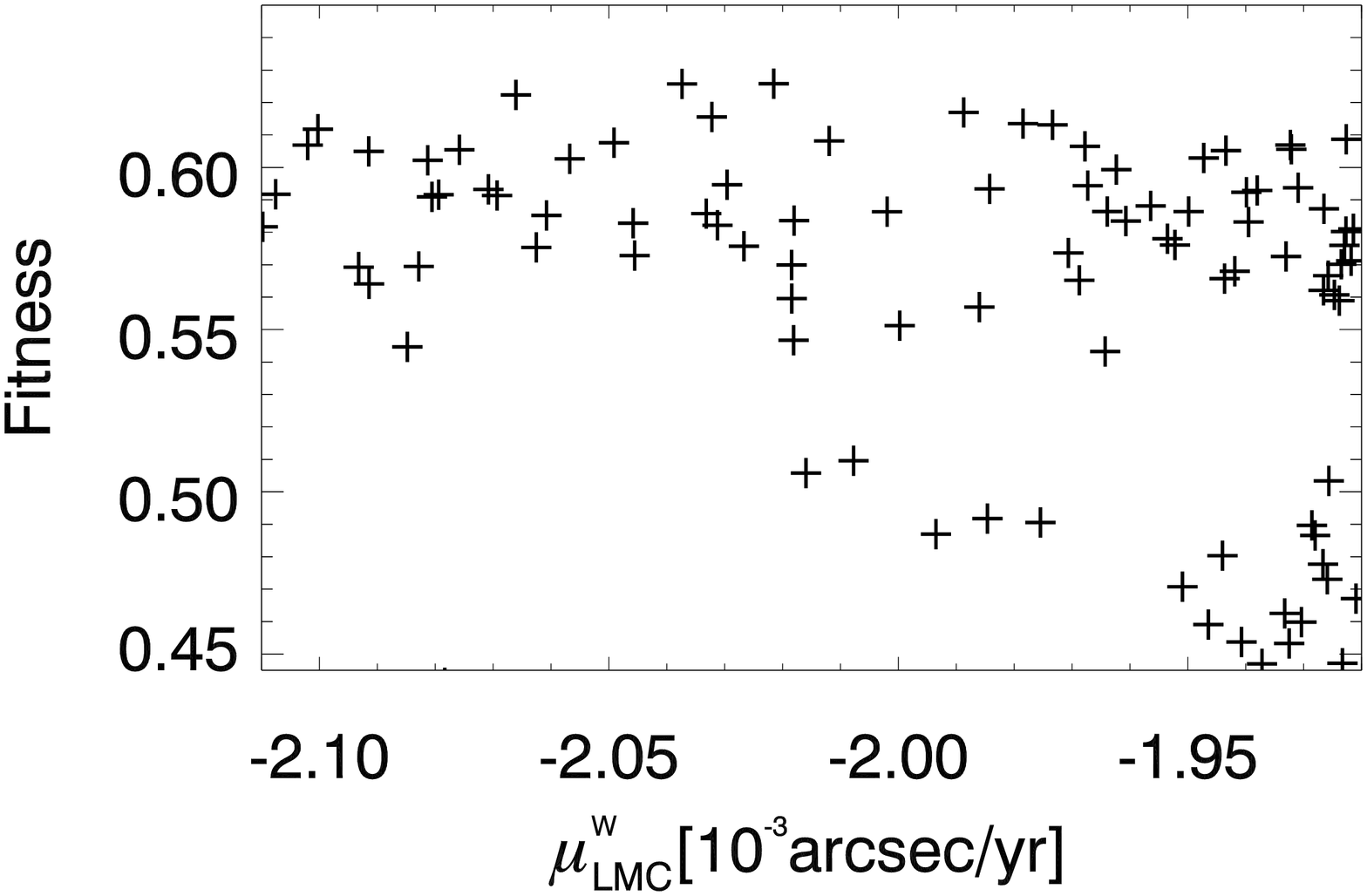}
\includegraphics[bb=38 0 490 750,angle=90,scale=.31, clip]{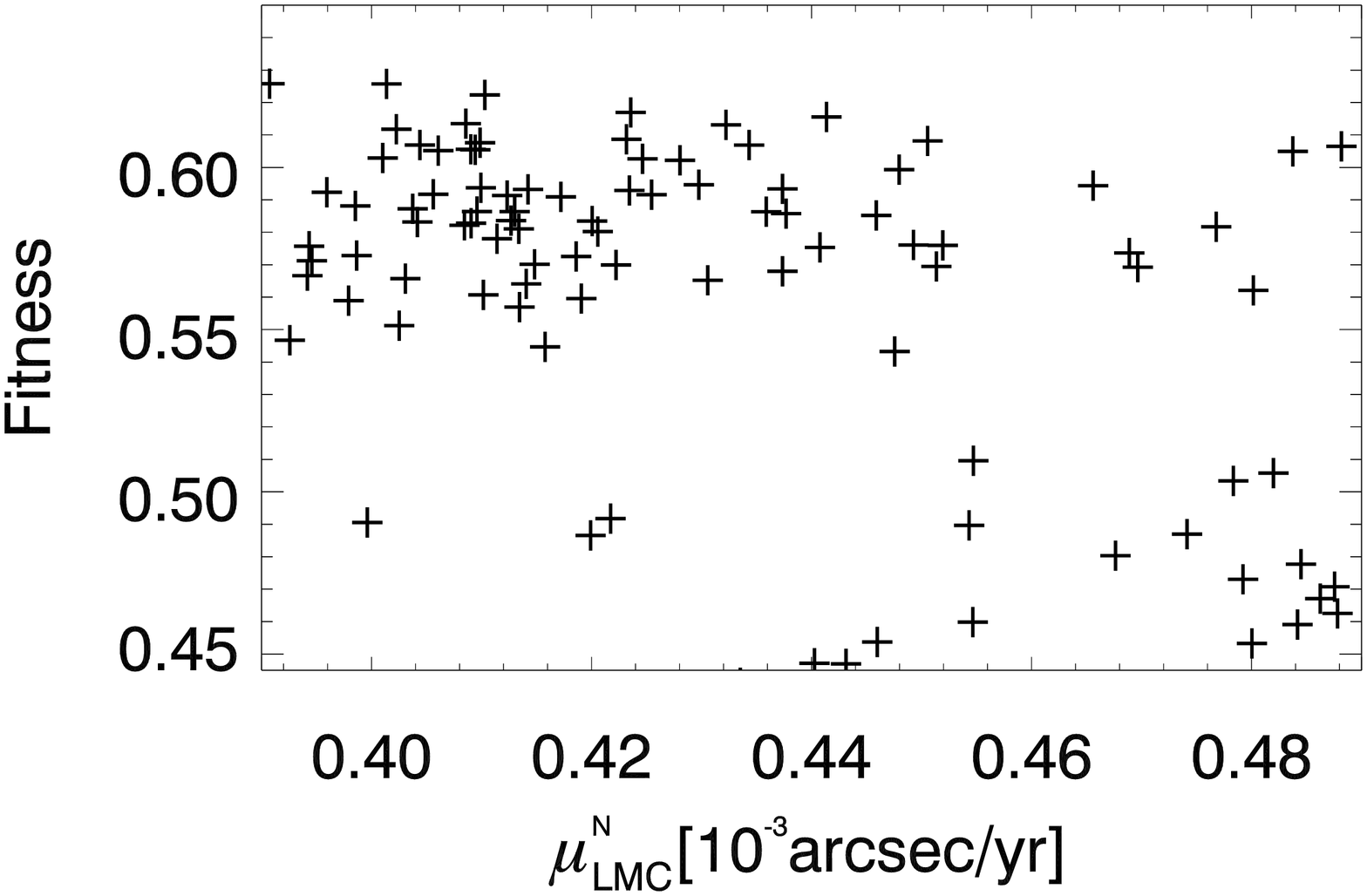}\\
\includegraphics[bb=38 0 490 750,angle=90,scale=.31, clip]{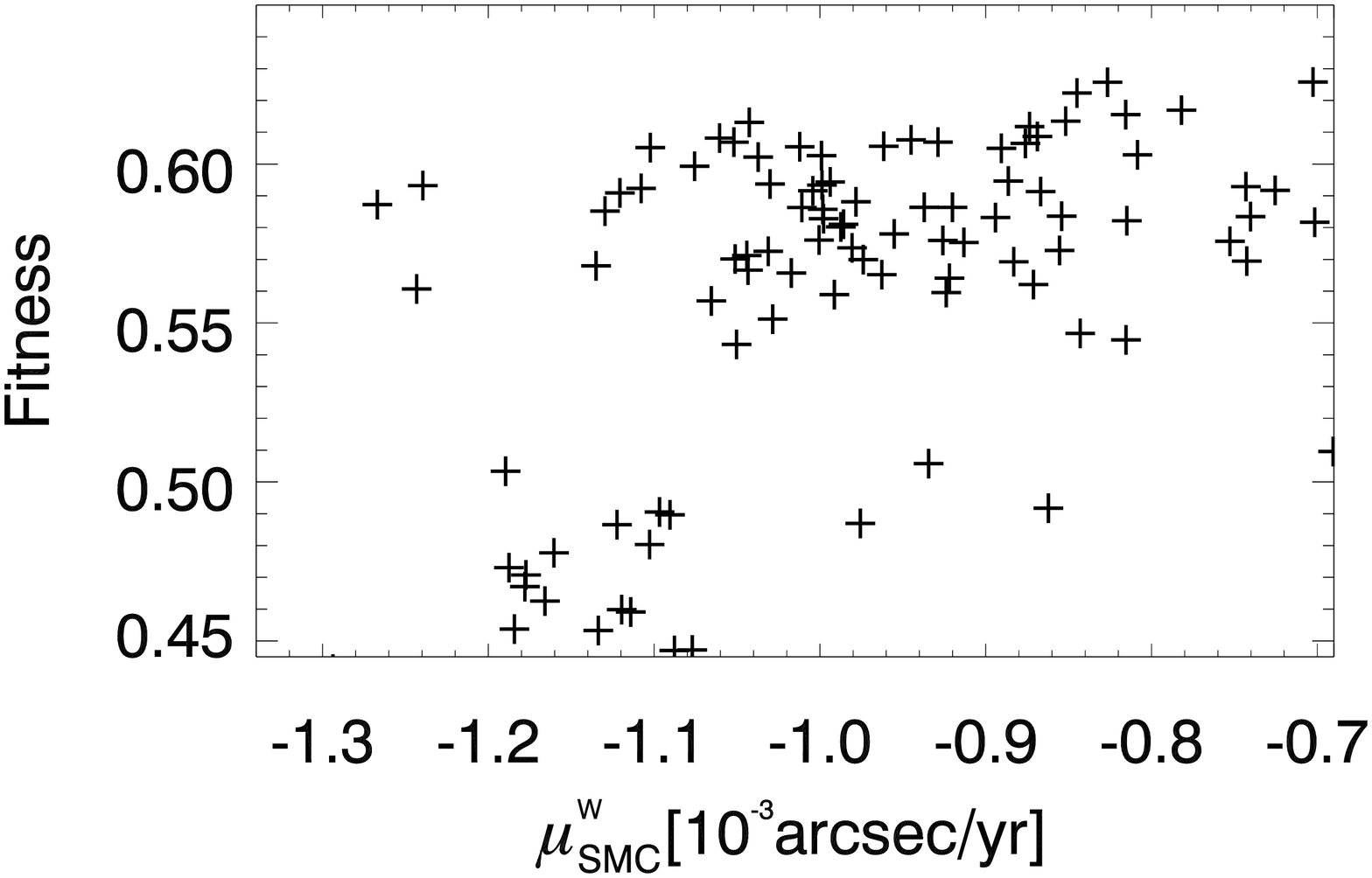}
\includegraphics[bb=38 0 490 750,angle=90,scale=.31, clip]{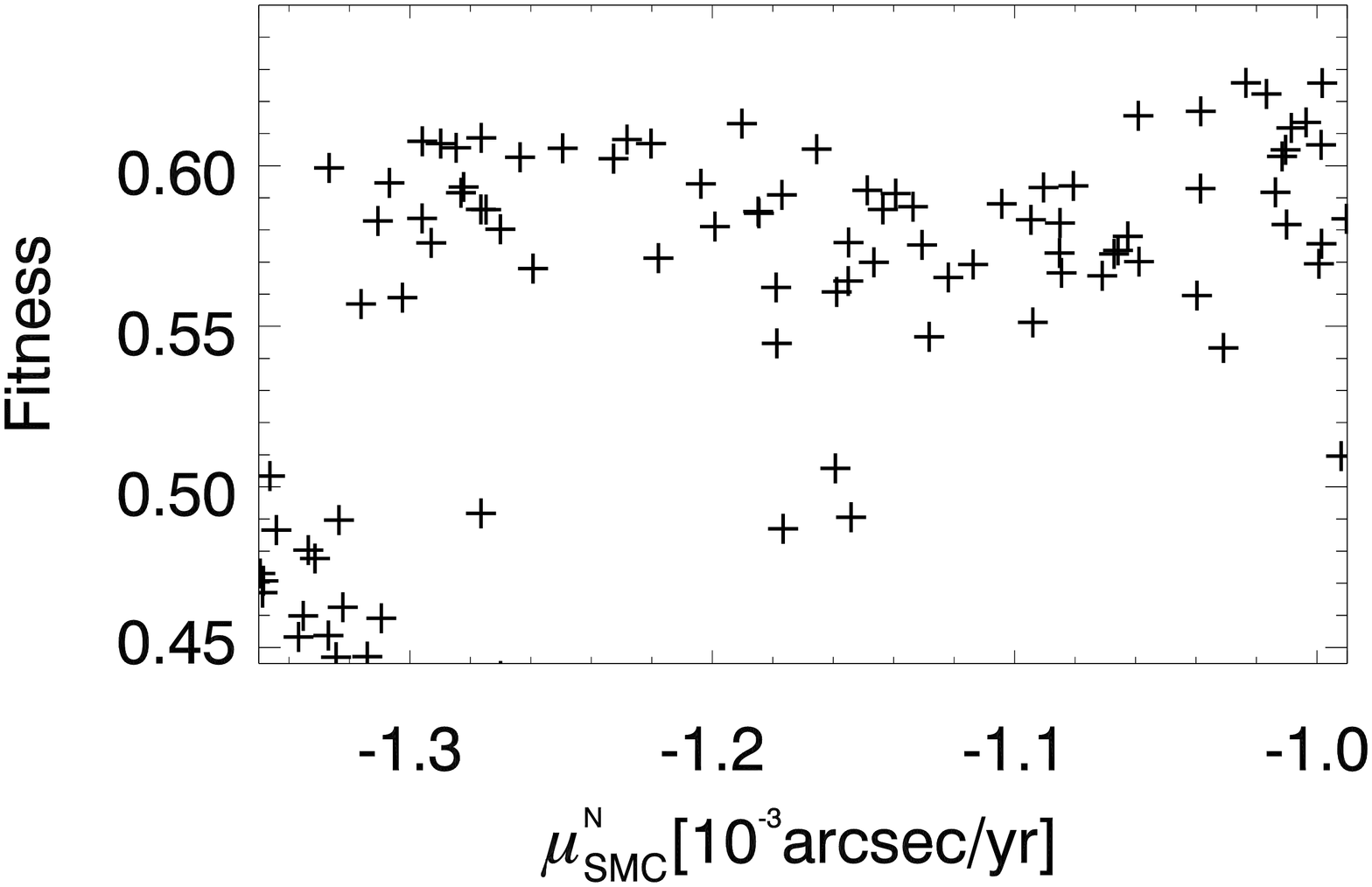}
\caption{Distribution of all high--fitness models of the Magellanic System identified by the genetic
algorithm. The upper row shows the fitness of the models as a function of the western (left plot) and
of the northern LMC proper motion component, respectively. The lower row offers the same relations
for the SMC.
\label{pm_fitness}}
\end{figure}
}
\clearpage

{\twocolumn
\begin{figure}
\includegraphics[bb=10 110 605 830,angle=90,scale=.30, clip]{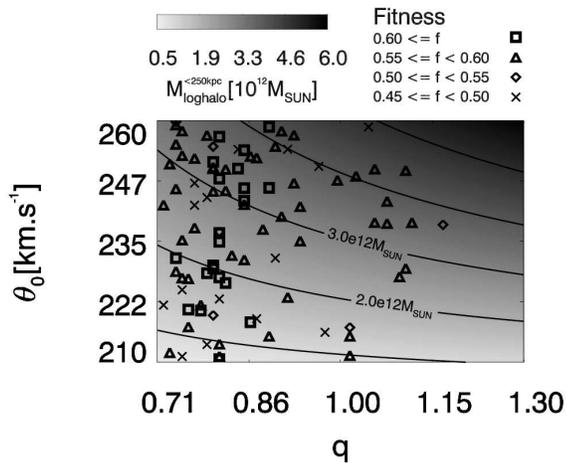}
\caption{Total mass of the Milky Way enclosed within the radius of 250\,kpc as a function
of the LSR circular velocity $\Theta_0$ and of the DM halo flattening $q$. All the high--fitness
models identified by the genetic algorithm search are over--plotted according to their $[q,\Theta_0]$
positions. While no trend exists regarding the dependence on $\Theta_0$, the models for oblate ($q<1$)
haloes are preferred.
\label{loghalomassmap}}
\end{figure}
}
\clearpage

{\twocolumn
\begin{figure}
\includegraphics[bb=30 60 490 745,angle=90,scale=.31, clip]{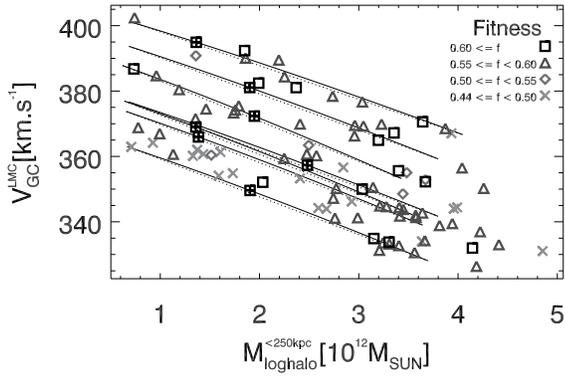}
\caption{The galactocentric velocity of the LMC as a function of the total mass
of the Milky Way. The models of the interaction between the Magellanic Clouds and
the Galaxy which were identified by the genetic algorithm are divided into
four groups regarding the value of their fitness. The curves (solid lines) corresponding
to the parametric equations~(\ref{parametric_mass}) and~(\ref{parametric_vel})
are plotted for several randomly selected models of the fitness $f>0.60$ (squares with crosses inside),
i.e. for different
combinations of the parameters $q$, $(\alpha_\mathrm{lmc},\delta_\mathrm{lmc})$, $(m-M)_\mathrm{lmc}$,
$(\mu_\mathrm{W}^\mathrm{lmc}, \mu_\mathrm{N}^\mathrm{lmc})$, and $\upsilon_\mathrm{rad}^\mathrm{lmc}$.
The dotted lines demonstrate the proximity of the parametric curves to a linear relation.
\label{lmcvel_solarmass}}
\end{figure}
}
\clearpage

{\onecolumn
\begin{figure}
\includegraphics[bb=40 25 550 660,angle=90,scale=.36, clip]{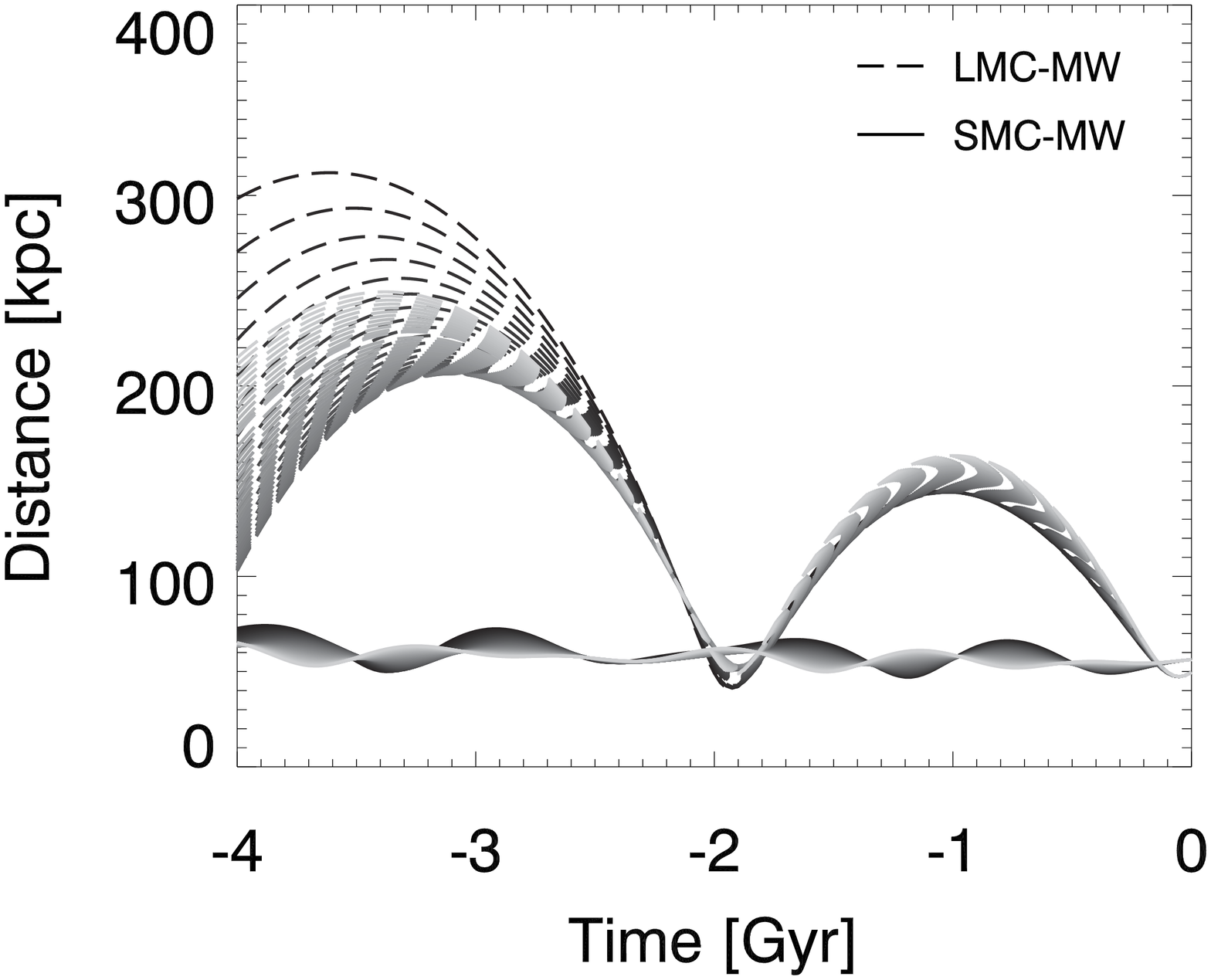}
\includegraphics[bb=40 25 550 660,angle=90,scale=.36, clip]{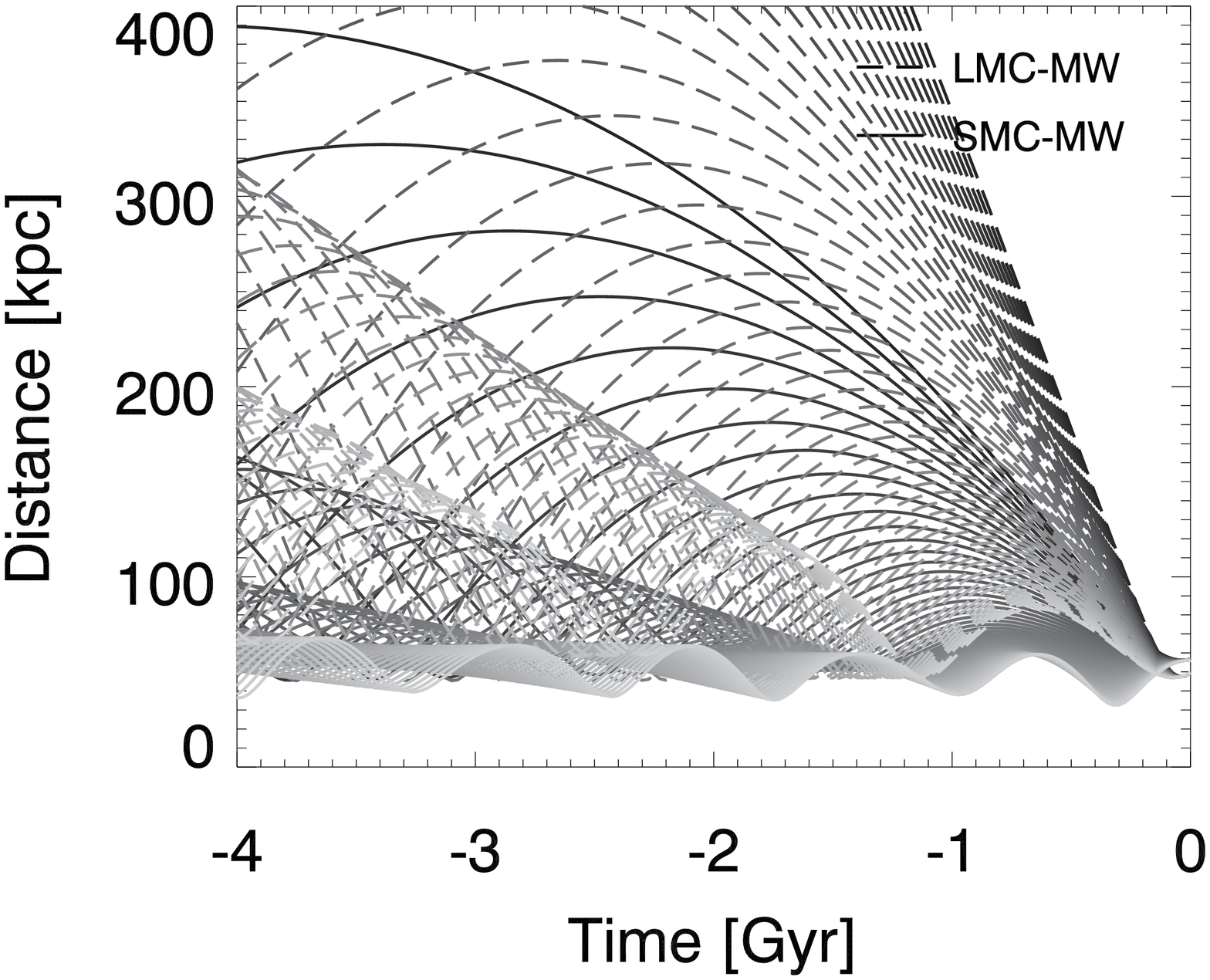}\\
\includegraphics[bb=20 40 620 675,angle=90,scale=.365, clip]{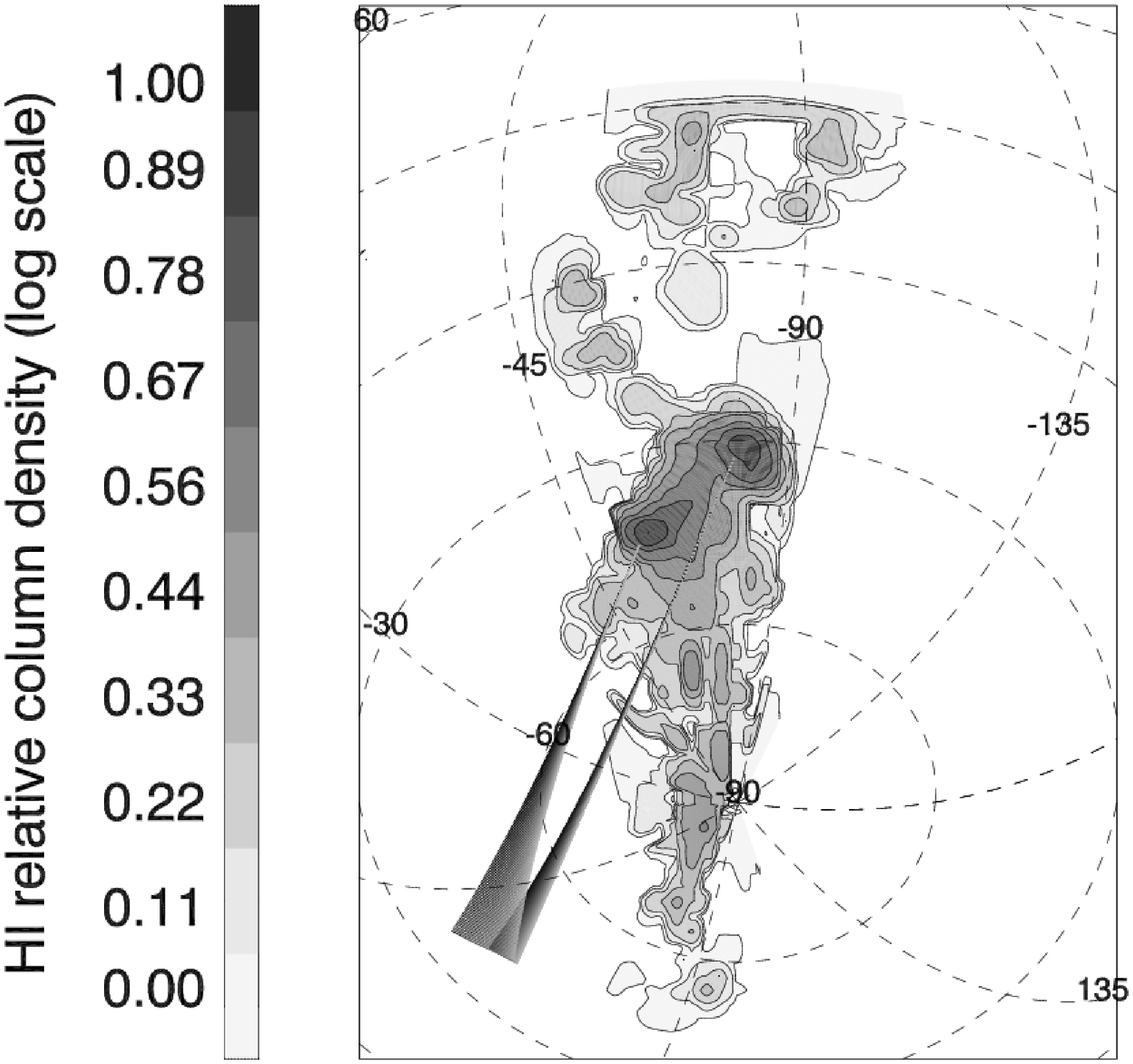}
\includegraphics[bb=20 40 620 675,angle=90,scale=.365, clip]{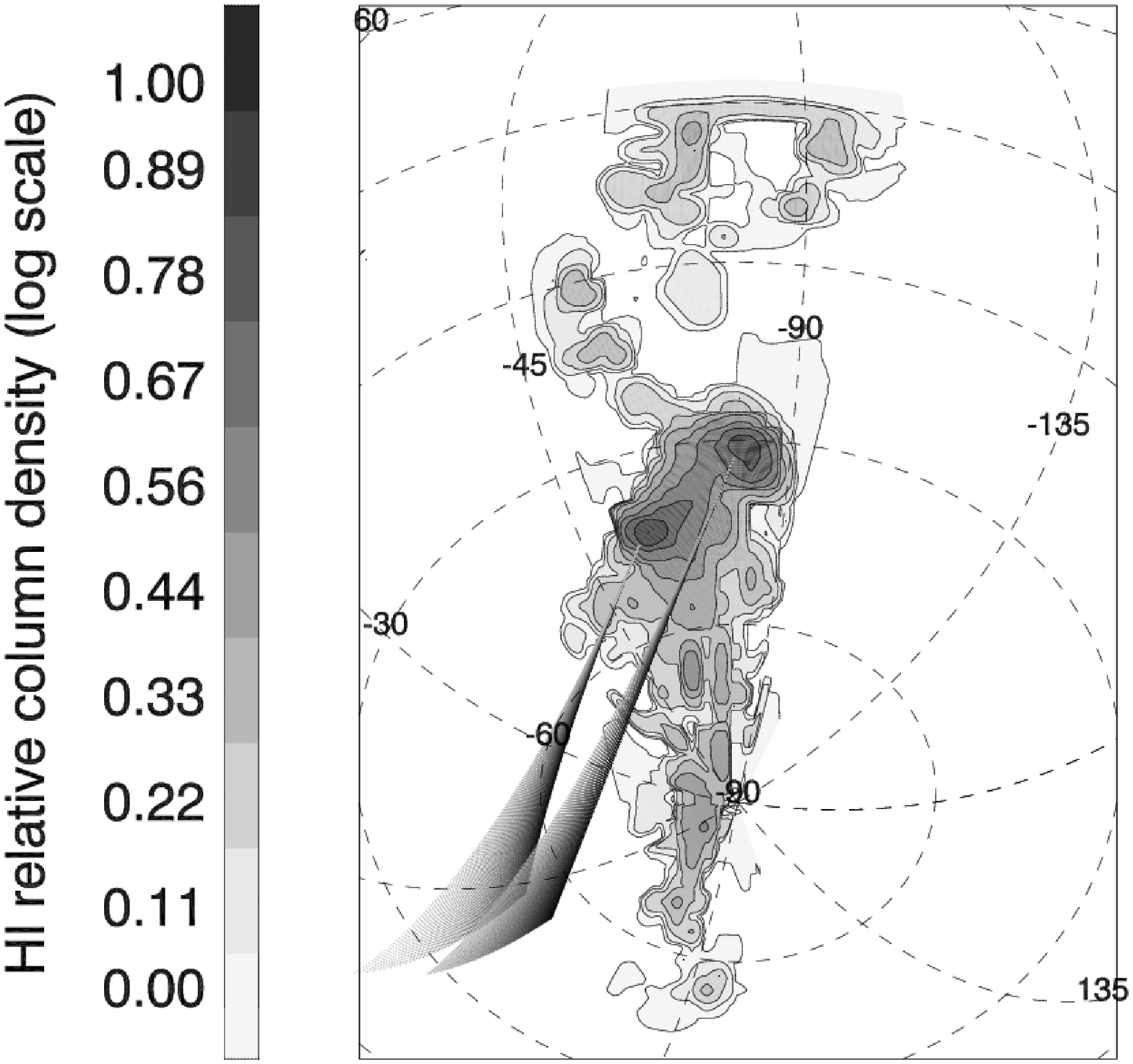}
\caption{Orbits of the Magellanic Clouds as functions of the Milky Way halo flattening
$q$ (left column) and of the LSR circular velocity $\Theta_0$, respectively. The parameter $q$ was
varied within the range of $\langle 0.71, 1.30\rangle$ while $\Theta_0$ was fixed to the value
of 238\,km\,s$^{-1}$. When the LSR circular velocity was the variable ($\Theta_0 = \langle 210, 260\rangle$),
the flattening $q=0.80$ was selected (right column).
The upper row depicts the past time dependence of the LMC--Galaxy and
the SMC--Galaxy distances. The plots in the lower row show the LMC (more to the right)
and SMC orbits over the last $300$\,Myr projected
to the plane of sky. Galactic coordinates are used. The color coding is such that the black to light gray
transition corresponds to changing the actual variable from its minimum to its maximum.
The contour maps shows the low resolution H\,I observations of the Magellanic Clouds and
associated structures by~\cite{Bruens05}.
\label{orbits_solarvel_q}}
\end{figure}
}
\clearpage

{\twocolumn
\begin{figure}
\includegraphics[bb=40 35 510 690,angle=90,scale=.34, clip]{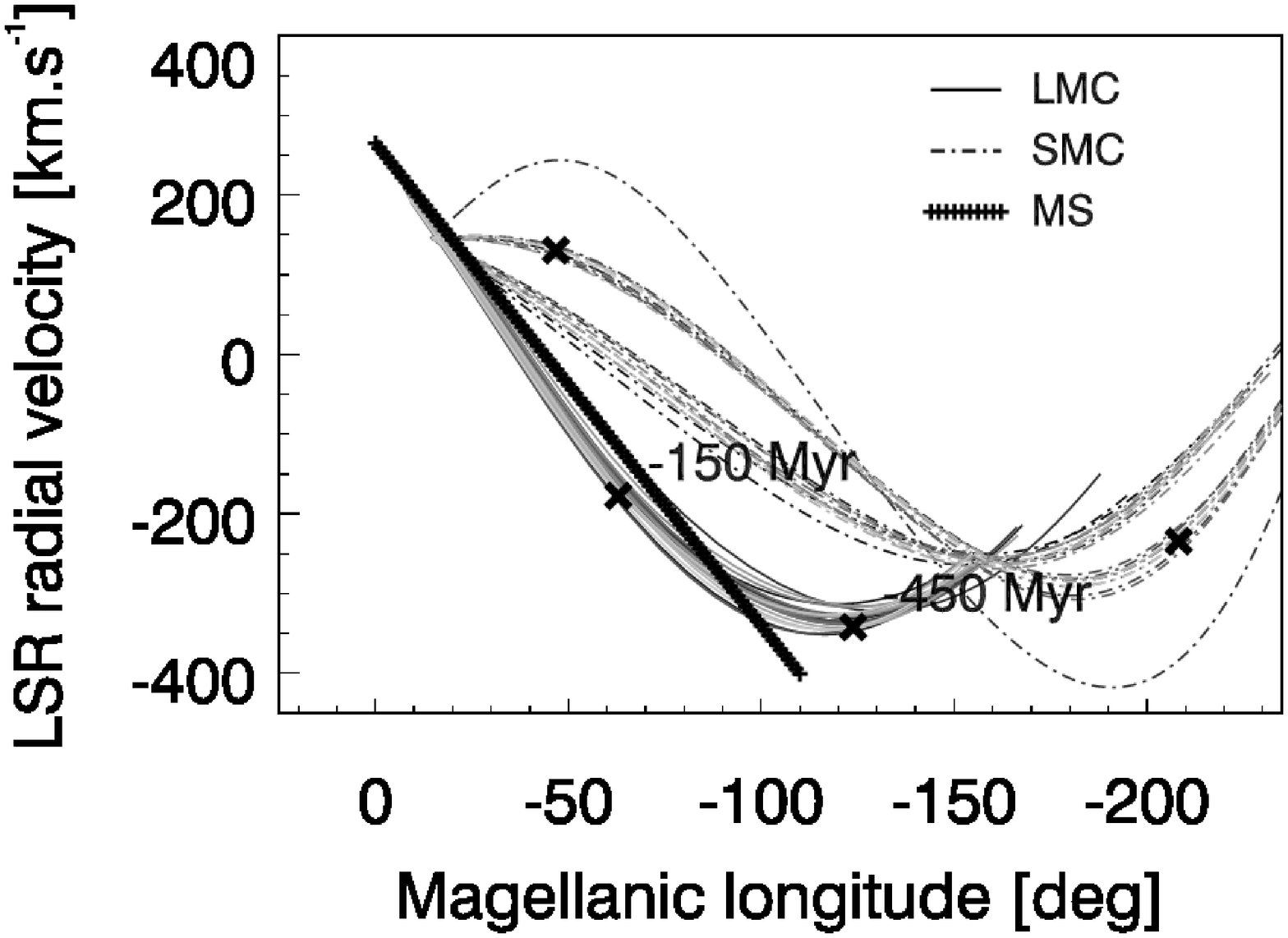}\\
\includegraphics[bb=40 35 510 690,angle=90,scale=.34, clip]{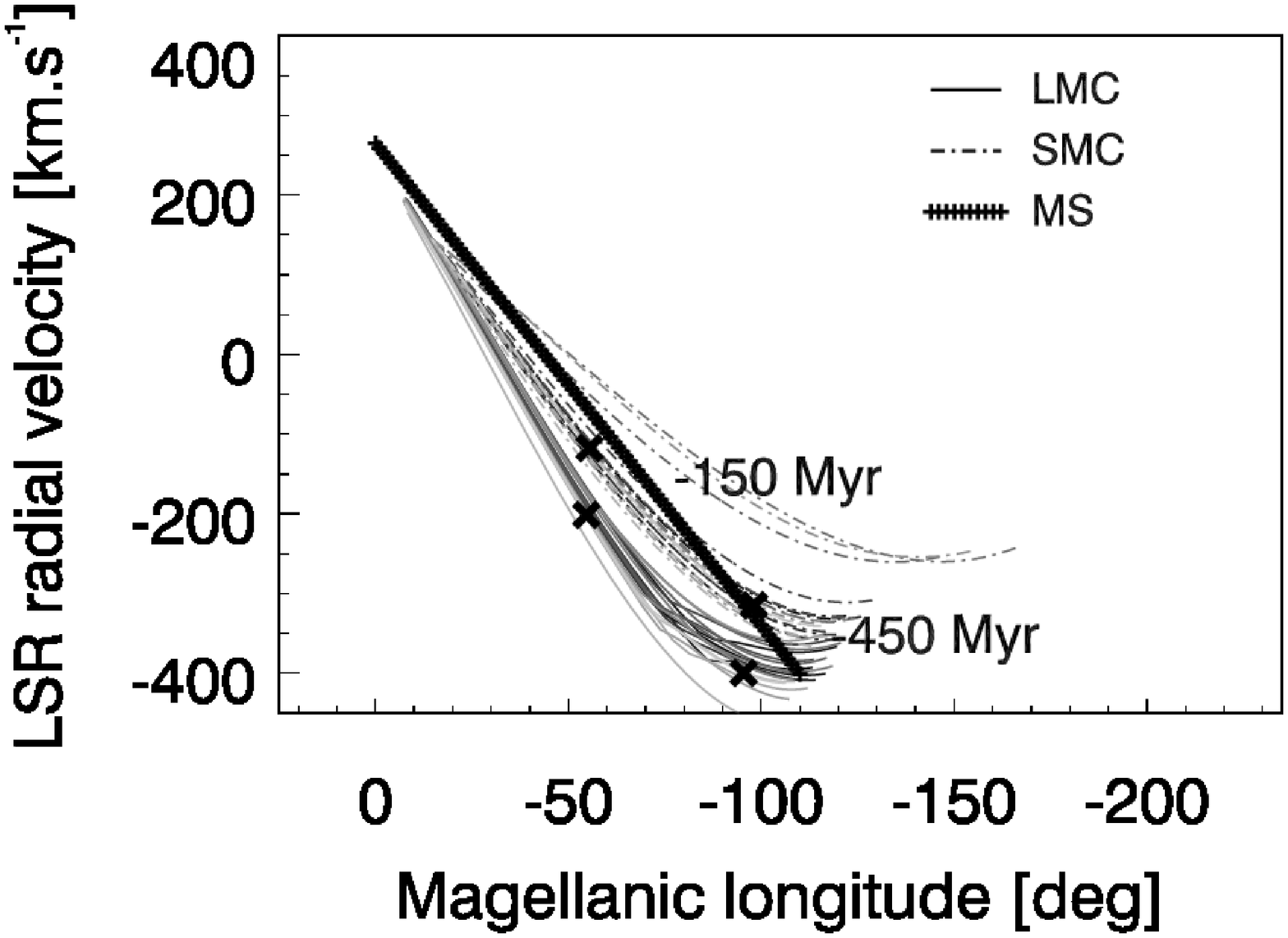}
\caption{LSR radial velocity profiles along the orbital tracks of the Magellanic
Clouds. The radial velocity is plotted as a function of the Magellanic longitude~\citep[see, e.g.][]{Bruens05}
for the models identified by the genetic algorithm. The best models of $f > 0.60$ (upper plot) and
the low--quality fits of $f < 0.50$ are shown, respectively. The past positions of the Magellanic Clouds
are marked at the times of $-150$\,Myr and $-450$\,Myr. The bold line stretched between
the Magellanic longitudes of $0^\circ$ and $-110^\circ$ corresponds to the observed
LSR radial velocity profile of the Magellanic Stream.
\label{orbits_radvel_maglong_700}}
\end{figure}
}
\clearpage

{\twocolumn
\begin{figure}
\includegraphics[bb=20 20 525 780,angle=90,scale=.28, clip]{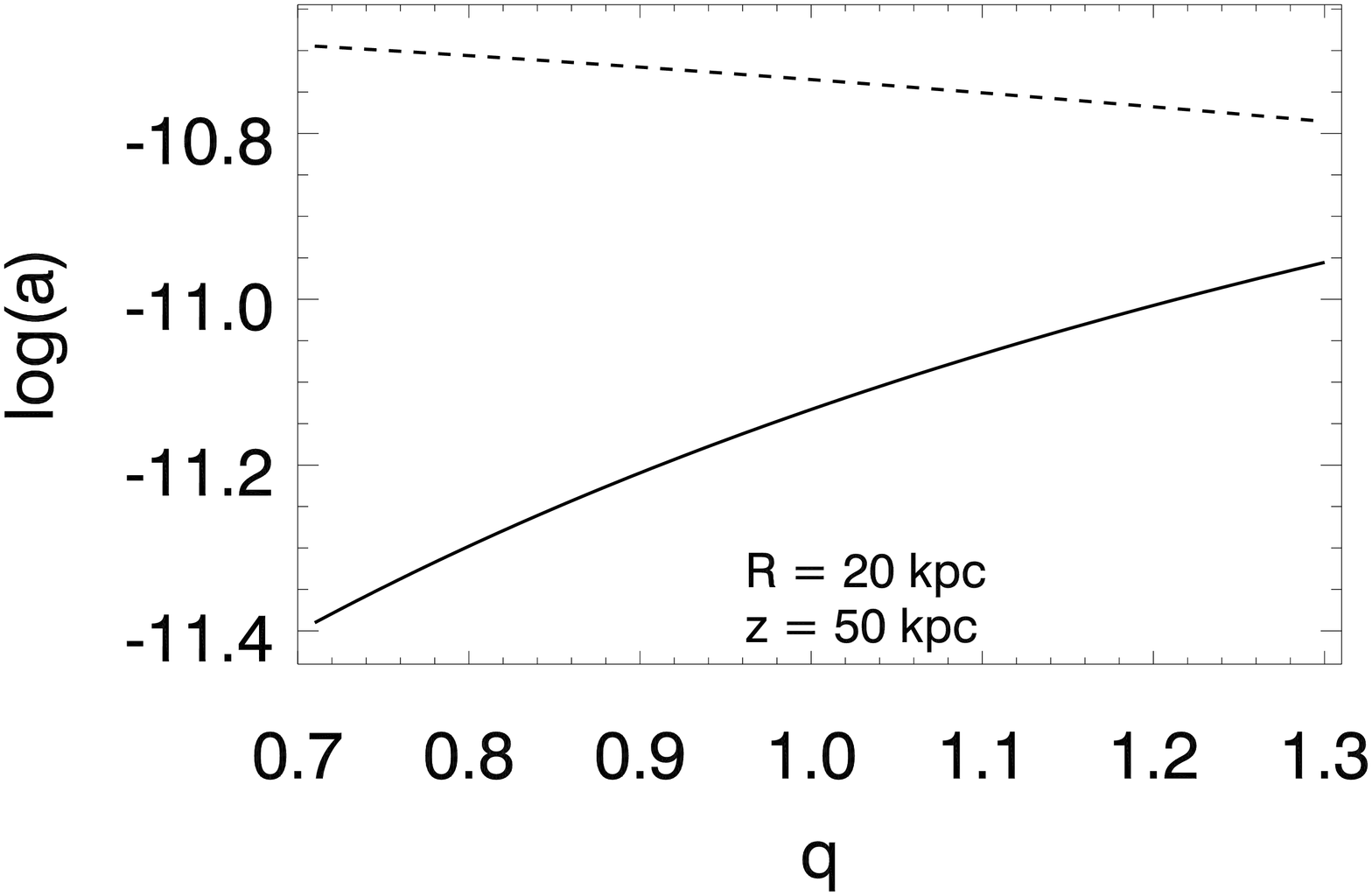}
\includegraphics[bb=20 20 525 780,angle=90,scale=.28, clip]{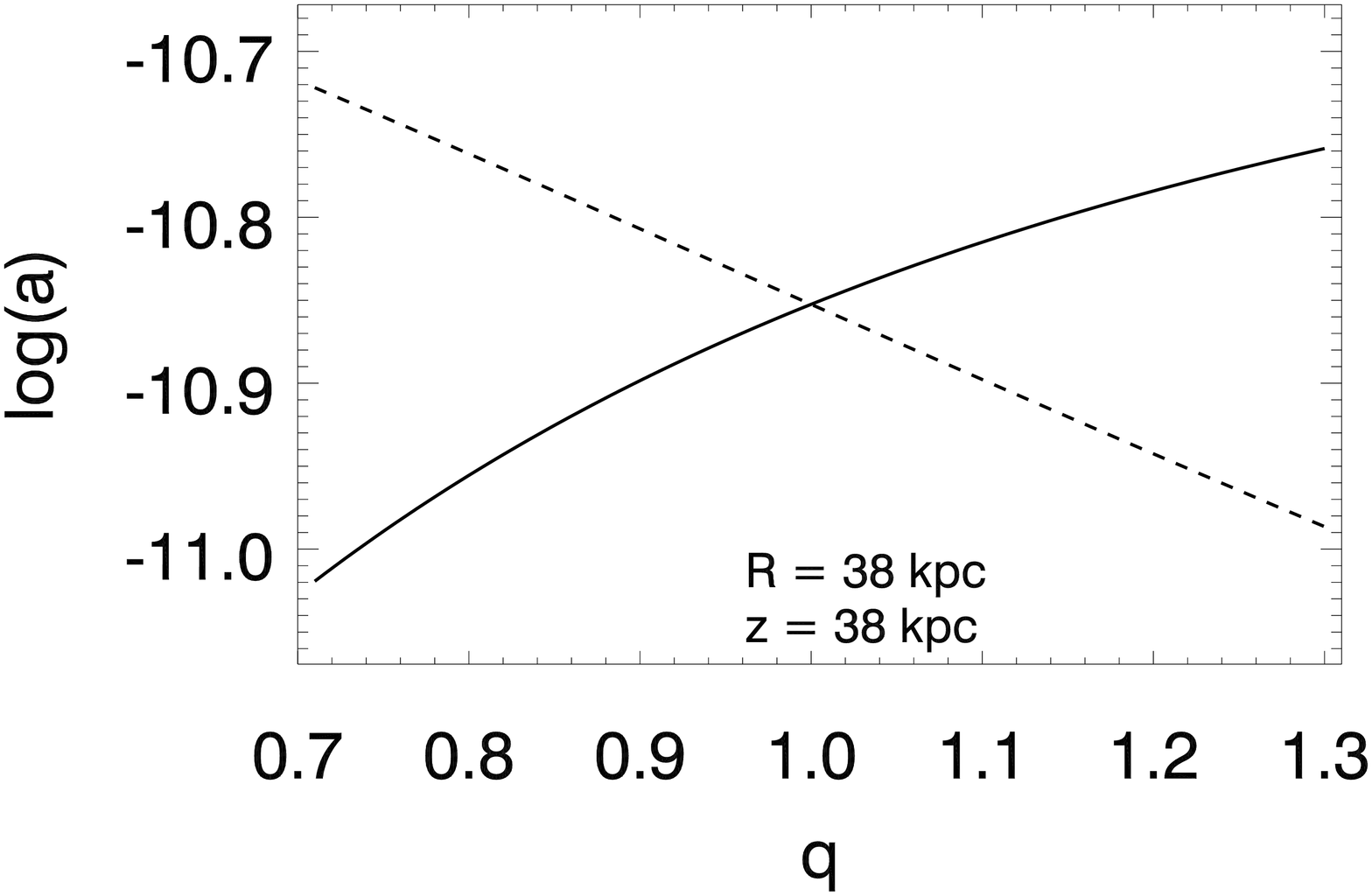}
\includegraphics[bb=20 20 525 780,angle=90,scale=.28, clip]{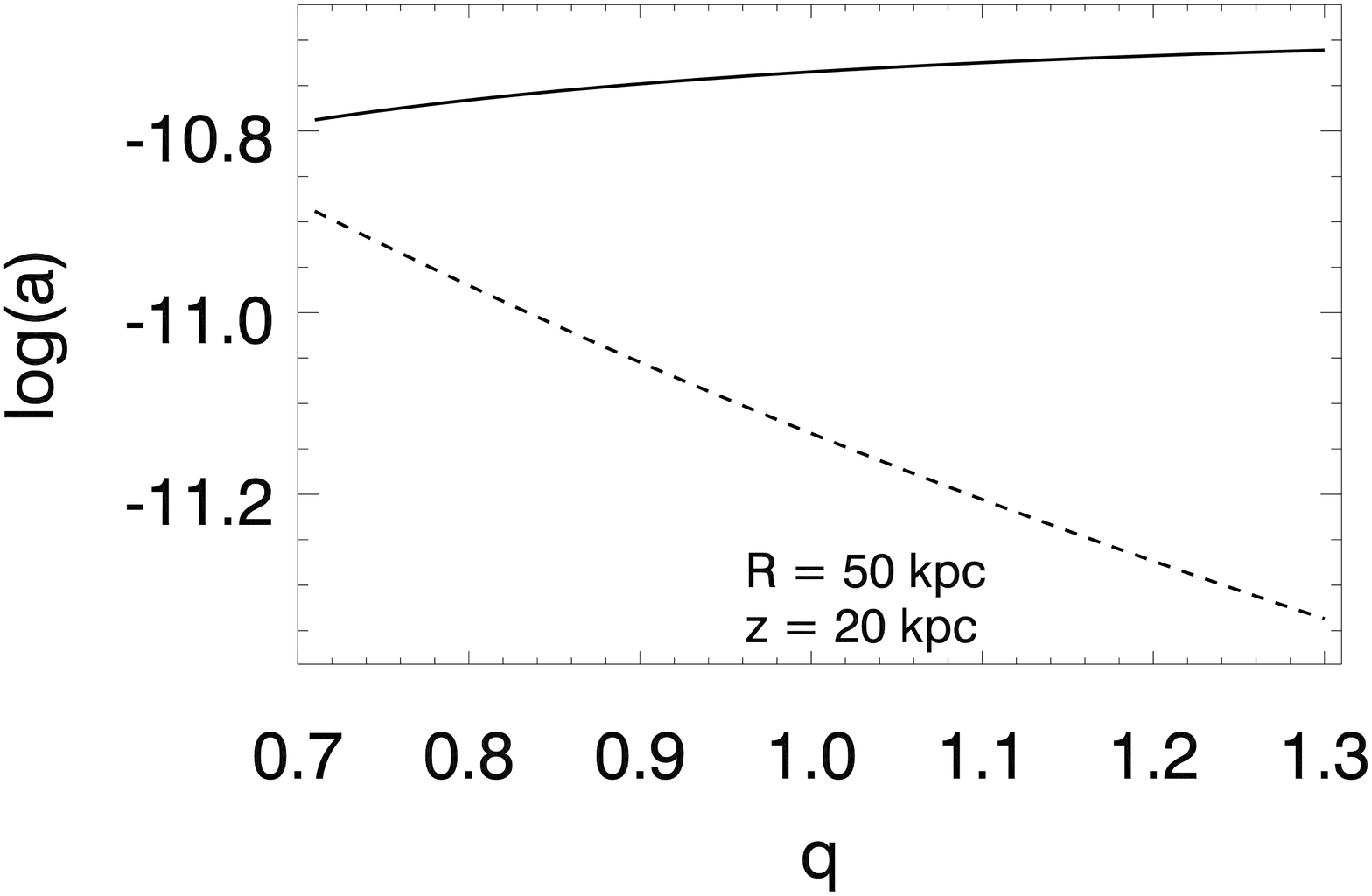}
\caption{Intensity of the gravitational field of the logarithmic halo. The radial (solid line)
and the $z$ components of the gravitational acceleration due to the axially symmetric
logarithmic halo of the Milky Way are plotted as functions of the halo flattening
parameter $q$. The values and the ratio of the components of the gravitational acceleration
depend on the position as well. The radial force increases if the halo flattening is increased
while the axial component of the gravitational field decreases in such a case.
\label{loghalo_force}}
\end{figure}
}
\clearpage

{\twocolumn
\begin{figure}
\includegraphics[bb=38 60 490 750,angle=90,scale=.31, clip]{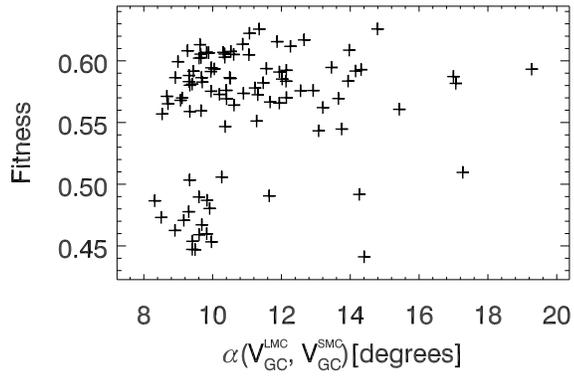}
\caption{Fitness of the models identified by our evolutionary optimizer is plotted as a function of the scalar product
of the current LMC and SMC velocity vectors. In most cases the deviation is smaller than 15$^\circ$.
\label{velocityscalarproduct}}
\end{figure}
}
\clearpage

{\twocolumn
\begin{figure}
\includegraphics[bb=40 15 570 660,angle=90,scale=.33, clip]{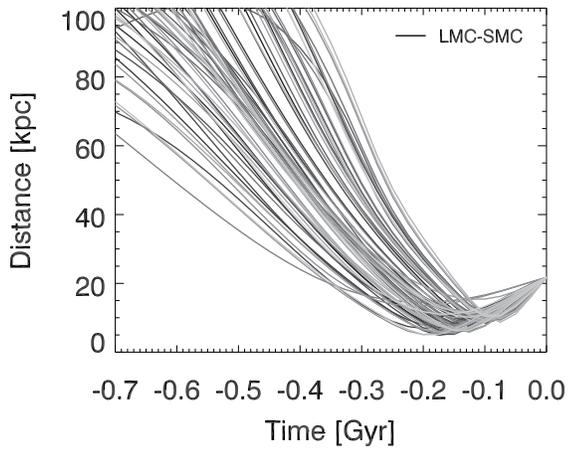}
\caption{Time dependence of the distance between the centers of mass of the Magellanic Clouds.
The function is plotted for the models of $f>0.55$ back to the time of $-0.7$\,Gyr. In all
cases the LMC--SMC distance is increasing at present implying a minimum of the distance
occurring in the past.
\label{reldist_700}}
\end{figure}
}
\clearpage

{\onecolumn
\begin{figure}
\includegraphics[bb=25 40 580 800,angle=90,scale=.31, clip]{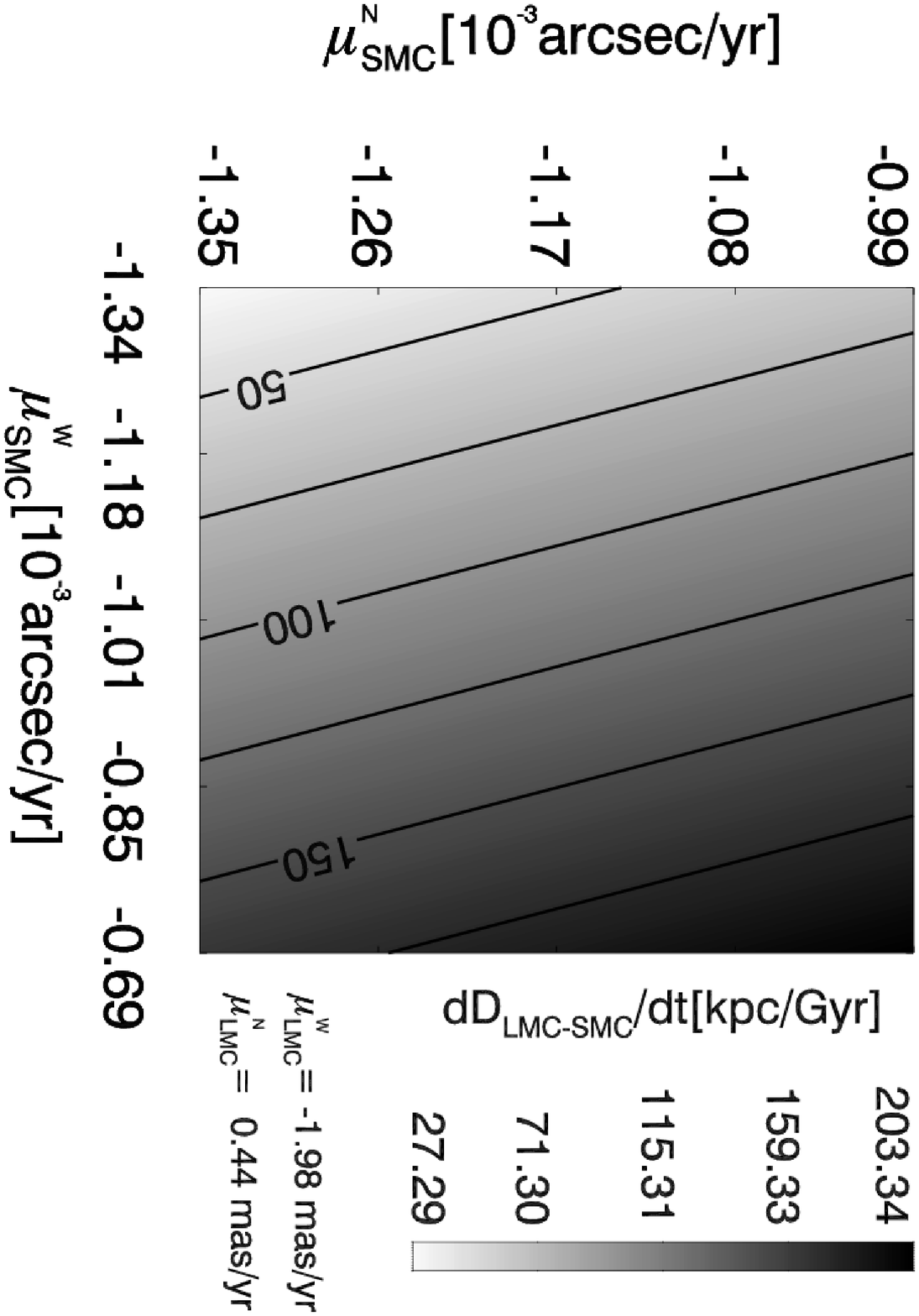}
\includegraphics[bb=25 40 580 800,angle=90,scale=.31, clip]{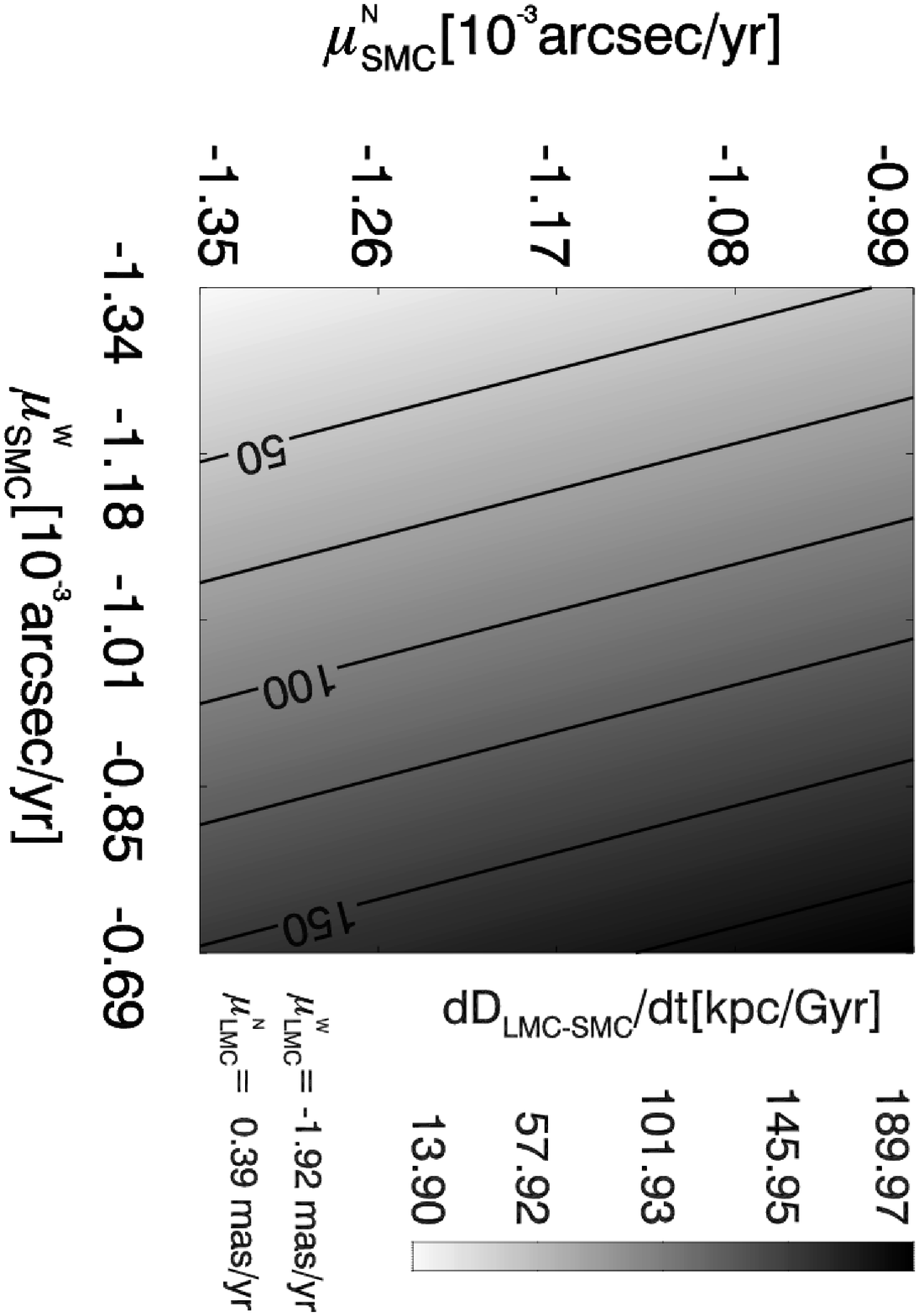}
\caption{Current rate at which the distance between the Magellanic Clouds changes with time. The function
$d D_\mathrm{l-s} / d t$ is plotted as a function of the
SMC proper motion components $\mu_\mathrm{W}$ and $\mu_\mathrm{N}$. The proper motion components of the LMC
correspond to those of a selected high--fidelity ($f>0.60$) model (left plot), and to the limiting values of the ranges
given by Equation~(\ref{hires_pm}), respectively. The minimum value for the western proper motion and the
maximum of the northern LMC proper motion component were chosen to achieve the minimum spatial velocity.
At present the rate $d D_\mathrm{l-s} / d t$ is positive. The distance between the Clouds is increasing.
\label{relvelmap}}
\end{figure}
}
\clearpage

{\twocolumn
\begin{figure}
\includegraphics[bb=40 15 570 660,angle=90,scale=.33, clip]{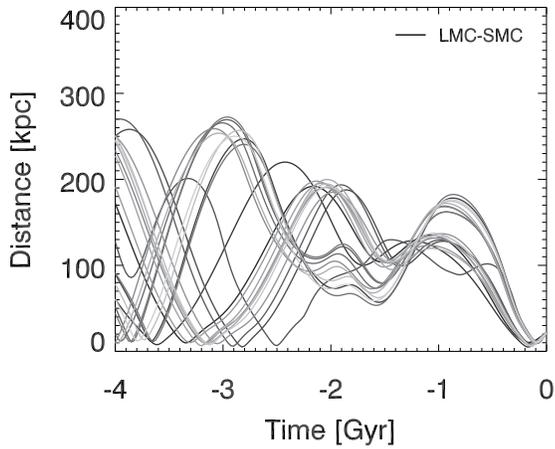}
\caption{Time dependence of the distance between the centers of mass of the Magellanic Clouds.
The function is plotted for the models of $f>0.60$ back to the time of $-4.0$\,Gyr. A close ($d \approx 15$\,kpc)
encounter between the LMC and the SMC occurred at $t<-2.5$\,Gyr in all cases despite
the fact that the Clouds cannot be considered gravitationally bound to each other until the recent past.
\label{reldist_4000}}
\end{figure}
}
\clearpage

{\onecolumn
\begin{figure}
\includegraphics[bb=30 0 555 730,angle=90,scale=.315, clip]{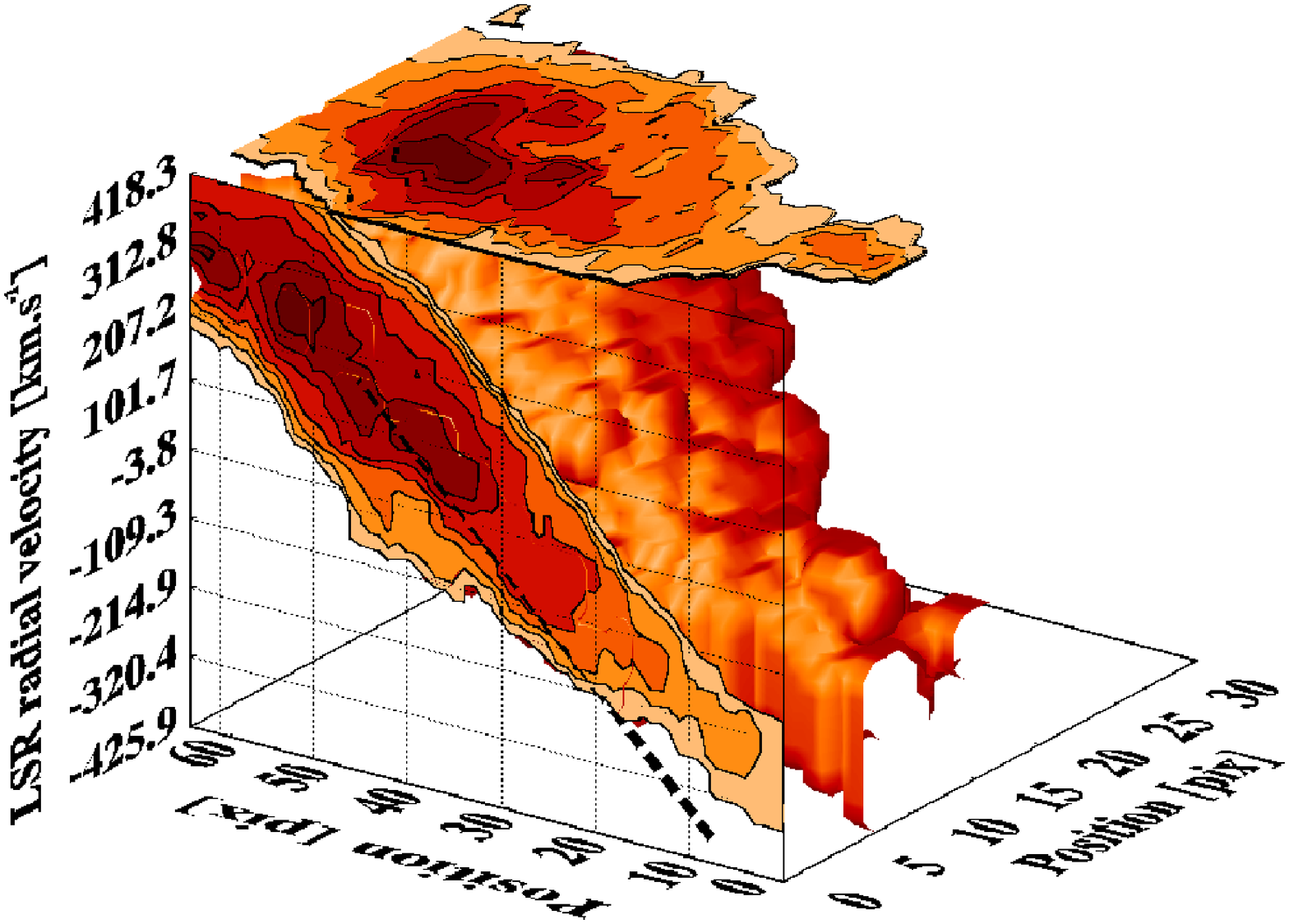}
\includegraphics[bb=30 0 555 730,angle=90,scale=.315, clip]{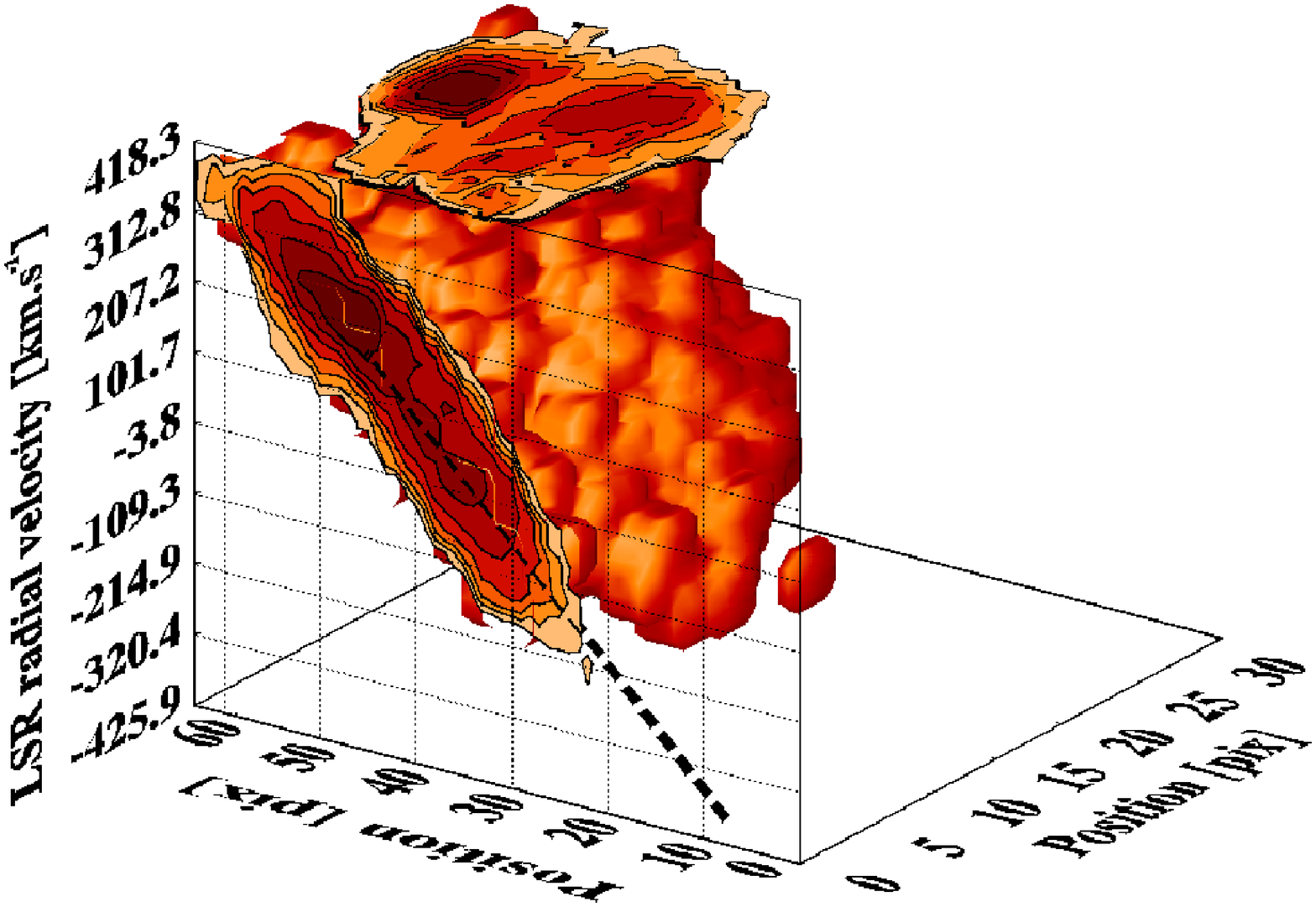}\\
\includegraphics[bb=35 20 550 765,angle=90,scale=.31, clip]{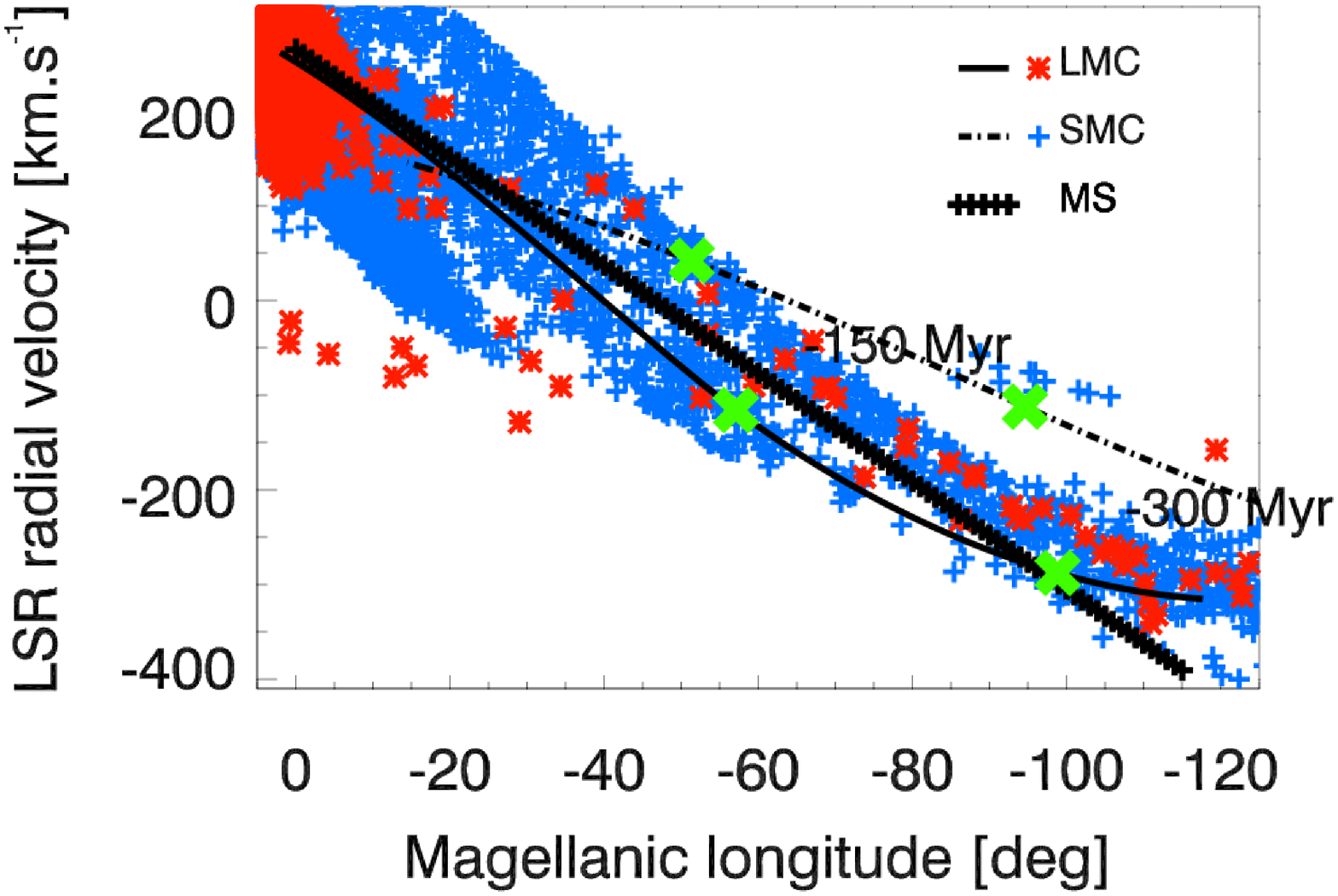}
\includegraphics[bb=35 20 550 765,angle=90,scale=.31, clip]{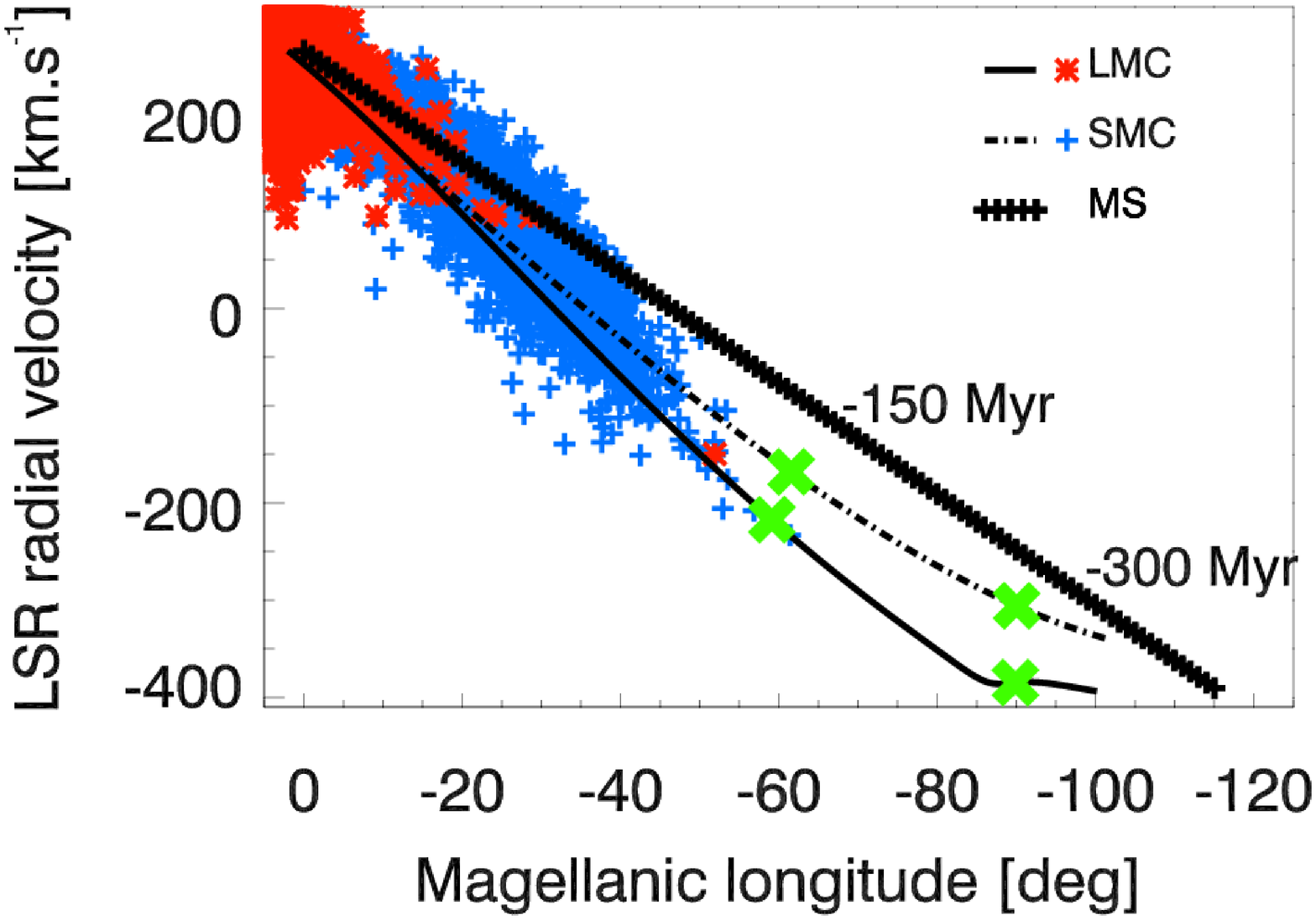}
\caption{The threshold of the fitness function. The plots on the left hand side present an example
of a satisfactory reproduction of the H\,I observations of the large--scale structures associated
with the Magellanic Clouds~\citep{Bruens05}. The plots in the right hand column illustrate the
typical output from a model for the LMC--SMC--Galaxy interaction that is not considered successful.
The upper row of this figure offers the simulated low--resolution 3\,D data--cube for both
models. The column density isosurface of $\Sigma = 10^{-4} \Sigma_\mathrm{max}$ is depicted,
together with the contour plots of the integrated relative column densities of H\,I
projected to the position--position and position--LSR radial velocity spaces, respectively.
The lower row provides a different view of the position versus LSR radial velocity dependence
for the modeled Magellanic Stream. Positions of all individual LMC/SMC particles are plotted.
Its LSR radial velocity profile is compared to
the mean profile of the observed Stream (thick black line). For the definition
of the Magellanic longitude, see~\cite{Wannier72}. Positions of the centers of mass
of the Clouds are indicated at the time of $-150$\,Myr and $-300$\,Myr, respectively.
\label{good_bad_model}}
\end{figure}
}
\clearpage

{\onecolumn
\begin{figure}
\includegraphics[bb=30 0 555 730,angle=90,scale=.3, clip]{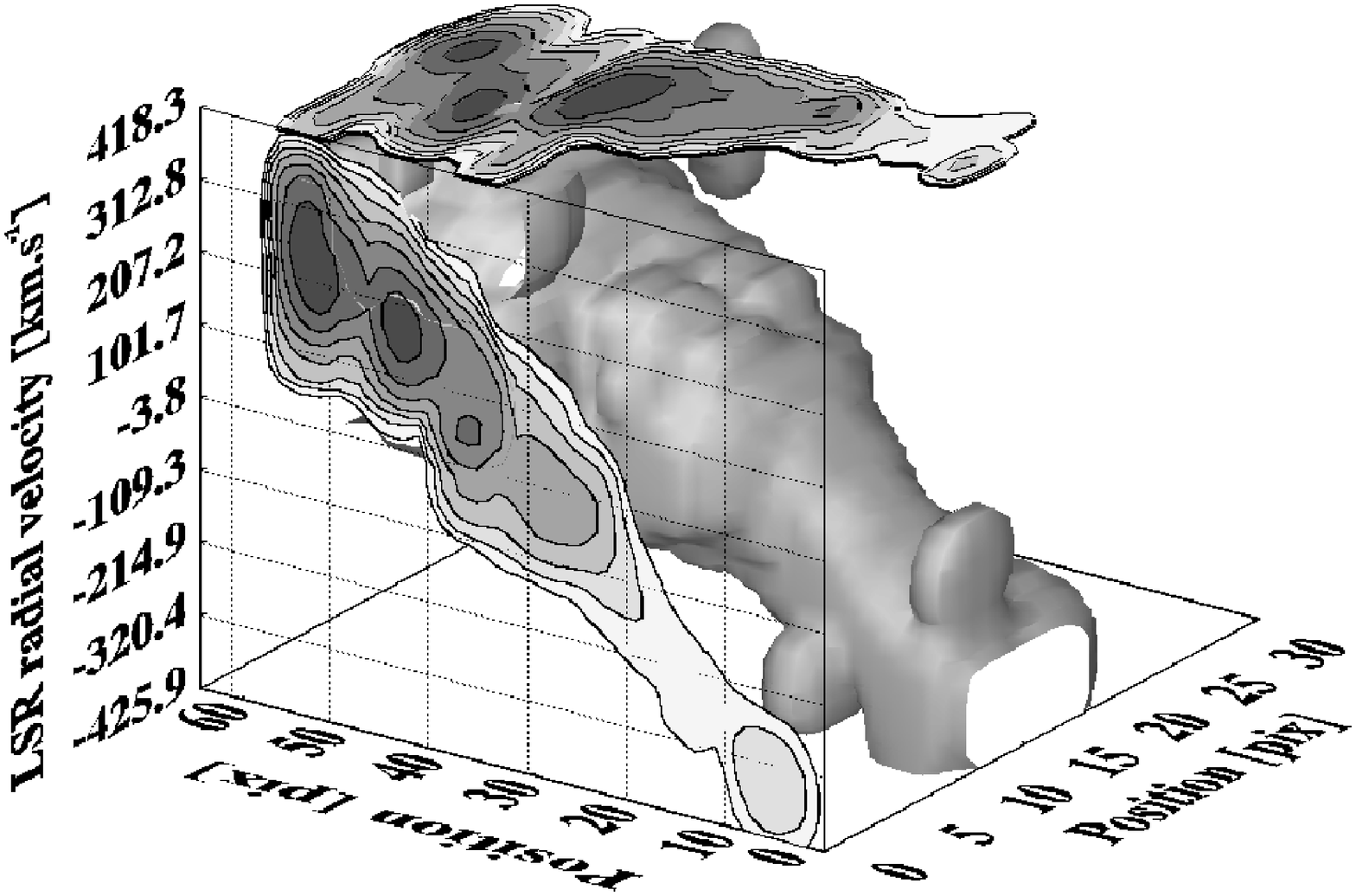}
\includegraphics[bb=30 0 555 730,angle=90,scale=.3, clip]{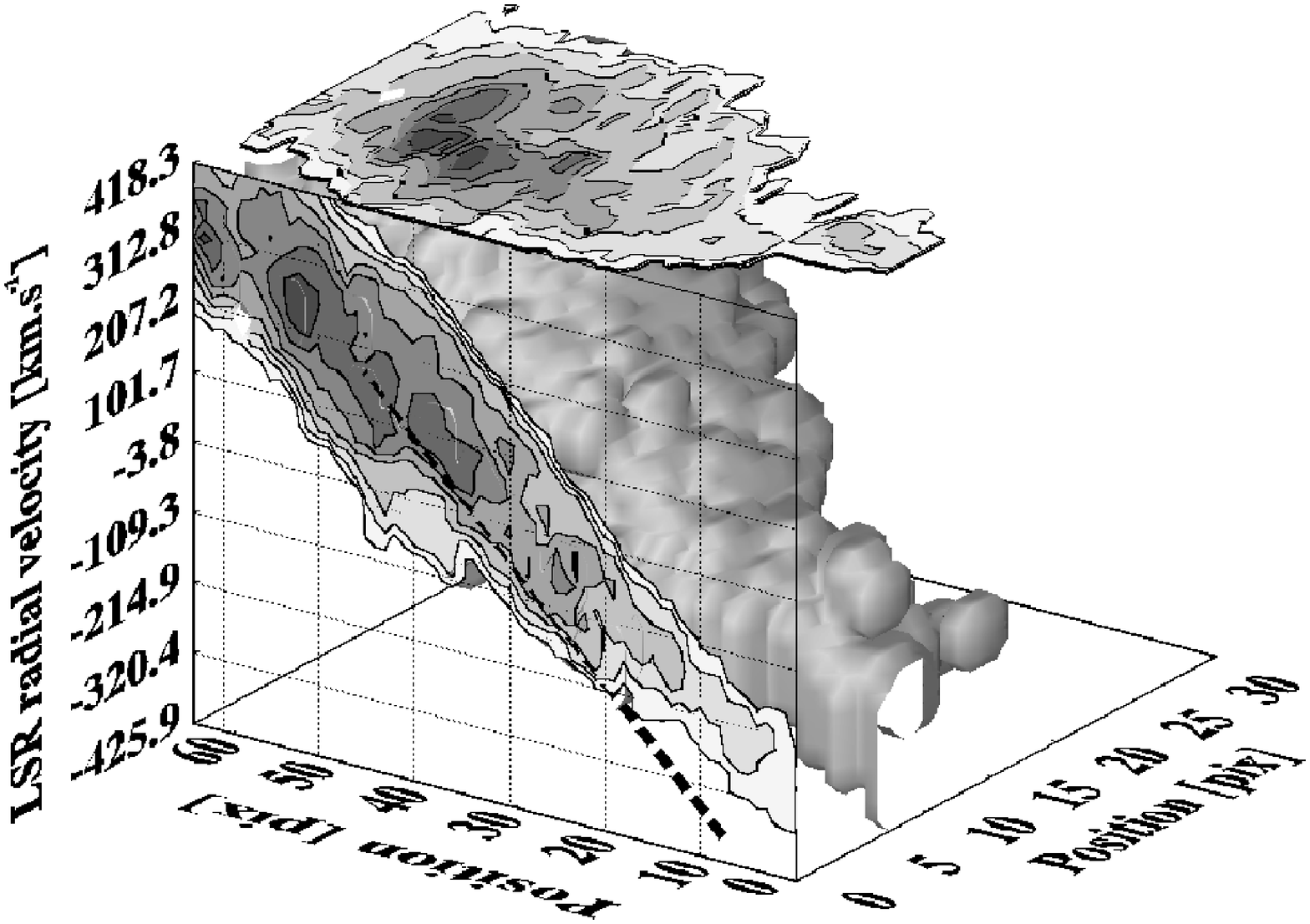}\\
\includegraphics[bb=40 50 550 715,angle=90,scale=.3, clip]{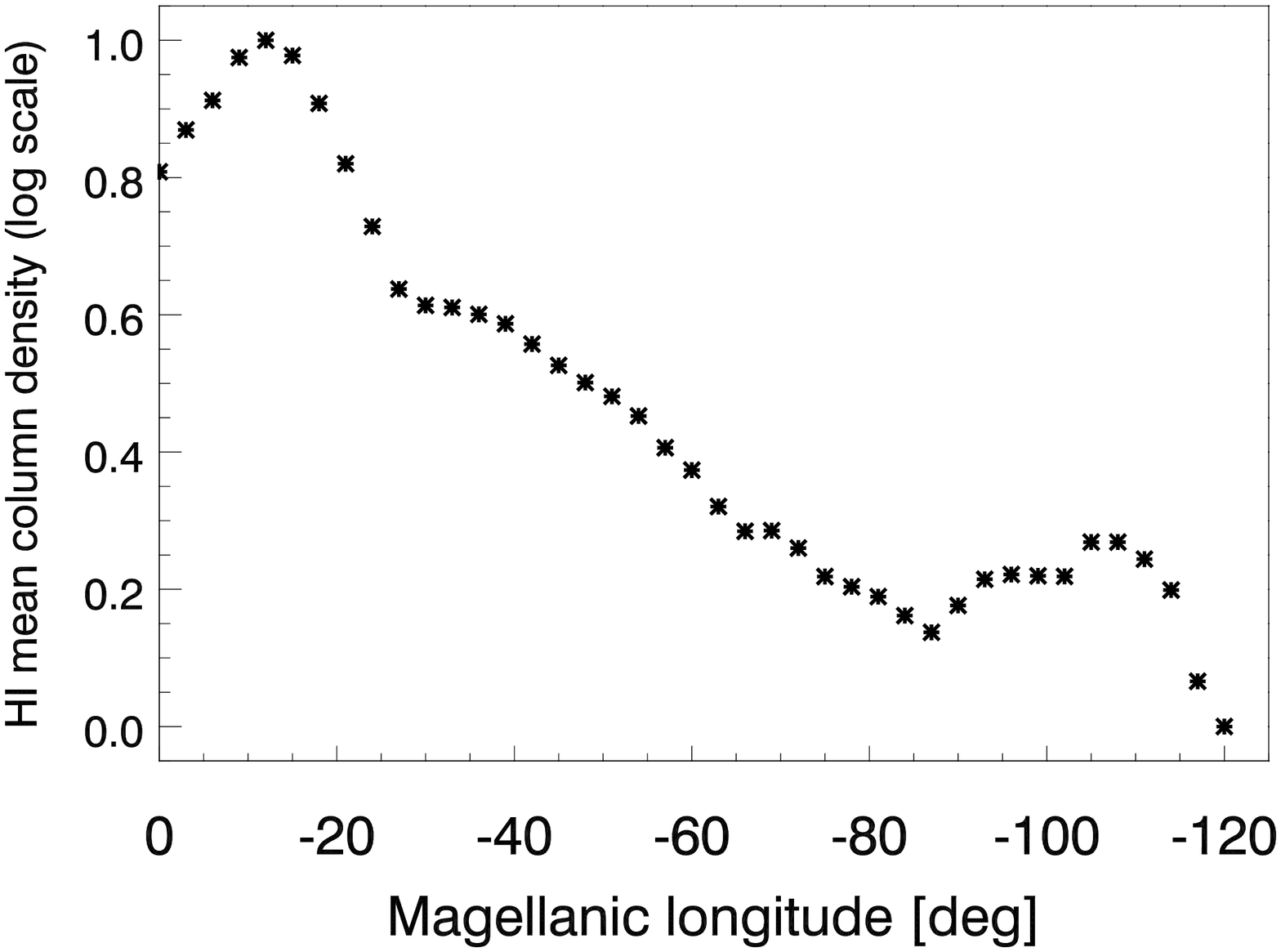}
\includegraphics[bb=40 50 550 715,angle=90,scale=.3, clip]{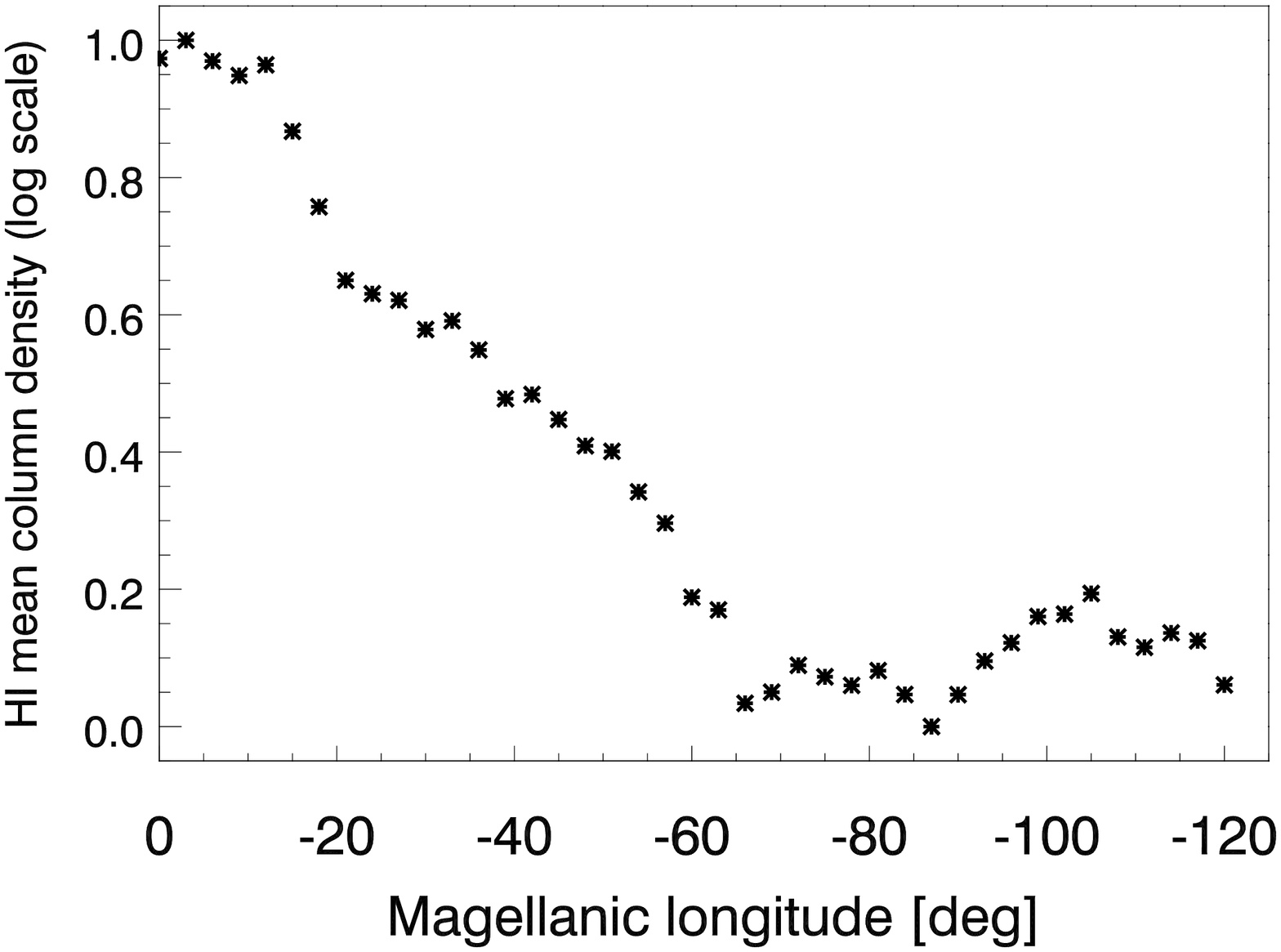}\\
\includegraphics[bb=20 40 620 675,angle=90,scale=.3, clip]{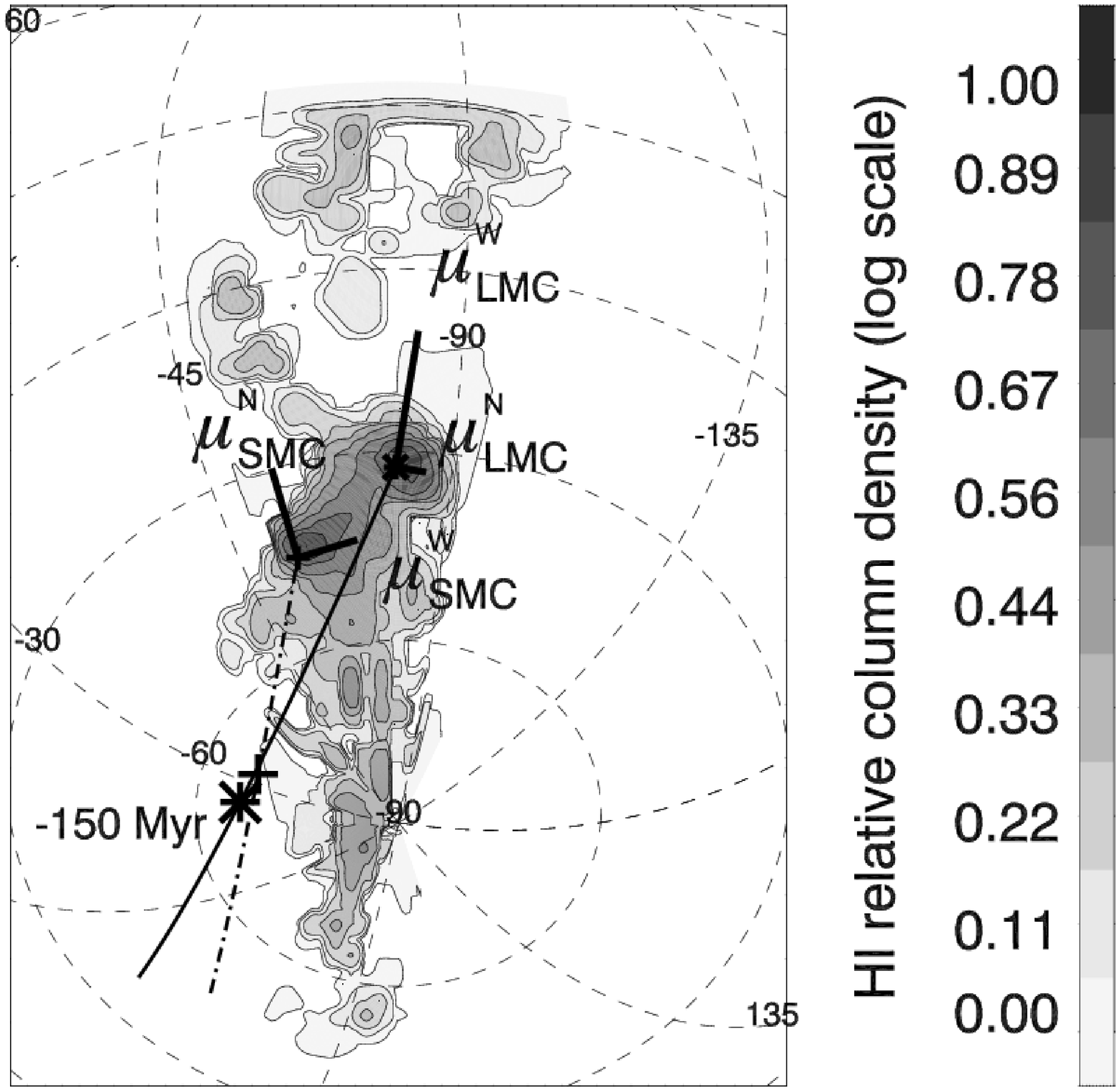}
\includegraphics[bb=20 40 620 675,angle=90,scale=.3, clip]{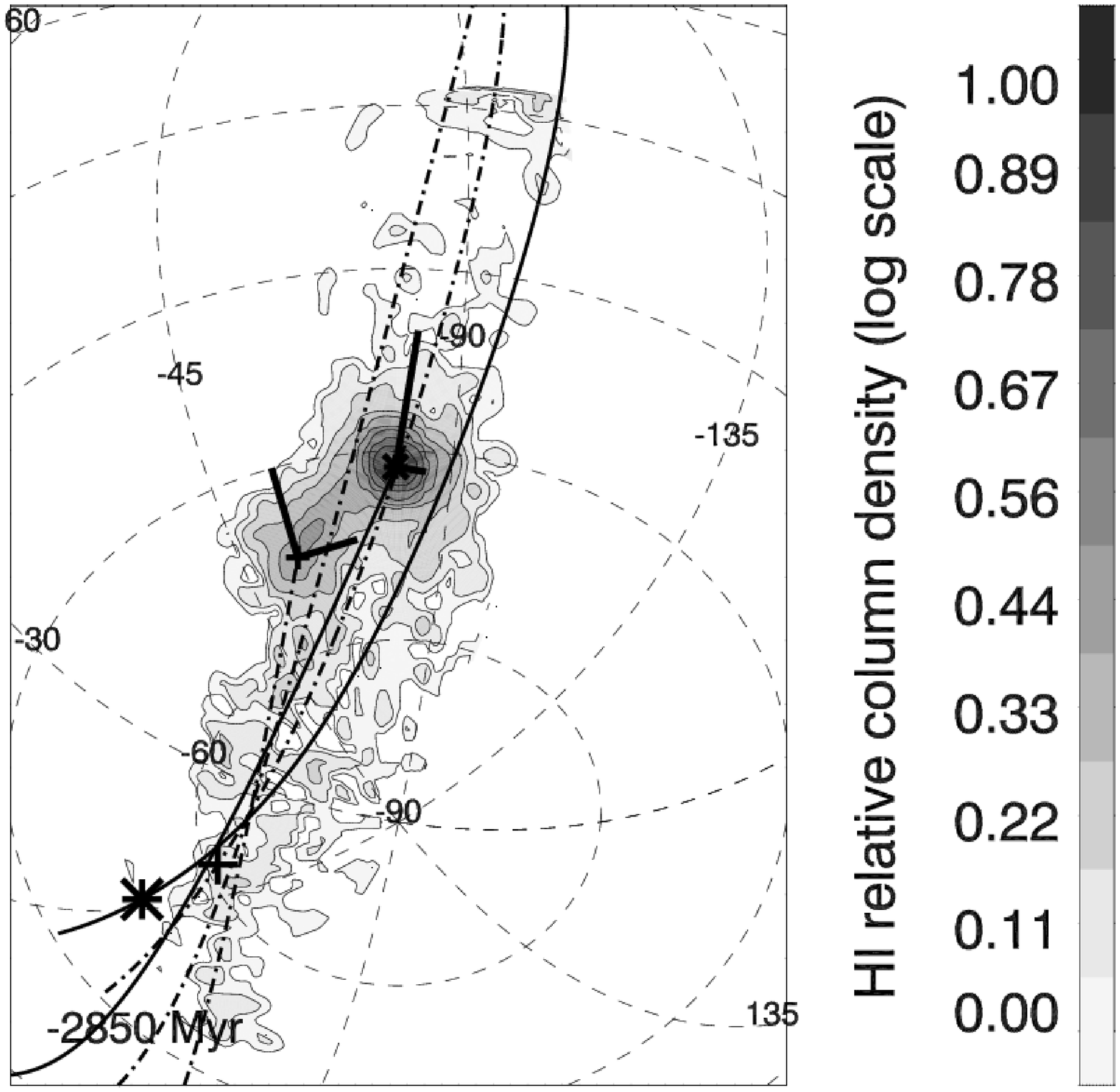}
\caption{Large--scale distribution of H\,I associated with the Magellanic Clouds -- observed versus simulated.
The comparison of the observed (left column) low--resolution H\,I data--cube for the extended Magellanic structures
(the Magellanic Stream, the Leading Arm) and of its modeled counterpart. The plots in the upper row
depict the H\,I column density isosurfaces of $\Sigma = 10^{-4} \Sigma_\mathrm{max}$,
and the contour plots of the integrated relative column densities of H\,I
projected to the position--position and position--LSR radial velocity spaces, respectively.
The middle row shows the mean column density of H\,I in the Stream as a function of the Magellanic longitude.
The third row offers a detailed view of the integrated H\,I column densities projected
to the plane of sky. The orbital tracks of the Clouds for the actual model are over--plotted over the last $300$\,Myr
(lower left plot) and over the last $3$\,Gyr, respectively.
The solid line corresponds to the past orbit of the LMC, while the dash--dotted line was used for the SMC.
The present western and norther proper motions for both the Clouds are indicated.
\label{model_0078}}
\end{figure}
}
\clearpage

{\twocolumn
\begin{figure}
\includegraphics[bb=10 60 495 710,angle=90,scale=.33, clip]{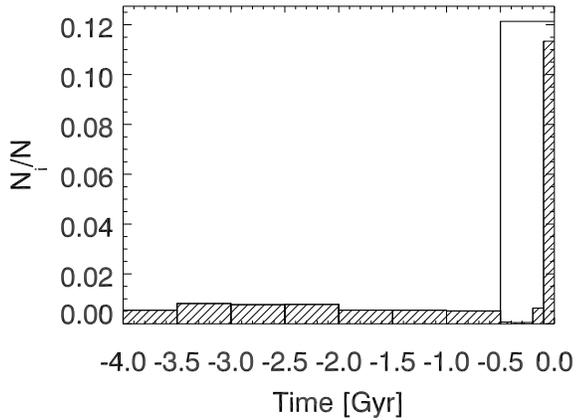}\\
\includegraphics[bb=25 135 610 655,angle=90,scale=.42, clip]{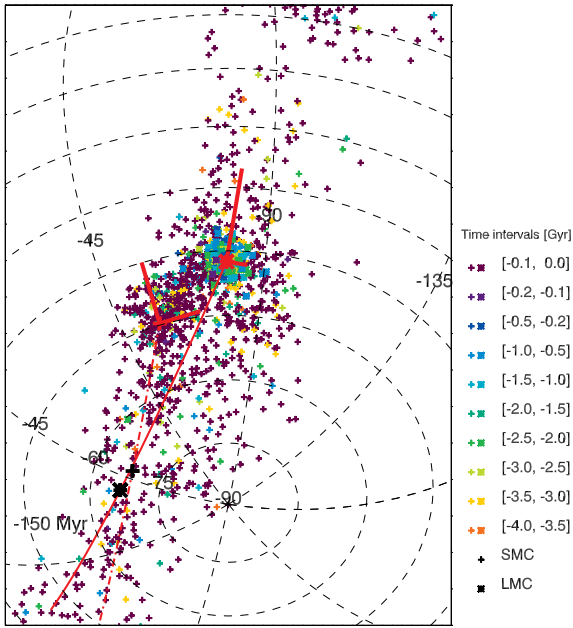}
\caption{Relative number of the LMC/SMC particles strongly disturbed due to the encounter
events in the LMC--SMC--Galaxy interaction for a high--fitness model of $f=0.61$
(see also Figure~\ref{model_0078}). The upper plot shows the relative counts of the
disturbed particles in eight bins of $500$\,Myr, starting at the time of $-4$\,Gyr. The last
bin is subdivided into five equal sections.
The lower plot depicts the present distribution of the LMC/SMC particles. They are color--coded according to the epoch when they were stripped
from the Clouds. The particles that remained bound to their galaxy of origin over the entire period of the simulation were not plotted.
The Magellanic Stream is a composition of filaments of different ages spread along the past orbits of the Magellanic Clouds.
\label{lostparticles}}
\end{figure}
}
\clearpage

{\twocolumn
\begin{figure}
\includegraphics[bb=20 10 550 815,angle=90,scale=.28, clip]{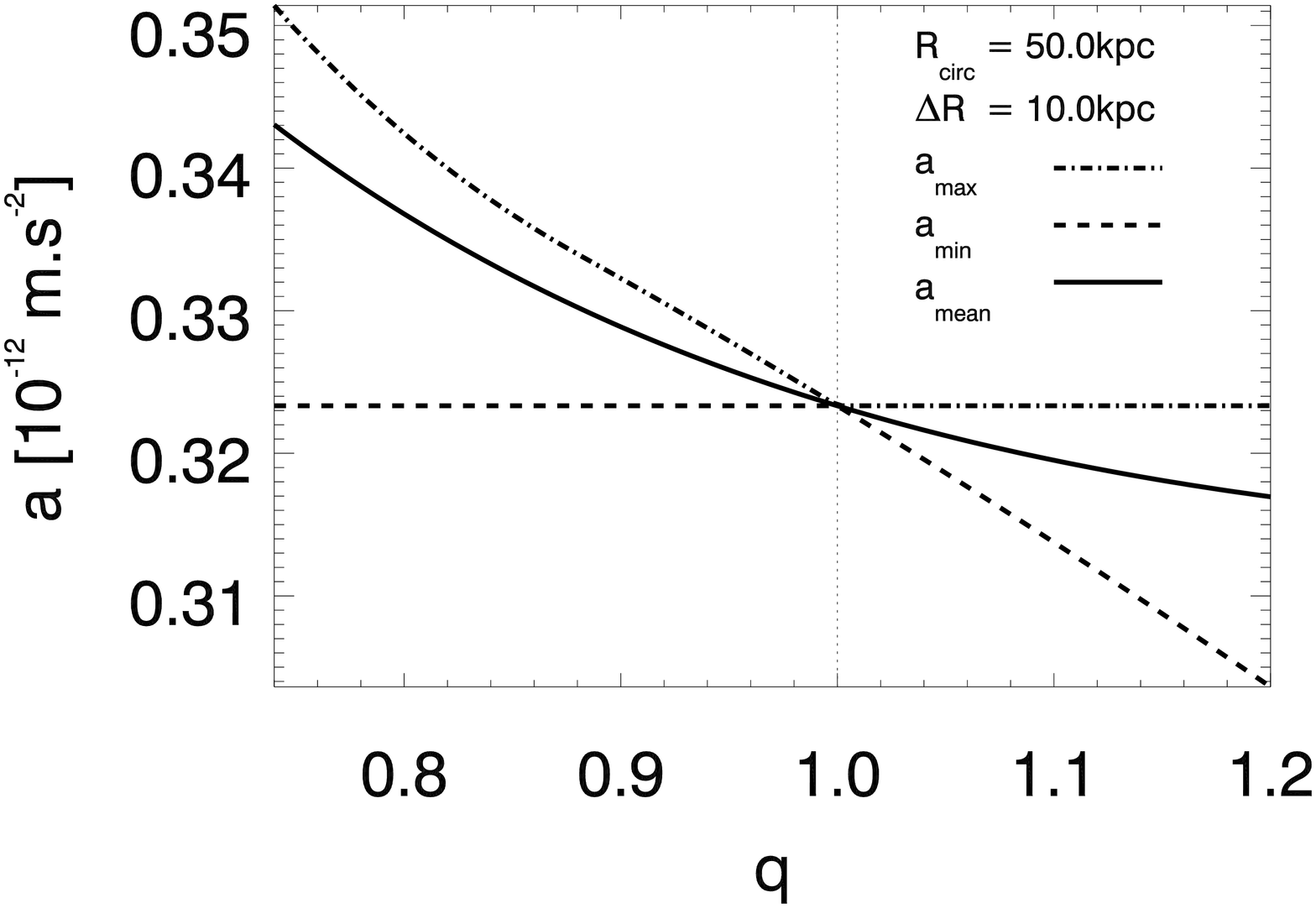}
\includegraphics[bb=20 10 550 815,angle=90,scale=.28, clip]{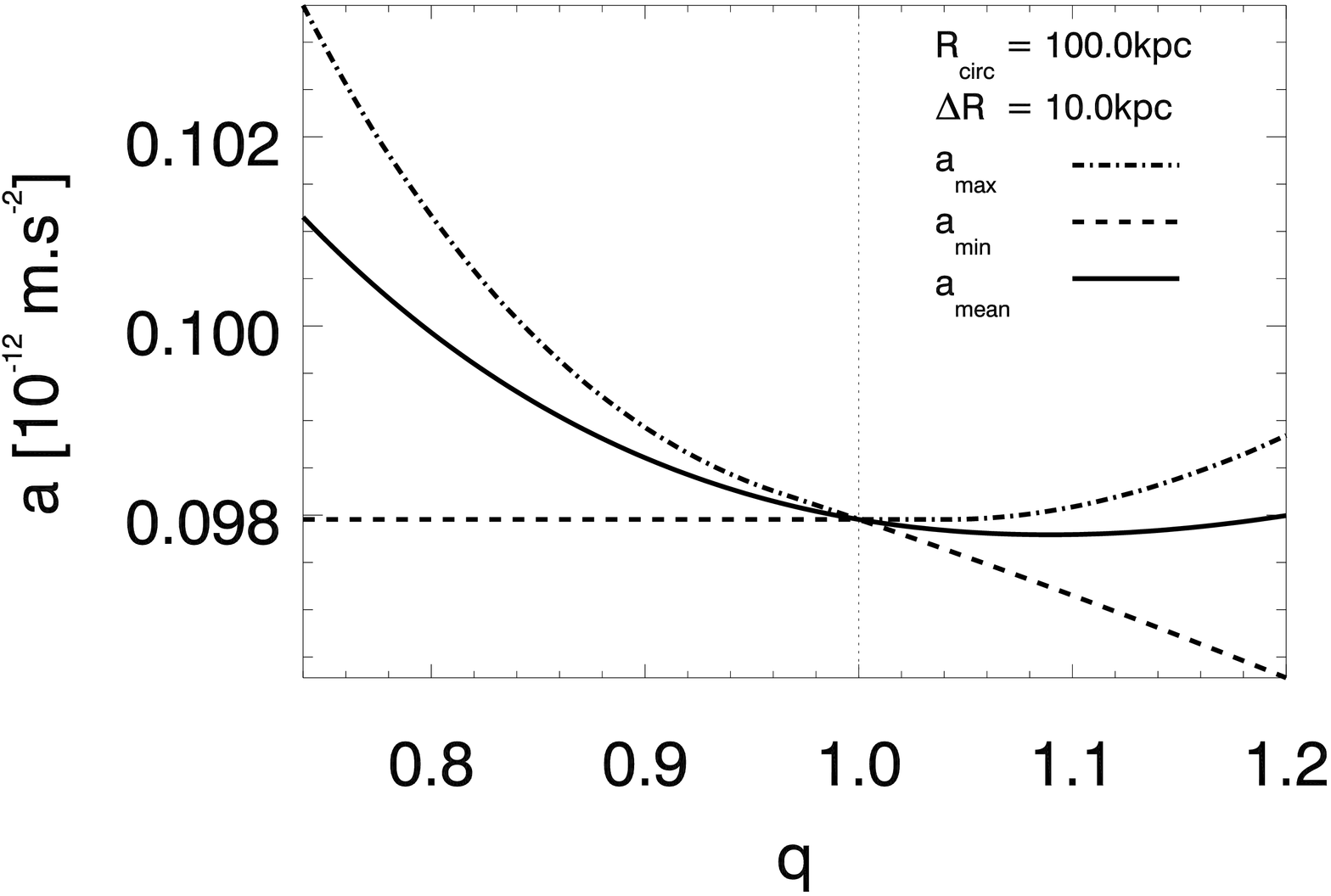}
\includegraphics[bb=20 10 550 815,angle=90,scale=.28, clip]{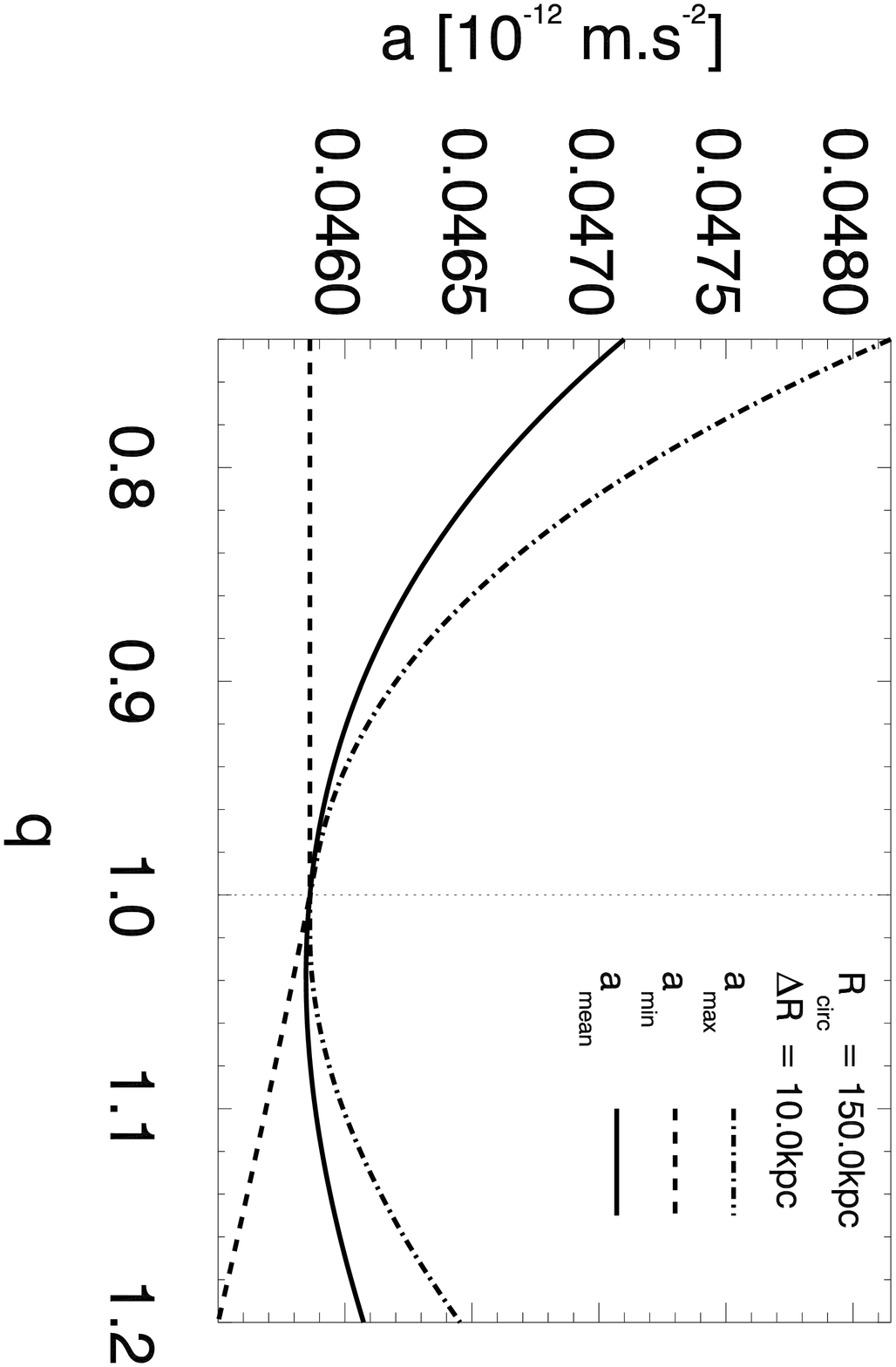}
\caption{Tidal acceleration for an axially symmetric logarithmic potential as a function of
its flattening parameter $q$. The values are calculated for two points on circular polar
orbits of a constant radial distance $\Delta r = 10$\,kpc. Three orbits
of the radii of 50\,kpc, 100\,kpc and 150\,kpc are shown, respectively.
The plots show the mean value, the maximum and also the minimum of the acceleration on the given orbits.
\label{loghalo_tides}}
\end{figure}
}
\clearpage


\end{document}